\documentclass[aps,prx,reprint,groupedaddress,longbibliography]{revtex4-1}

\usepackage{amsmath}
\usepackage{graphicx}
\usepackage{hyperref}
\usepackage{mathtools}
\usepackage{float}
\usepackage{stmaryrd}
\usepackage{algorithm}
\usepackage{algpseudocode}
\usepackage{wasysym}
\usepackage{bm}

\hypersetup{colorlinks,citecolor=blue,linkcolor=blue,urlcolor=blue}

\newcommand{\ket}[1]{\ensuremath{|#1\rangle}}

\newcommand{\bra}[1]{\ensuremath{\left\langle#1\right|}}
\newcommand{\tr}[1]{\ensuremath{\textrm{tr}\big(#1\big)}}
\renewcommand{\vec}[1]{\bm{#1}}

\DeclarePairedDelimiterX{\norm}[1]{\lVert}{\rVert}{#1}

\begin{document}

\title{Engineering Topological Models \\with a General-Purpose Symmetry-to-Hamiltonian Approach}

\author{Eli Chertkov, Benjamin Villalonga, and Bryan K. Clark}
\affiliation{Institute for Condensed Matter Theory and Department of Physics, University of Illinois at Urbana-Champaign, Urbana, Illinois 61801, USA}

\begin{abstract}
Symmetry is at the heart of modern physics.
Phases of matter are classified by symmetry breaking, topological phases are characterized by non-local symmetries, and point group symmetries are critical to our understanding of crystalline materials.
Symmetries could then be used as a criterion to engineer quantum systems with targeted properties.
Toward that end, we have developed a novel approach, the symmetric Hamiltonian construction (SHC), that takes as input symmetries, specified by integrals of motion or discrete symmetry transformations, and produces as output all local Hamiltonians consistent with these symmetries (see \href{https://github.com/ClarkResearchGroup/qosy}{github.com/ClarkResearchGroup/qosy} for our open-source code). This approach builds on the slow operator method [\href{http://dx.doi.org/10.1103/PhysRevE.92.012128}{PRE \textbf{92}, 012128}].
We use our new approach to construct new Hamiltonians for topological phases of matter.

Topological phases of matter are exotic quantum phases with potential applications in quantum computation.
In this work, we focus on two types of topological phases of matter: superconductors with Majorana zero modes and $Z_2$ quantum spin liquids. 
In our first application of the SHC approach, we analytically construct a large and highly tunable class of superconducting Hamiltonians with Majorana zero modes with a given targeted spatial distribution. 
This result lays the foundation for potential new experimental routes to realizing Majorana fermions. 
In our second application, we find new $Z_2$ spin liquid Hamiltonians on the square and kagome lattices.
These new Hamiltonians are not sums of commuting operators nor frustration-free and, when perturbed appropriately (in a way that preserves their $Z_2$ spin liquid behavior), exhibit level-spacing statistics that suggest non-integrability.
This result demonstrates how our approach can automatically generate new spin liquid Hamiltonians with interesting properties not often seen in solvable models.
\end{abstract}

\maketitle

\section{Introduction}

Symmetry is central to our understanding of the phases of matter seen in nature. Many phase transitions, such as those of liquids, magnets, or superconductors, can be described by the spontaneous breaking of symmetry according to Landau's theory \cite{Landau1958}. Moreover, the existence of space group and point group symmetries in crystalline phases of matter highly influences the formation of order in these systems. Even topological phases of matter, exotic quantum phases that are notable for their lack of order, have non-trivial topological symmetries that give rise to their exotic properties. 

We propose a new approach for studying quantum phases of matter based on symmetries.
Our new approach, the symmetric Hamiltonian construction (SHC), is an algorithm that takes as input symmetries and produces as output Hamiltonians that obey those symmetries.
The SHC is an example of an inverse method, a method for generating models from data.
Inverse methods are widely used throughout machine learning, such as in deep learning \cite{LeCun2015}. In physics, they have been used in classical systems to design interaction potentials that stabilize crystalline and magnetic order \cite{inverseopt,DiStasio2013,Marcotte2013,Chertkov2016} and in quantum systems to design or reconstruct Hamiltonians from eigenstates or density matrices \cite{Chertkov2018,Qi2019,Greiter2018,Bairey2019,Bairey2019_2} as well as to build single-body Hamiltonians compatible with a given symmetry group \cite{Varjas2018}.
The SHC algorithm extends ideas developed in the slow operator method \cite{Kim2015} and is quite general.
The symmetries provided as input can be either integrals of motion, which can generate continuous symmetries, or discrete symmetry transformations.
Example symmetries include particle number conservation, $SU(2)$ symmetry, and point group symmetries, as well as more exotic topological symmetries such as those that we consider in this work.
The Hamiltonians produced as output can be interacting as well as non-interacting; and can be made to commute or anticommute with the input symmetry operators.
Our numerical implementation of the SHC is publicly available as the QOSY: Quantum Operators from Symmetry \texttt{Python} package \cite{qosy}.

In this work, we use the SHC to construct new Hamiltonians for two topological systems: superconductors with Majorana zero modes and $Z_2$ quantum spin liquids.  We engineer superconducting Hamiltonians that commute with Majorana zero mode operators whose spatial distributions are specified as input.  Separately, we construct multiple new interacting $Z_2$ quantum spin liquid Hamiltonians on different lattice geometries that commute with topologically non-trivial Wilson loop operators.

Majorana zero modes are integrals of motion that occur in certain superconductors that exhibit the statistics of Majorana fermions \cite{DasSarma2015}, an exotic class of fermions that are their own antiparticles \cite{Majorana1937}.
In addition to their importance to fundamental physics, Majorana fermions, which are non-Abelian anyons, have potential applications as the building blocks for qubits in fault-tolerant quantum computers \cite{Nayak2008,DasSarma2015}.
Many experiments have attempted to observe Majorana fermions by engineering particular superconducting Hamiltonians, such as the Kitaev chain \cite{Kitaev2001}, that are theoretically known to host Majorana zero modes \cite{Beenakker2013}.
To expedite the experimental search for Majorana fermions, it is desirable to expand the small list of superconducting Hamiltonians known to possess Majorana zero modes.
For this reason, we apply the SHC to design new examples of such Hamiltonians.
We successfully construct a large family of local, highly tunable superconducting Hamiltonians that commute with Majorana zero modes that can be distributed arbitrarily in space.
Many of these Hamiltonians have the potential to be realized in experiment.

Quantum spin liquids are exotic magnets in which spins do not order even at zero temperature due to quantum fluctuations \cite{Savary2017}.
Gapped quantum spin liquids are topologically ordered, meaning that they exhibit anyonic quasiparticles, ground state degeneracy that depends on the topology of the underlying lattice, and non-local symmetries.
In particular, gapped $Z_2$ spin liquids host Abelian anyons and a particular set of symmetries known as Wilson loops, non-local loop operators with non-trivial topological properties.
Despite significant study, there are few known model Hamiltonians that exhibit the physics of $Z_2$ quantum spin liquids. 
Some exactly solvable Hamiltonians, such as the toric code \cite{Kitaev2003} and related models \cite{Gottesman1997,Kitaev1998,Kitaev2002,Kitaev2003,Levin2005,Bullock2007}, are often sums of commuting operators while other candidate $Z_2$ spin liquid Hamiltonians \cite{Yan2011,Carrasquilla2015} are difficult to solve numerically.  

% % Roger Melko paper, kagome, S+S+ interaction, kagome Heisenberg, mention
Using the SHC, we find new families of $Z_2$ spin liquid Hamiltonians for spins on the square and kagome lattices. 
The Hamiltonians that we discover are not sums of commuting operators, are not frustration-free, and can possess local and non-local integrals of motion.
We find that these Hamiltonians, perturbed in an appropriate way, exhibit GOE level-spacing statistics in particular quantum number sectors, suggesting that they could be non-integrable.
These models provide new, interesting examples of $Z_2$ topological order in spin systems.

The remainder of this paper is organized as follows. 
Section~\ref{sec:methods} describes the SHC method.
Section~\ref{sec:mzmresults} details our construction of superconducting Hamiltonians that commute with Majorana zero modes.
Section~\ref{sec:z2spinliquids} describes the new $Z_2$ spin liquids that we found by searching for Hamiltonians that commute with Wilson loops.
We conclude in Section~\ref{sec:conclusions}.

\section{The symmetric Hamiltonian construction method} \label{sec:methods}

\begin{figure}
    \begin{center}
    \includegraphics[width=0.45\textwidth]{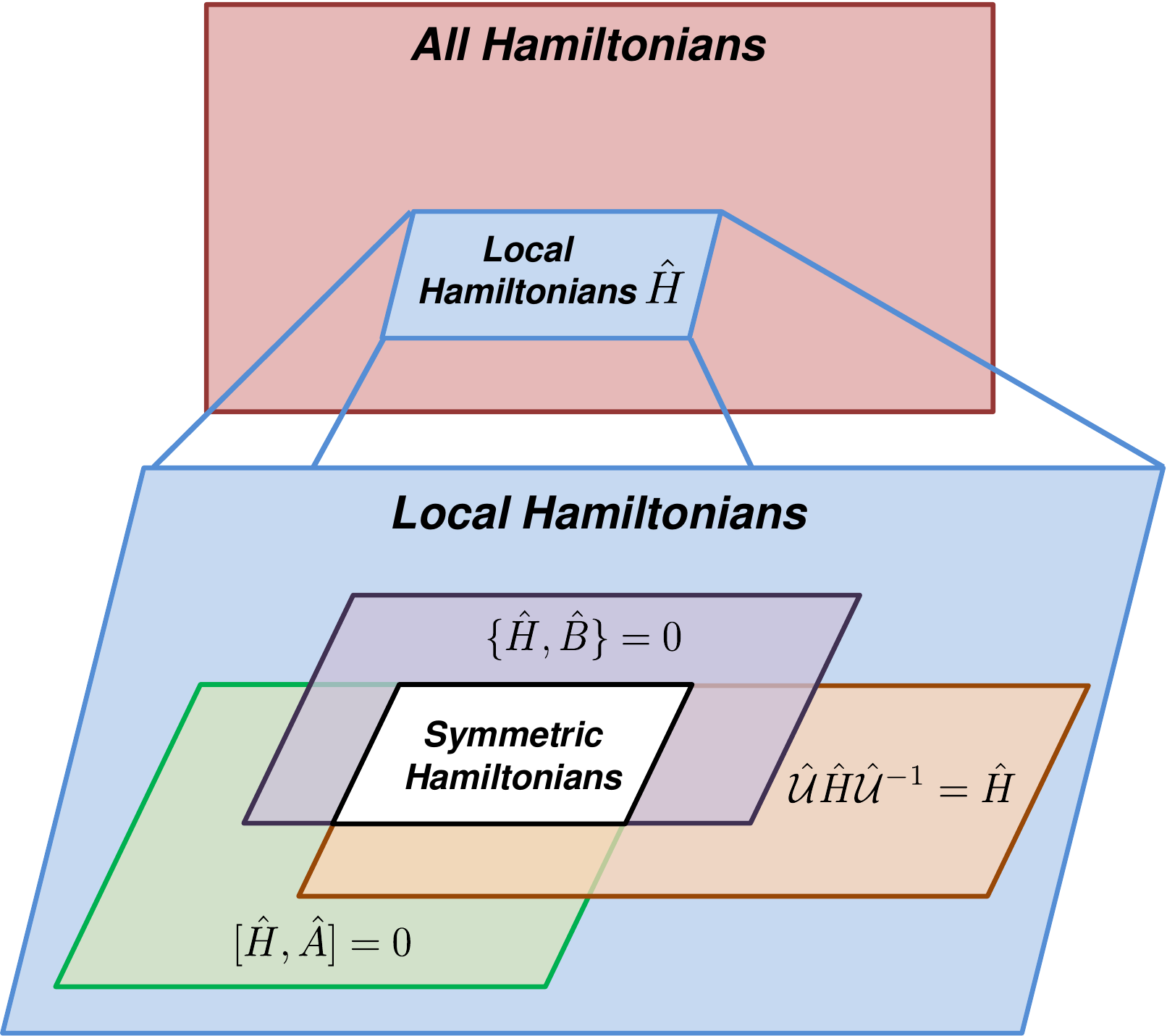} 
    \end{center}
    \caption{The set of all possible Hamiltonians is a large vector space (shown in red). We are interested in a small subspace of this space that consists of local Hamiltonians (shown in blue). Local Hamiltonians with particular symmetries, such as Hamiltonians that commute with the integral of motion $\hat{A}$ or anti-commute with the antisymmetry $\hat{B}$ or are invariant under the symmetry transformation $\hat{\mathcal{U}}$, are also vector spaces (shown in green, purple, and orange, respectively). The symmetric Hamiltonian construction takes as input a list of symmetries, such as $\hat{A},\hat{B},$ and $\hat{\mathcal{U}}$, and produces as output the space of local symmetric Hamiltonians that obey all of these symmetries (shown in white).}
    \label{fig:hamiltonianspaces}
\end{figure}

In this section, we describe the symmetric Hamiltonian construction (SHC) procedure for generating Hamiltonians with desired symmetries (see Fig.~\ref{fig:hamiltonianspaces}), which includes integrals of motion (Sec.~\ref{sec:iom}) and symmetry transformations (Sec.~\ref{sec:invariant}). In Sec.~\ref{sec:superoperators}, we describe how these calculations can be interpreted as finding the ``ground states'' of super-operator ``Hamiltonians.''  

Our numerical implementation of these methods is publicly available as the QOSY: Quantum Operators from Symmetry \texttt{Python} package \cite{qosy}, whose features are briefly discussed in Appendix~\ref{sec:qosy}.

\subsection{Constructing Hamiltonians with desired integrals of motion}
\label{sec:iom}

Here we present our method for constructing Hamiltonians with a desired integral of motion, i.e., a Hermitian operator $\hat{\mathcal{O}}$ that commutes with the Hamiltonian $[\hat{H},\hat{\mathcal{O}}]=0$. This method takes in as input an integral of motion $\hat{\mathcal{O}}$ and produces as output Hamiltonians $\hat{H}$ that commute with $\hat{\mathcal{O}}$. Our \emph{inverse} method for finding these symmetric Hamiltonians is based on the slow operator method \cite{Kim2015}, a forward method for finding integrals of motion from Hamiltonians.

In the inverse method, our goal is to find a Hamiltonian that minimizes the norm of the commutator
\begin{align}
\varepsilon \equiv \norm{[\hat{H}, \hat{\mathcal{O}}]}^2 \label{eq:comNormSlowOp}
\end{align}
subject to the constraint that $\norm{\hat{H}}=1$ (which can always be achieved with a trivial normalization), where $\norm{\hat{\mathcal{O}}}^2 \equiv  \tr{\hat{\mathcal{O}}^\dagger \hat{\mathcal{O}}}/\tr{\hat{I}}$ is the Frobenius norm and $\hat{I}$ is the identity operator. 

In the slow operator forward method, the goal is to find \emph{integrals of motion} that minimize Eq.~(\ref{eq:comNormSlowOp}) for a given Hamiltonian. Numerically, this has been done using exact diagonalization \cite{Kim2015,OBrien2016,Lin2017,Pancotti2018}, matrix product operators \cite{Pancotti2018,Kells2018}, and other tensor networks \cite{Wahl2019}. In these contexts, the Hamiltonians studied were non-integrable models and the minimal commutator norms $\varepsilon$ discovered were often small, but not exactly zero, making the optimized operators approximate integrals of motion. Generically, we expect the inverse problem to be easier than the forward one. This is because integrals of motion can be highly non-local while physical Hamiltonians should be local. Since local operators make up a much smaller space of operators, it is much easier to search for local Hamiltonians than non-local integrals of motion. In Section~\ref{sec:z2spinliquids}, for particular $\hat{\mathcal{O}}$, we are able to efficiently find Hamiltonians for which $\varepsilon \approx 10^{-16}$ in finite-size systems.

Instead of representing the operators $\hat{H}$ and $\hat{\mathcal{O}}$ in Eq.~(\ref{eq:comNormSlowOp}) as matrices or tensor networks, we find it useful to expand \emph{both} of these operators in a basis of \emph{operator strings} $\hat{\mathcal{S}}_1,\ldots,\hat{\mathcal{S}}_d$ that span a $d$-dimensional space of Hermitian operators. In particular, we consider three different types of operator strings, Pauli strings, Fermion strings, and Majorana strings, to represent all possible spin-$1/2$ and fermionic operators.

For a system of $n$ spin-$1/2$ qubits, we define a $4^n$-dimensional basis of \emph{Pauli strings}:
\begin{align}
\hat{\mathcal{S}}_a = \hat{\sigma}_{1}^{t_1} \otimes \cdots \otimes \hat{\sigma}_{n}^{t_n} \label{eq:pauliStringBasis}
\end{align}
where $a=1,\ldots,4^n$ is a unique index for the operator, $t_i \in \{0,1,2,3\}$, $\hat{\sigma}^{0}=\hat{I}$, and $\hat{\sigma}^1,\hat{\sigma}^2,\hat{\sigma}^3$ are Pauli matrices.

For a system of $n$ fermions, we likewise define a $4^n$-dimensional basis of \emph{Fermion strings}, which come in three types:
\begin{align}
\hat{\mathcal{S}}_a = 
\begin{cases}
\hat{c}^\dagger_{i_1} \cdots \hat{c}^\dagger_{i_m} \hat{c}_{i_m} \cdots \hat{c}_{i_1}, \\
\hat{c}^\dagger_{i_1} \cdots \hat{c}^\dagger_{i_m} \hat{c}_{j_l} \cdots \hat{c}_{j_1} + \textrm{\textrm{H.c.}},\\
i\hat{c}^\dagger_{i_1} \cdots \hat{c}^\dagger_{i_m} \hat{c}_{j_l} \cdots \hat{c}_{j_1} + \textrm{H.c.}, 
\end{cases}
\label{eq:fermionStringBasis}
\end{align}
where $\hat{c}^\dagger_j$ and $\hat{c}_j$ are fermionic creation and anhillation operators, $1 \leq i_1 < \ldots < i_m \leq n, 1\leq j_1 < \ldots \leq j_l \leq n$, $0 \leq l \leq m \leq n$, and the indices are lexicographically ordered so that $(j_1,\ldots,j_l) < (i_1,\ldots,i_m)$.

Finally, for fermions, we also define a $4^n$-dimensional basis of \emph{Majorana strings}:
\begin{align}
\hat{\mathcal{S}}_a = i^{\sigma_a} \hat{\tau}_1^{t_1}\cdots \hat{\tau}_n^{t_n} \label{eq:majoranaStringBasis}
\end{align}
where $t_i \in \{0,1,2,3\}$,  $(\hat{\tau}_i^0,\hat{\tau}_i^1,\hat{\tau}_i^2,\hat{\tau}_i^3)=(\hat{I},\hat{a}_i,\hat{b}_i,\hat{d}_i)$, and $\sigma_a \in \{0,1\}$ is chosen to make $\hat{\mathcal{S}}_a$ Hermitian. The operators $\hat{a}_i \equiv \hat{c}_i + \hat{c}_i^\dagger, \hat{b}_i \equiv -i(\hat{c}_i-\hat{c}^\dagger_i)$ are Majorana fermions and $\hat{d}_i \equiv -i\hat{a}_i\hat{b}_i = \hat{I} - 2\hat{c}^\dagger_i \hat{c}_i$ is a fermion parity operator. A discussion of the properties of these three operator string bases is provided in Appendices~\ref{sec:pauli_basis},~\ref{sec:fermion_basis},~and~\ref{sec:majorana_basis}.

We perform our SHC calculations directly in a basis of operator strings. The Hamiltonian (unknown) and integral of motion (known) can be written in terms of operator strings as $\hat{H} = \sum_a J_a \hat{\mathcal{S}}_a$ and $\hat{\mathcal{O}} = \sum_b g_b \hat{\mathcal{S}}_b$, respectively, where $J_a$ (unknown) and $g_b$ (known) are real coupling constants. Our goal is to find $J_a$ such that the commutator norm is zero: $\varepsilon=0$. The commutator between two operator strings can be expanded in terms of other operator strings
\begin{align}
[\hat{\mathcal{S}}_a, \hat{\mathcal{S}}_b] = \sum_{c} f_{ab}^c \hat{\mathcal{S}}_c
\end{align}
where $f_{ab}^c=-f_{ba}^c$ are (basis-dependent) structure constants. Importantly, for bases of Pauli strings and Majorana strings, the structure constants $f_{ab}^c$ are highly sparse and easy to compute algebraically without representing the operator strings as matrices or tensor networks (see Appendices~\ref{sec:pauli_basis}~and~\ref{sec:majorana_basis} for details). Computing the structure constants for Fermion strings, however, is not efficient, so we instead perform computations in the Majorana string basis and convert back and forth to the Fermion string basis as needed.

Using the structure constants, we define the Liouvillian matrix $(L_{\hat{\mathcal{O}}})_{ca} \equiv \sum_b g_b f_{ab}^c$, which describes how operator strings commute with the known integral of motion: $[\hat{\mathcal{S}}_a, \hat{\mathcal{O}}] = \sum_c (L_{\hat{\mathcal{O}}})_{ca} \hat{\mathcal{S}}_c$.

Finally, using the Liouvillian matrix, we can define the \emph{commutant matrix} $C_{\hat{\mathcal{O}}}$ \cite{OBrien2016}, the central quantity that we will work with in the inverse method \footnote{As discussed in the Appendix, this is the form of the commutant matrix for an orthonormal basis of operator strings, such as the Pauli string or Majorana string bases.}:
\begin{align}
C_{\hat{\mathcal{O}}} \equiv L_{\hat{\mathcal{O}}}^\dagger L_{\hat{\mathcal{O}}}. \label{eq:commutantMatrix}
\end{align}
The commutator norm can then be written as a quadratic form involving the commutant matrix
\begin{align}
\varepsilon = J^T C_{\hat{\mathcal{O}}} J \label{eq:comNorm}
\end{align}
where $J$ is the vector of coupling constants of the Hamiltonian. The commutant matrix $C_{\hat{\mathcal{O}}}$ is Hermitian and positive semi-definite, making its eigenvalues real and non-negative.

A normalized eigenvector $J$ of $C_{\hat{\mathcal{O}}}$ with eigenvalue $\varepsilon$ corresponds to a Hamiltonian $\hat{H}=\sum_a J_a \hat{\mathcal{S}}_a$ whose commutator norm with $\hat{\mathcal{O}}$ is $\norm[\big]{[\hat{H},\hat{\mathcal{O}}]}^2=\varepsilon$. This indicates that the operators $\hat{H}$ and $\hat{\mathcal{O}}$ exactly commute when $J$ is a null vector, i.e., an eigenvector with zero eigenvalue ($\varepsilon = 0$), of the commutant matrix. Therefore, a null vector $J$ corresponds to a Hamiltonian $\hat{H}$ with $\hat{\mathcal{O}}$ as an integral of motion. Moreover, since any linear combination of null vectors is also a null vector, we see that the null space of $C_{\hat{\mathcal{O}}}$ corresponds to an entire vector space of Hamiltonians with the desired integral of motion $\hat{\mathcal{O}}$.

We now can see that the inverse method for finding Hamiltonians with a desired integral of motion amounts to finding the null space, or zero modes, of the commutant matrix. As we will discuss in Sec.~\ref{sec:superoperators}, computing the null space of the commutant matrix can be done in a number of ways. The simplest way to do this is to explicitly construct the commutant matrix and diagonalize it numerically for a finite-dimensional basis of operator strings. This is feasible when searching for Hamiltonians since they can be represented as local operators, which can be spanned by relatively low-dimensional spaces of local operator strings.

For a fixed $d$-dimensional space of operators, it is interesting to consider the possible dimensionality of the commutant matrix's null space. If the null space of $C_{\hat{\mathcal{O}}}$ is one-dimensional, then there is a unique Hamiltonian $\hat{H}$ in the chosen space that commutes with $\hat{\mathcal{O}}$. If the null space dimension is greater than one, then there are many Hamiltonians $\hat{H}_1,\hat{H}_2,\ldots$ in that space that, in any linear combination, commute with $\hat{\mathcal{O}}$. If the null space dimension is zero, then there is no Hamiltonian that exactly commutes with $\hat{\mathcal{O}}$ in the chosen space. Nonetheless, since the eigenvalues of $C_{\hat{\mathcal{O}}}$ correspond to commutator norms, the smallest eigenvalue eigenvector of $C_{\hat{\mathcal{O}}}$ corresponds to the Hamiltonian in the space that is ``closest'' to commuting with $\hat{\mathcal{O}}$~\footnote{This is related to the original motivation to name the forward method the \emph{slow operator method}, since energy eigenstates exhibit \emph{slowly} varying expectation values over these quasi-integrals of motion.}.

We also generalize this inverse method to construct Hamiltonians with desired \emph{antisymmetries}, i.e., Hamiltonians $\hat{H}$ that \emph{anti-commute} with a desired operator $\hat{\mathcal{O}}$. To find Hamiltonians that anti-commute with $\hat{\mathcal{O}}$, we look for Hamiltonians that minimize the anti-commutator norm 
\begin{align}
\bar{\varepsilon} = ||\{\hat{H},\hat{\mathcal{O}}\}||^2 = J^T \bar{C}_{\hat{\mathcal{O}}} J
\end{align}
where $\bar{C}_{\hat{\mathcal{O}}} = \bar{L}_{\hat{\mathcal{O}}}^\dagger \bar{L}_{\hat{\mathcal{O}}}$ is the \emph{anti-commutant matrix}, $(\bar{L}_{\hat{\mathcal{O}}})_{ca} \equiv \sum_{b} g_b \bar{f}_{ab}^c$, and $\{\hat{\mathcal{S}}_a, \hat{\mathcal{S}}_b\} \equiv \sum_{c} \bar{f}_{ab}^c \hat{\mathcal{S}}_c$. Similar to the method described above, finding Hamiltonians that anti-commute with $\hat{\mathcal{O}}$ amounts to finding the null space of the anti-commutant matrix $\bar{C}_{\hat{\mathcal{O}}}$.

Finally, we note that the inverse method described in this section is directly related to the recently developed eigenstate-to-Hamiltonian construction (EHC) algorithm \cite{Chertkov2018,Qi2019,Greiter2018} for constructing Hamiltonians from eigenstates. In fact, as pointed out in Ref.~\onlinecite{Qi2019}, if the integral of motion corresponds to a pure state density matrix, $\hat{\mathcal{O}}=\ket{\psi}\bra{\psi}$, then the commutant matrix of Eq.~(\ref{eq:commutantMatrix}) is proportional to the quantum covariance matrix $\bra{\psi} \hat{\mathcal{S}}_a \hat{\mathcal{S}}_b \ket{\psi} - \bra{\psi} \hat{\mathcal{S}}_a\ket{\psi} \bra{\psi} \hat{\mathcal{S}}_b\ket{\psi}$, which is the central quantity computed in the EHC algorithm. Moreover, if the integral of motion is a mixed state density matrix, then the SHC method is closely related to the method described in Ref.~\onlinecite{Bairey2019} for learning Hamiltonians from local measurements.

\subsection{Constructing Hamiltonians that are invariant under desired symmetry transformations}
\label{sec:invariant}

Next, we detail how to construct Hamiltonians that are invariant under desired discrete symmetry transformations, i.e., unitary operators $\hat{\mathcal{U}}_g$ associated with a finite group $g \in G$ that leave the Hamiltonian invariant $\hat{\mathcal{U}}_g \hat{H} \hat{\mathcal{U}}_g^{-1} = \hat{H}$. Since $[\hat{H}, \hat{\mathcal{U}}_g]=0$, in principle the method from Sec.~\ref{sec:iom} could be applied. However, generically, $\hat{\mathcal{U}}_g$ will be a sum of many operator strings, making such calculations intractable in the basis of operator strings. Instead, in this section, we describe two alternative efficient approaches that take as input a finite symmetry group $G$ made of symmetry transformations, such as space group, time-reversal, and charge-conjugation symmetries, and produce as output Hamiltonians invariant under the action of these transformations.

In the first approach, we directly symmetrize our basis of operator strings to produce a new basis of symmetric operators \cite{Varjas2018}
\begin{align}
\hat{\mathcal{S}}_a' \equiv \sum_{g \in G} g \cdot \hat{\mathcal{S}}_a
\end{align}
where $g \cdot \hat{\mathcal{S}}_a$ is the group action of the element $g$ on the operator string $\hat{\mathcal{S}}_a$, which maps that operator string to a linear combination of other operator strings. This is similar to a symmetrization procedure performed by Ref.~\onlinecite{Varjas2018}. For finite groups of order $|G|$ and bases of operator strings of dimension $d$, we can enumerate over all operator strings and perform this calculation in time $d|G|$. When performing such calculations, one needs to take care to ignore non-symmetrizable operator strings \footnote{Consider the Majorana string $\hat{\mathcal{S}}_1 \equiv i\hat{a}_1 \hat{a}_2$ and the symmetry group of permutations $G = \{(12), (21)\}$. Upon symmetrization, this string is $\hat{\mathcal{S}}_1' \equiv (12) \cdot i\hat{a}_1 \hat{a}_2 + (21) \cdot i\hat{a}_1 \hat{a}_2 = i\hat{a}_1 \hat{a}_2 + i\hat{a}_2 \hat{a}_1 = i\hat{a}_1 \hat{a}_2 - i\hat{a}_1 \hat{a}_2 = 0$, making it unsymmetrizable.} and not include linearly dependent symmetrized operators. In this new symmetrized basis, any linear combination $\hat{H} = \sum_a J_a \hat{\mathcal{S}}_a'$ of the symmetrized operators is a Hamiltonian that is invariant under the transformations in the symmetry group $G$.

In the second approach, we analyze the spectrum of the representations of the generators of the symmetry group $G$ in the space of operator strings. In particular, consider an element $g \in G$ that can be represented by a unitary (or anti-unitary) operator $\hat{\mathcal{U}}_g$ acting on the usual Hilbert space of states. The action of the symmetry transformation $g$ on an operator string is given by conjugation with $\hat{\mathcal{U}}_g$ and can be expanded in terms of other operator strings:
\begin{align}
g \cdot \hat{\mathcal{S}}_a \equiv \hat{\mathcal{U}}_g \hat{\mathcal{S}}_a \hat{\mathcal{U}}_g^{-1} = \sum_{b} (D_g)_{ba} \hat{\mathcal{S}}_b. \label{eq:symmetrySa}
\end{align}
The matrix $D_g$ is the representation of the symmetry transformation $g$ on the space of operator strings and for common symmetry operators, such as space group symmetries, particle-hole symmetry, and time-reversal symmetry, can be straight-forwardly computed as we discuss in Appendix~\ref{sec:representations}. From Eq.~(\ref{eq:symmetrySa}), we see that a Hamiltonian $\hat{H}=\sum_a J_a \hat{\mathcal{S}}_a$ transforms under the symmetry as \footnote{Antiunitary operators, such as time-reversal, are antilinear transformations. Since we require $\hat{H}$ to be Hermitian, the $J_a$ are real and are unaffected by the antilinearity of $\hat{\mathcal{U}}_g$.}
\begin{align}
g \cdot \hat{H} = \hat{\mathcal{U}}_g \hat{H} \hat{\mathcal{U}}_g^{-1} = \sum_{a} J_a \hat{\mathcal{U}}_g \hat{\mathcal{S}}_a \hat{\mathcal{U}}_g^{-1} = \sum_{a,b} (D_g)_{ba} J_a  \hat{\mathcal{S}}_b. \label{eq:gH}
\end{align}
From Eq.~(\ref{eq:gH}), we see that $g \cdot \hat{H} = \pm\hat{H}$ when $J$ is an eigenvector of $D_g$ with eigenvalue $\pm 1$. The $(+1)$-eigenvalue eigenvectors correspond to the coupling constants of Hamiltonians with $g$ as a symmetry so that $[\hat{H},\hat{\mathcal{U}}_g]=0$. Likewise, the $(-1)$-eigenvalue eigenvectors correspond to Hamiltonians with $g$ as an \emph{antisymmetry} so that $\{\hat{H},\hat{\mathcal{U}}_g\}=0$. Therefore, we see that we can find Hamiltonians that are invariant under a desired symmetry transformation $g$ by finding the $(+1)$-eigenvectors of the representation $D_g$, and likewise for antisymmetries the $(-1)$-eigenvectors of $D_g$. To find Hamiltonians that are invariant under all of the symmetry transformations in the symmetry group $G$, one can compute the intersection of the $(+1)$-eigenspaces of the representations of the group's generators.

\subsection{Constructing Hamiltonians with desired symmetries by finding the ground states of superoperators} \label{sec:superoperators}

In the methods described above in Secs.~\ref{sec:iom}~and~\ref{sec:invariant}, we perform calculations in a Hilbert space of Hermitian operators. The commutant matrix $C_{\hat{\mathcal{O}}}$ and representation $D_g$ can be interpreted as \emph{superoperators} acting on operators in this space. From this perspective, we can frame the inverse problem of constructing Hamiltonians with a set of desired symmetries and antisymmetries as finding the ground state of a particular Hermitian, positive semi-definite ``superoperator Hamiltonian.''

In particular, if we desire to construct Hamiltonians that commute with integrals of motion $\hat{\mathcal{O}}_1,\ldots,\hat{\mathcal{O}}_{N_1}$, anti-commute with $\hat{\mathcal{O}}_1',\ldots,\hat{\mathcal{O}}_{N_2}'$, are invariant under symmetry transformations $g_1,\ldots,g_{M_1}$ and anti-invariant under $g_1',\ldots,g_{M_2}'$, then we can do so by finding the ``ground states'' of the superoperator Hamiltonian
\begin{align}
\mathcal{H} \equiv \sum_{i=1}^{N_1} C_{\hat{\mathcal{O}}_i} + \sum_{j=1}^{N_2} \bar{C}_{\hat{\mathcal{O}}_j'} &+ \sum_{k=1}^{M_1} \left[I-\frac{1}{2}(D_{g_k} + D_{g_k}^\dagger)\right] \nonumber \\
&+ \sum_{l=1}^{M_2} \left[I+\frac{1}{2}(D_{g_l'} + D_{g_l'}^\dagger)\right]
\end{align}
which is Hermitian and positive semi-definite by construction, where $I$ is the identity superoperator. If we find ground states of $\mathcal{H}$ that have zero ``energy,'' then we have found Hamiltonians that obey all of the desired symmetries at once. If the ground state energy is non-zero, then the energy indicates how much the discovered Hamiltonian fails to commute (or anti-commute) with the given symmetries.

The simplest way to find the ground states of the superoperator Hamiltonian $\mathcal{H}$ is to write it as a matrix in a basis of operators and perform exact diagonalization on the matrix, e.g., using the Lanczos algorithm. Consider performing such a calculation in the Pauli string basis. Rather than working in the full $4^n$-dimensional space of operators for a system of $n$ spins, it is convenient to consider a much smaller basis of range-$R$ $k$-local Pauli strings. Range-$R$ $k$-local Pauli strings are Pauli strings made from a product of $k$ (non-Identity) Pauli matrices on sites separated spatially by at most the maximum distance $R$. For example, the space of range-$2$ $3$-local Pauli strings on a 1D chain includes all possible spin Hamiltonians with three-site interactions between nearest and next nearest neighbor sites. In Sections~\ref{sec:mzmresults}~and~\ref{sec:z2spinliquids}, we obtain our results by exactly diagonalizing $\mathcal{H}$ in range-$R$ $k$-local bases of Majorana strings and Pauli strings.

Finally, we note that many other well-developed methods can be used to find the ground states of $\mathcal{H}$. For example, one can represent the superoperator $\mathcal{H}$ as a matrix product operator \cite{Zwolak2004,Kells2018} and perform DMRG to find its ground state(s) or use methods such as variational Monte Carlo or other forms of quantum Monte Carlo to do so. In attempting this, one should keep in mind that, even though the notion of locality might be kept in $\mathcal{H}$, there is no known guarantee for its gapped/gapless nature, which might hinder the applicability of these methods.

\section{Hamiltonians with zero modes} \label{sec:mzmresults}

In this section, we use the SHC to analytically design non-interacting and interacting Hamiltonians that commute with a desired pair of zero modes that can be distributed arbitrarily in space. First, we present some theoretical background on zero modes and Majorana zero modes (MZMs). Then, we describe our general framework for constructing these Hamiltonians out of ``bond operator'' building blocks, which are two-site operators that involve fermionic hopping, pairing, and chemical potentials. Finally, we provide examples of how this procedure can produce various Hamiltonians that commute with zero modes.  We start by finding $p$-wave superconducting Hamiltonians that exactly commute with either exponentially-decaying MZMs, Gaussian-distributed MZMs, or zero modes with complicated spatial distributions.  Then we give examples of  $s$-wave superconducting Hamiltonians and interacting Hamiltonians that commute with MZMs.

\noindent \textbf{Background.} A zero mode $\hat{\gamma}$ is a Hermitian operator that \cite{Fendley2012,Alicea2016,Fendley2016}: 
\begin{enumerate}
    \item commutes with the Hamiltonian: $[\hat{\gamma},\hat{H}]=0$
    \item squares to identity: $\hat{\gamma}^2=\hat{I}$
    \item anticommutes with fermion parity: $\{\hat{\gamma},(-1)^{\hat{N}}\}=0$
\end{enumerate}
where $\hat{H}$ is a Hamiltonian that conserves fermion parity $[\hat{H}, (-1)^{\hat{N}}]=0$, where $(-1)^{\hat{N}}\equiv \prod_{j} (\hat{I}-2\hat{n}_j)$. Property 1 indicates that a zero mode is a symmetry of the system; property 2 indicates that its eigenvalues are $\pm 1$; and property 3 indicates that each state $\ket{\psi_+}$ with definite fermion parity $\eta=\pm 1$ comes paired with an orthogonal state $\ket{\psi_-}\equiv \hat{\gamma}\ket{\psi_+}$ with opposite parity $-\eta$. For fermion-parity-conserving Hamiltonians, the opposite parity states $\ket{\psi_+}$ and $\ket{\psi_-}$ are degenerate energy eigenstates.

Generically, zero modes come in pairs. For $M$ pairs of zero modes, the operators $\hat{\gamma}^{(1)},\ldots,\hat{\gamma}^{(2M)}$ all commute with the Hamiltonian and satisfy the anticommutation relations
\begin{align}
\{\hat{\gamma}^{(m)}, \hat{\gamma}^{(n)}\} = 2\delta_{mn}. \label{eq:zm_anticommutation}
\end{align}
These $2M$ zero modes lead to a $2^{M}$-fold degeneracy in each of the eigenstates of the Hamiltonian. By pairing zero modes into complex fermions $\hat{f}_{m}\equiv( \hat{\gamma}^{(2m-1)}+i\hat{\gamma}^{(2m)})/2$ for $m=1,\ldots,M$, we can see that the number operators $\hat{f}_{m}^\dagger \hat{f}_{m}$ commute with the Hamiltonian and each other. Therefore, these operators are simultaneously diagonalizable and the Hamiltonian eigenstates can be labeled by the occupation numbers $0,1$ for each of the $M$ number operators \cite{Nayak2008}. From properties 1-3, we can deduce that these $2^M$ many-body states are degenerate in energy and that the $\hat{f}_{m}^\dagger$ operators create single-quasiparticle modes of zero energy. Finally, we mention that \emph{Majorana} zero modes (MZMs) are zero modes that are spatially localized and well-separated from one another \cite{DasSarma2015}. These properties allow such zero modes to exhibit the non-Abelian statistics of Ising anyons \cite{Nayak2008}.

In this work, we will consider zero modes of the form
\begin{align}
\hat{\gamma}^{(1)} &= \sum_{j} \alpha_j^{(1)} \hat{a}_j + \beta_j^{(1)} \hat{b}_j, \nonumber \\
\hat{\gamma}^{(2)} &= \sum_{j} \alpha_j^{(2)} \hat{a}_j + \beta_j^{(2)} \hat{b}_j, \label{eq:genericzeromodes}
\end{align}
where $\alpha_j^{(1)},\beta_j^{(1)}$ and $\alpha_j^{(2)},\beta_j^{(2)}$ are real parameters that specify the ``amplitudes'' of the zero modes on site $j$ (or, more generically, orbital $j$). To ensure that $\hat{\gamma}^{(1)}$ and $\hat{\gamma}^{(2)}$ are zero modes with the properties mentioned above, the parameters are constrained such that $(\hat{\gamma}^{(1)})^2=(\hat{\gamma}^{(2)})^2=\hat{I}$ and $\{\hat{\gamma}^{(1)},\hat{\gamma}^{(2)}\}=0$. This implies that the zero modes are orthonormal,
\begin{align}
||\hat{\gamma}^{(1)}||^2=\sum_{j} (\alpha_j^{(1)})^2 + (\beta_j^{(1)})^2=1, \nonumber \\
||\hat{\gamma}^{(2)}||^2=\sum_{j} (\alpha_j^{(2)})^2 + (\beta_j^{(2)})^2=1, \nonumber \\
\langle \hat{\gamma}^{(1)}, \hat{\gamma}^{(2)} \rangle=\sum_{j} \alpha_j^{(1)} \alpha_j^{(2)} + \beta_j^{(1)} \beta_j^{(2)}=0. \label{eq:zeromodeconstraint}
\end{align}

\begin{figure}
    \begin{center}
    \includegraphics[width=0.26\textwidth]{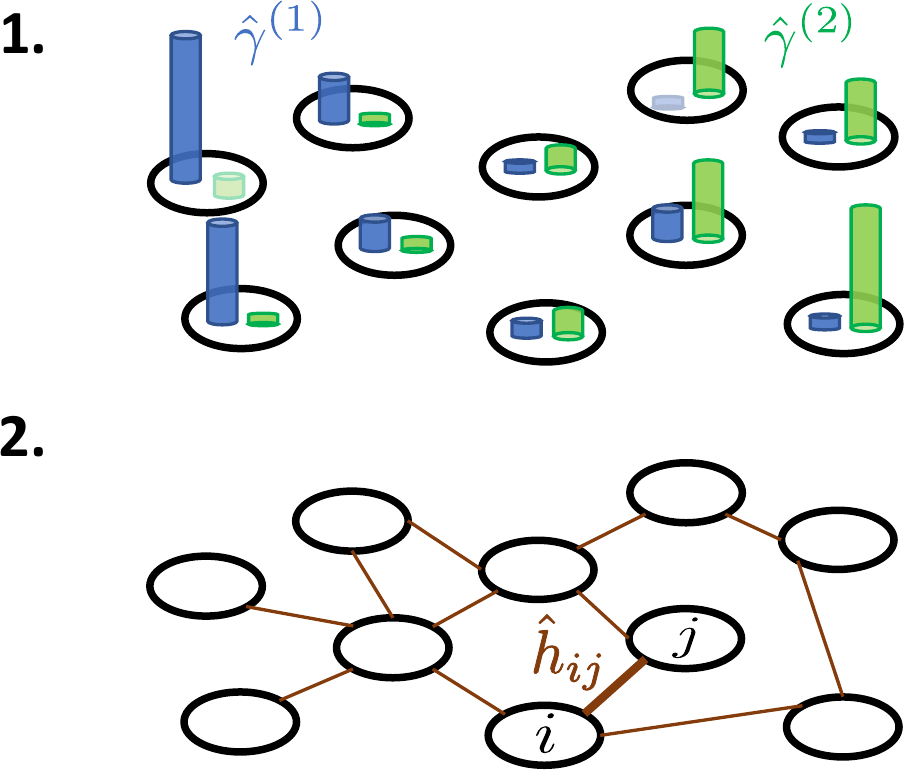}
    \end{center}
    \caption{Steps for constructing zero mode Hamiltonians on an arbitrary graph. \textbf{1.} Specify the spatial distributions of the zero modes $\hat{\gamma}^{(1)}$ and $\hat{\gamma}^{(2)}$ by choosing their amplitudes $\alpha_j^{(1)},\beta_j^{(1)}$ and $\alpha_j^{(2)},\beta_j^{(2)}$, respectively, on the vertices $j$ of the graph. \textbf{2.} Draw a set of edges between the vertices until the vertices and edges form a connected graph. The resulting graph represents a family of Hamiltonians of the form $\hat{H}_{ZM}=\sum_{ij} J_{ij} \hat{h}_{ij}$, where $J_{ij} \neq 0$ on the $(i,j)$ edges and $\hat{h}_{ij}$ are bond operators specified by Eq.~(\ref{eq:bondoperator}). These Hamiltonians exactly commute with $\hat{\gamma}^{(1)}$ and $\hat{\gamma}^{(2)}$ and no other zero modes of the form of Eq.~(\ref{eq:genericzeromodes}).}
    \label{fig:zmframework}
\end{figure}

\noindent \textbf{Framework for designing zero mode Hamiltonians.} 
We present a novel framework, illustrated in Fig.~\ref{fig:zmframework}, for designing Hamiltonians that commute with a desired pair of zero modes $\hat{\gamma}^{(1)}$ and $\hat{\gamma}^{(2)}$ that can be distributed arbitrarily in space. Given a spatial distribution of the two zero modes specified by the values of $\alpha_j^{(1)},\beta_j^{(1)}$ and $\alpha_j^{(2)},\beta_j^{(2)}$ in Eq.~(\ref{eq:genericzeromodes}), we output a family of Hamiltonians 
\begin{align}
\hat{H}_{ZM} &= \sum_{ij} J_{ij} \hat{h}_{ij}, \label{eq:HZM1}
\end{align}
that commute with the two zero modes and are built from Hermitian bond operators 
\begin{align}
\hat{h}_{ij}&=\left[(\tilde{t}^R_{ij} + i\tilde{t}^I_{ij})\hat{c}^\dagger_i \hat{c}_j + (\tilde{\Delta}^R_{ij} + i\tilde{\Delta}^I_{ij})\hat{c}^\dagger_i \hat{c}^\dagger_j + \textrm{H.c.}\right] \nonumber \\
&\quad\quad+ \tilde{\mu}_{i}^{(ij)} \hat{n}_i + \tilde{\mu}_{j}^{(ij)} \hat{n}_j, \label{eq:bondoperator}
\end{align}
which act only on sites $i$ and $j$. The $J_{ij}$ are real parameters that independently scale each bond operator and are arbitrary up to the constraint that they must be non-zero on a connected graph of bonds $\{(i,j)\}$. This constraint guarantees that the Hamiltonians \emph{do not commute with any other zero modes that are linear combinations of the Majorana fermions $\hat{a}_j$ and $\hat{b}_j$ other than $\hat{\gamma}^{(1)}$ and $\hat{\gamma}^{(2)}$} \footnote{There is an additional subtlety, discussed in Appendix~\ref{sec:derivation_zm}, that arises when $(\alpha_i^{(1)}, \beta_i^{(1)},\alpha_j^{(1)}, \beta_j^{(1)}) \propto (\alpha_i^{(2)}, \beta_i^{(2)},\alpha_j^{(2)}, \beta_j^{(2)})$. In this case, the $\hat{h}_{ij}$ bond operator does not uniquely isolate the $\hat{\gamma}^{(1)}, \hat{\gamma}^{(2)}$ zero modes as the only zero modes that commute with $\hat{H}_{ZM}$. Such bond operators can be added to the Hamiltonian $\hat{H}_{ZM}$, but they do not count towards the connected graph constraint.}. The family of Hamiltonians in Eq.~(\ref{eq:HZM1}) includes Hamiltonians on various lattices and graphs such as square lattices, kagome lattices, tetrahedral lattices, trees, and aperiodic tilings. 

We were able to determine the analytic form of the bond operators $\hat{h}_{ij}$ and their properties by computing the commutant matrices $C_{\hat{\gamma}^{(1)}},C_{\hat{\gamma}^{(2)}},$ and $C_{\hat{h}_{ij}}$, as described in Appendix~\ref{sec:derivation_zm}. The parameters of the bond operators $\hat{h}_{ij}$ depend on the zero mode amplitudes $\alpha_j^{(1)},\beta_j^{(1)}$ and $\alpha_j^{(2)},\beta_j^{(2)}$ on orbitals $i$ and $j$ and take the form
\begin{align}
\tilde{t}^R_{ij} &\equiv  - \alpha_i^{(1)}  \beta_j^{(2)} + \beta_i^{(1)}\alpha_j^{(2)} - \alpha_j^{(1)}  \beta_i^{(2)}  + \beta_j^{(1)} \alpha_i^{(2)}  , \nonumber \\
\tilde{t}^I_{ij} &\equiv - \alpha_i^{(1)}  \alpha_j^{(2)} - \beta_i^{(1)}  \beta_j^{(2)} + \alpha_j^{(1)} \alpha_i^{(2)}  + \beta_j^{(1)} \beta_i^{(2)} , \nonumber \\
\tilde{\Delta}^R_{ij} &\equiv - \alpha_i^{(1)}  \beta_j^{(2)} -\beta_i^{(1)} \alpha_j^{(2)}  + \alpha_j^{(1)}  \beta_i^{(2)}  + \beta_j^{(1)} \alpha_i^{(2)} , \nonumber \\
\tilde{\Delta}^I_{ij} &\equiv + \alpha_i^{(1)}  \alpha_j^{(2)} - \beta_i^{(1)}  \beta_j^{(2)} - \alpha_j^{(1)} \alpha_i^{(2)}  + \beta_j^{(1)} \beta_i^{(2)} , \nonumber \\
\tilde{\mu}_{i}^{(ij)} &\equiv 2 (\alpha_j^{(1)}  \beta_j^{(2)} - \beta_j^{(1)} \alpha_j^{(2)}), \nonumber \\
\tilde{\mu}_{j}^{(ij)} &\equiv 2 (\alpha_i^{(1)}  \beta_i^{(2)} - \beta_i^{(1)} \alpha_i^{(2)}). \label{eq:bondoperatorparams}
\end{align}
For the chemical potential parameters $\tilde{\mu}_{i}^{(ij)}$ and $\tilde{\mu}_{j}^{(ij)}$, we add an $(ij)$ superscript to make clear that these chemical potentials on sites $i$ and $j$ are associated with the bond operator $\hat{h}_{ij}$ and not another bond operator, such as $\hat{h}_{ik}$ or $\hat{h}_{kj}$. Also, importantly, we choose an ordering convention for our fermionic operators and require that $i < j$ for each bond operator $\hat{h}_{ij}$ to be consistent with our convention.

Each bond operator $\hat{h}_{ij}$ individually commutes with $\hat{\gamma}^{(1)}$ and $\hat{\gamma}^{(2)}$, though overlapping bond operators do not commute with one another: $[\hat{h}_{ij}, \hat{h}_{jk}] \neq 0$. This makes the $\hat{H}_{ZM}$ Hamiltonians analogous to frustration-free Hamiltonians, which by definition have ground states that are simultaneously the ground states of each local term in the Hamiltonian~\cite{Katsura2015,Jevtic2017}.

We can rewrite Eq.~(\ref{eq:HZM1}) as
\begin{align}
\hat{H}_{ZM} &= \sum_{ij}\left[(t^R_{ij} + it^I_{ij})\hat{c}^\dagger_i \hat{c}_j + (\Delta^R_{ij} + i\Delta^I_{ij})\hat{c}^\dagger_i \hat{c}^\dagger_j + \textrm{H.c.}\right] \nonumber \\
&\quad\quad+ \sum_{j} \mu_{j} \hat{n}_j \label{eq:HZM2}
\end{align}
whose parameters take on the values 
\begin{align}
t_{ij}^R = J_{ij}\tilde{t}_{ij}^R, \quad t_{ij}^I = J_{ij}\tilde{t}_{ij}^I, \quad \Delta_{ij}^R = J_{ij}\tilde{\Delta}_{ij}^R, \quad \Delta_{ij}^I = J_{ij}\tilde{\Delta}_{ij}^I, \label{eq:tdeltaforms}
\end{align}
and
\begin{align}
\mu_{j} = \sum_{i<j} J_{ij}\tilde{\mu}_{j}^{(ij)} + \sum_{i>j} J_{ji} \tilde{\mu}_{j}^{(ji)}. \label{eq:muform}
\end{align}
The chemical potential on a site $j$ is the sum of the chemical potentials contributed by each bond operator. The need for two sums is due to our choice of convention that $i < j$ for each bond operator $\hat{h}_{ij}$.

To simplify the remaining discussion in our examples, we consider the restricted class of zero modes with $\beta_j^{(1)}=\alpha_j^{(2)}=0$ for all $j$:
\begin{align}
\hat{\gamma}^{(1)} = \sum_j \alpha_j \hat{a}_j, \quad\quad \hat{\gamma}^{(2)} = \sum_j \beta_j \hat{b}_j, \label{eq:psiijkc}
\end{align}
where we relabeled $\alpha_j^{(1)} \rightarrow \alpha_j$ and $\beta_j^{(2)} \rightarrow \beta_j$. One reason to consider this class of zero modes is that it contains the zero modes of the Kitaev chain \cite{Kitaev2001}. Another reason is that, upon normalization, these zero modes automatically satisfy Eq.~(\ref{eq:zeromodeconstraint}).

For the zero modes of Eq.~(\ref{eq:psiijkc}), the bond operator can be simplified to
\begin{align}
\hat{h}_{ij} &\equiv \tilde{t}_{ij}\left(\hat{c}^\dagger_i \hat{c}_j + \textrm{H.c.} \right) + \tilde{\Delta}_{ij} \left(\hat{c}^\dagger_i \hat{c}_j^\dagger + \textrm{H.c.} \right) \nonumber \\
&\quad+ \tilde{\mu}_{i}^{(ij)}\hat{n}_i + \tilde{\mu}_{j}^{(ij)}\hat{n}_j \label{eq:hij}
\end{align}
where
\begin{align}
&\tilde{t}_{ij} \equiv 1 + \frac{\alpha_j/\alpha_i}{\beta_j/\beta_i}, \quad \tilde{\Delta}_{ij} \equiv 1 - \frac{\alpha_j/\alpha_i}{\beta_j/\beta_i} \nonumber \\
&\tilde{\mu}_{i}^{(ij)} \equiv - 2\alpha_j/\alpha_i,\quad\,\, \tilde{\mu}_{j}^{(ij)} \equiv - 2\beta_i/\beta_j. \label{eq:hijparams}
\end{align}

\noindent \textbf{Example: Exponentially decaying Majorana zero modes.} As a first example, we discuss how to use our framework to construct $p$-wave 1D chain Hamiltonians with exponentially decaying Majorana zero modes. We will see that the models we construct this way are closely related to the Kitaev chain, which has nearly exponentially decaying MZMs \cite{Kitaev2001}.

As input, we choose two MZMs exponentially localized at each edge of a 1D chain of $L$ spinless fermions:
\begin{align}
\hat{\gamma}^{(1)} \propto \sum_{j=1}^{L} e^{-j/\xi} \hat{a}_{j}, \quad \hat{\gamma}^{(2)} \propto \sum_{j=1}^{L} e^{-(L-j)/\xi} \hat{b}_{j}, \label{eq:psiAB}
\end{align}
where $\xi > 0$ is a correlation length in units of the lattice spacing.

For a bond between the sites $i$ and $j$ separated by a distance $d=|i - j|$, the parameters of the bond operator Eq.~(\ref{eq:hijparams}) become
\begin{align*}
\tilde{t}_{ij} = 1 + e^{-2d/\xi}, \, \tilde{\Delta}_{ij} = 1 - e^{-2d/\xi}, \, \tilde{\mu}_{i}^{(ij)} = \tilde{\mu}_{j}^{(ij)} = - 2e^{-d/\xi}.
\end{align*}
The bond operators with these specific parameters define a large family of Hamiltonians $\hat{H}_{ZM}=\sum_{ij}J_{ij} \hat{h}_{ij}$ that exactly commute with the zero modes of Eq.~(\ref{eq:psiAB}).

First, we will focus on a simple, interesting subspace of these zero mode Hamiltonians. In particular, we consider those Hamiltonians constrained to have constant nearest neighbor ($d=1$) hopping $-t$ on a 1D chain. We implement this constraint by choosing $J_{ij}=-t\delta_{i,j-1}/\tilde{t}_{ij}$ for each bond $(i,j)$. Given this constraint and our input zero modes, we find a unique Hamiltonian that commutes with the desired zero modes
\begin{align}
\hat{H}_{\textrm{exp}}^{(1)} &= \sum_{ij}J_{ij}\hat{h}_{ij} = \sum_{j=1}^{L-1} -(t/\tilde{t}_{j,j+1})\hat{h}_{j,j+1} \nonumber \\
&= \sum_{j=1}^{L-1} (-t\hat{c}^\dagger_j \hat{c}_{j+1} - \Delta^{(1)} \hat{c}^\dagger_j \hat{c}^\dagger_{j+1} + \textrm{H.c.}) - \sum_{j=1}^{L}\mu^{(1)}_j \hat{n}_j, \label{eq:Hexp1}
\end{align}
where 
\begin{align}
\Delta^{(1)}/t &\equiv \tanh(1/\xi), \nonumber \\
\mu^{(1)}_j/t &\equiv \begin{cases} -1/\cosh(1/\xi) & j=1,L \\ -2/\cosh(1/\xi) & \textrm{otherwise}. \end{cases} \label{eq:Hexp1params}
\end{align}
Note that this Hamiltonian almost matches the Kitaev chain \cite{Kitaev2001} with pairing $\Delta/t=\tanh(1/\xi)$ and chemical potential $\mu/t=-2/\cosh(1/\xi)$, though slightly differs at the edges $j=1,L$ of the chain: $\hat{H}_{\textrm{exp}}^{(1)} = \hat{H}_{KC} + \frac{\mu}{2}\hat{n}_1 + \frac{\mu}{2}\hat{n}_L$.

Another interesting subspace of this large family of Hamiltonians is the space of Hamiltonians with $d$-th neighbor bonds for $1<d<L/2$. For this subspace, we choose the constraint that $J_{ij}=-t\delta_{i,j-d}/\tilde{t}_{ij}$. Under this constraint, we can use the bond operators to construct the following Hamiltonians
\begin{align}
\hat{H}_{\textrm{exp}}^{(d)} = \sum_{j=1}^{L-d} (-t\hat{c}^\dagger_j \hat{c}_{j+d} - \Delta^{(d)} \hat{c}^\dagger_j \hat{c}^\dagger_{j+d} + \textrm{H.c.}) - \sum_{j=1}^{L}\mu^{(d)}_j \hat{n}_j, \label{eq:expH}
\end{align}
where 
\begin{align}
\Delta^{(d)}/t &\equiv \tanh(d/\xi), \nonumber \\
\mu^{(d)}_j/t &\equiv \begin{cases} -1/\cosh(d/\xi) & \min(j-1,L-j) < d \\ -2/\cosh(d/\xi) & \textrm{otherwise}. \end{cases}
\end{align}
Interestingly, all Hamiltonians of Eq.~(\ref{eq:expH}), in any linear combination, exactly commute with the exponentially decaying MZMs of Eq.~(\ref{eq:psiAB}). However, note that only the nearest-neighbor $d=1$ Hamiltonian connects together all of the sites in the 1D chain, while the $d>1$ Hamiltonians form disconnected graphs. Therefore, Hamiltonians of the form $\sum_{d=1}^{L/2-1} J_d \hat{H}_{\textrm{exp}}^{(d)}$ with $J_1 \neq 0$ commute with exactly two zero modes, while those with $J_1 = 0$ potentially commute with more than two.

\noindent \textbf{Example: Gaussian-distributed Majorana zero modes.} Next, we construct Hamiltonians that commute with MZMs that are spatially localized as Gaussians of width $\sigma$ centered at positions $\vec{x}_1$ and $\vec{x}_2$. We provide as input zero modes with the amplitudes
\begin{align*}
\alpha_{\vec{x}} \propto \exp\left(-(\vec{x}-\vec{x}_{1})^2/2\sigma^2\right), \, \beta_{\vec{x}} \propto \exp\left(-(\vec{x}-\vec{x}_{2})^2/2\sigma^2\right)
\end{align*}
where we replaced the site label $j$ with its spatial coordinate $\vec{x}$ in a lattice so that $\alpha_j,\beta_j \rightarrow \alpha_{\vec{x}},\beta_{\vec{x}}$. For concreteness, we will consider a 1D chain lattice and 2D square lattice, but the same construction applies to arbitrary lattices in any dimension.

The $\alpha_{\vec{x}},\beta_{\vec{x}}$ parameters determine the coefficients of the bond operator $\hat{h}_{ij} \rightarrow \hat{h}_{\vec{x},\vec{x}+\vec{\delta}}$ that connects the site $i$ at position $\vec{x}$ to site $j$ at position $\vec{x}+\vec{\delta}$. For Gaussian-distributed zero modes, the parameters of the bond operator $\hat{h}_{\vec{x},\vec{x}+\vec{\delta}}$ of Eq.~(\ref{eq:hij}) satisfy
\begin{align}
\frac{\tilde{\Delta}_{\vec{x},\vec{x}+\vec{\delta}}}{\tilde{t}_{\vec{x},\vec{x}+\vec{\delta}}} &\equiv \tanh\left((\vec{x}_2 - \vec{x}_1)\cdot \vec{\delta}/2\sigma^2\right) \nonumber \\
\frac{\tilde{\mu}_{\vec{x}}^{(\vec{x},\vec{x}+\vec{\delta})}}{\tilde{t}_{\vec{x},\vec{x}+\vec{\delta}}} &\equiv -\frac{2e^{-[(\vec{x}-\vec{x}_1)\cdot \vec{\delta} + \vec{\delta}^2]/2\sigma^2}}{1 + e^{-(\vec{x}_2-\vec{x}_1)\cdot \vec{\delta}/\sigma^2}} \nonumber \\
\frac{\tilde{\mu}_{\vec{x}+\vec{\delta}}^{(\vec{x},\vec{x}+\vec{\delta})}}{\tilde{t}_{\vec{x},\vec{x}+\vec{\delta}}} &\equiv -\frac{2e^{[(\vec{x}-\vec{x}_2)\cdot \vec{\delta} + \vec{\delta}^2]/2\sigma^2}}{1 + e^{-(\vec{x}_2-\vec{x}_1)\cdot \vec{\delta}/\sigma^2}}. \label{eq:gaussianparams}
\end{align}
An interesting property to note is that ${\tilde{\Delta}_{\vec{x},\vec{x}+\vec{\delta}}}/{\tilde{t}_{\vec{x},\vec{x}+\vec{\delta}}}$ is actually independent of position $\vec{x}$ and only depends on the displacement of the MZMs $\vec{x}_2-\vec{x}_1$ and the direction of the bond $\vec{\delta}$. 

\begin{figure}
    \begin{center}
    \includegraphics[width=0.45\textwidth]{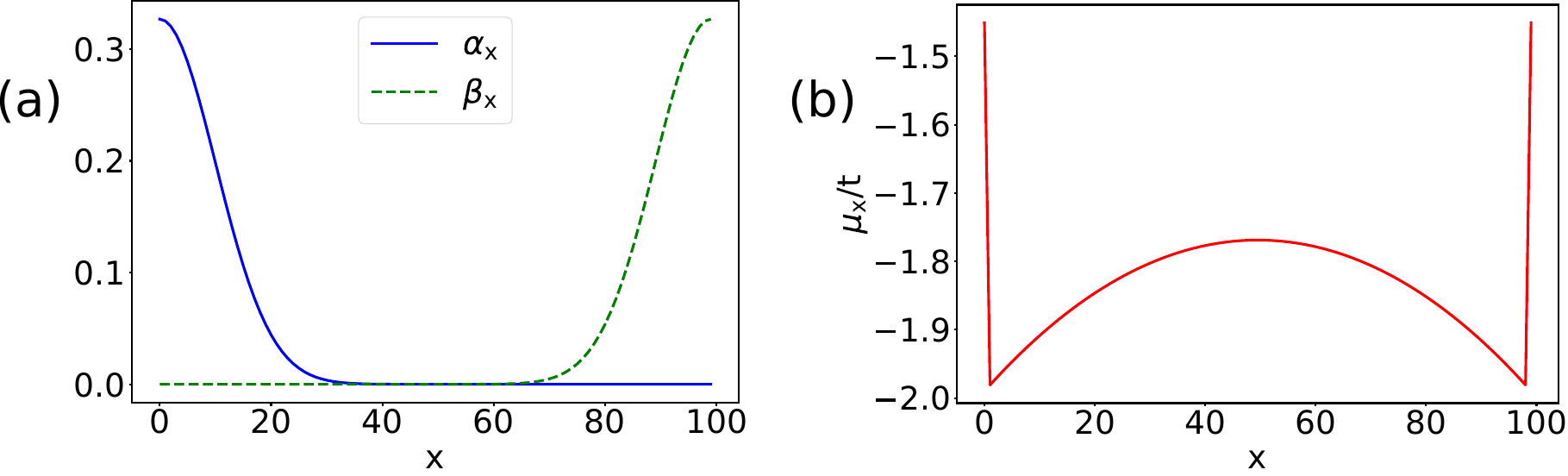}
    \end{center}
    \caption{(a) Two 1D Majorana zero modes $\hat{\gamma}^{(1)}=\sum_{x}\alpha_{x}\hat{a}_{x}$ and $\hat{\gamma}^{(2)}=\sum_{x}\beta_{x}\hat{b}_{x}$ of the form of Eq.~(\ref{eq:psiijkc}) with Gaussian profiles of width $\sigma=10$, on a $100$-site chain. (b) The spatial distribution of the chemical potential $\mu_x$ for the Hamiltonian Eq.~(\ref{eq:HMZM1D}) with a constant pairing $\Delta_{x,x+1}/t \approx 0.4582$ that commutes with the two Majorana zero modes.}
    \label{fig:mzms1d}
\end{figure}

\begin{figure}
    \begin{center}
    \includegraphics[width=0.45\textwidth]{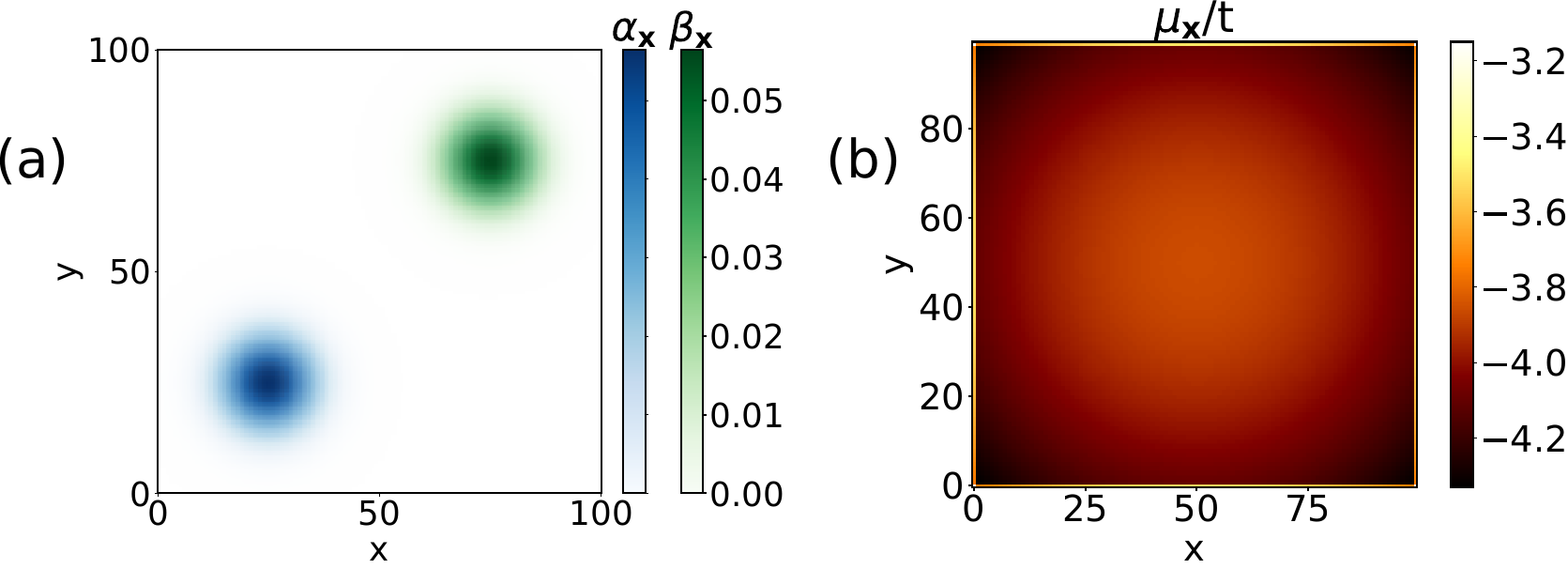} 
    \end{center}
    \caption{(a) Two 2D Majorana zero modes $\hat{\gamma}^{(1)}=\sum_{\vec{x}}\alpha_{\vec{x}}\hat{a}_{\vec{x}}$ and $\hat{\gamma}^{(2)}=\sum_{\vec{x}}\beta_{\vec{x}}\hat{b}_{\vec{x}}$ with Gaussian profiles of width $\sigma = 10$ localized at two corners of a $100 \times 100$ square lattice. (b) The spatial distribution of the chemical potential $\mu_{\vec{x}}$ for the Hamiltonian Eq.~(\ref{eq:HMZM2D}) with constant pairing $\Delta_{\vec{x},\vec{x}+\hat{\vec{x}}}/t=\Delta_{\vec{x},\vec{x}+\hat{\vec{y}}}/t\approx 0.2449$ that commutes with the two Majorana zero modes.}
    \label{fig:mzms2d}
\end{figure}

By arranging these bond operators uniformly onto the nearest neighbor bonds of a 1D chain or 2D square lattice, i.e., choosing $J_{\vec{x},\vec{y}} = -t\delta_{\vec{x},\vec{y}-\vec{\delta}}/\tilde{t}_{\vec{x},\vec{y}}$, we construct the following two Hamiltonians that commute with the two Gaussian-distributed zero modes
\begin{align}
\hat{H}_{1D} &= \sum_{x=1}^{L-1}\left[ -t \hat{c}^\dagger_{x} \hat{c}_{x+1} - \Delta_{x,x+1} \hat{c}^\dagger_{x} \hat{c}^\dagger_{x+1} + \textrm{H.c.}\right] - \sum_{x=1}^{L} \mu_x \hat{n}_x, \label{eq:HMZM1D} \\
\hat{H}_{2D} &= \sum_{\vec{x}}\sum_{\vec{\delta}=\hat{\vec{x}}, \hat{\vec{y}}} \left[ -t\hat{c}^\dagger_{\vec{x}} \hat{c}_{\vec{x}+\vec{\delta}} - \Delta_{\vec{x},\vec{x}+\vec{\delta}}\hat{c}^\dagger_{\vec{x}} \hat{c}^\dagger_{\vec{x}+\vec{\delta}} + \textrm{H.c.}\right] \nonumber \\
&\quad\quad- \sum_{\vec{x}}\mu_{\vec{x}} \hat{n}_{\vec{x}}, \label{eq:HMZM2D}
\end{align}
where the pairings and chemical potentials can be generated from Eq.~(\ref{eq:gaussianparams}) by applying Eqs.~(\ref{eq:tdeltaforms})-(\ref{eq:muform}). To illustrate, we show the chemical potential distributions for a 100-site 1D chain and a $100 \times 100$ 2D square lattice with Gaussian Majorana zero modes in Figs.~\ref{fig:mzms1d}~and~\ref{fig:mzms2d}, respectively. Again, there is nothing special about chains, square lattices, 1D, or 2D. The same Hamiltonian as Eq.~(\ref{eq:HMZM2D}), but in different dimensions and with different lattice vectors $\vec{\delta}$, would also commute exactly with two Gaussian-distributed MZMs. Even more generally, a similar construction for any arbitrarily connected graph is straightforward, as discussed above.

\noindent \textbf{Example: Zero modes with complicated spatial distributions.} Using the $\hat{h}_{ij}$ bond operators, we can also design Hamiltonians with zero modes that have highly non-trivial spatial distributions.

\begin{figure}
    \begin{center}
    \includegraphics[width=0.48\textwidth]{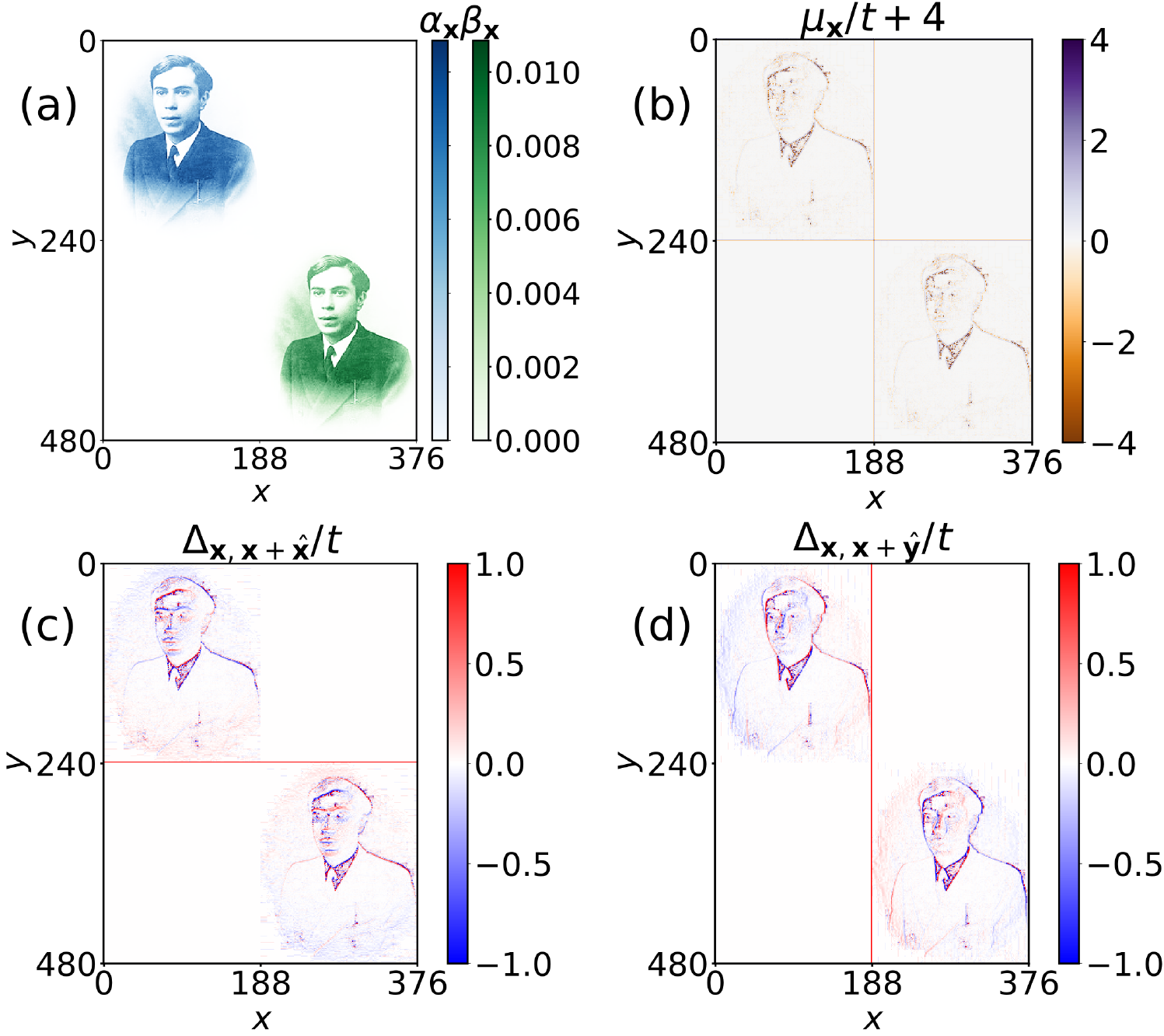}
    \end{center}
    \caption{(a) Two Majorana zero modes shaped like Ettore Majorana. The (b) chemical potential $\mu_{\vec{x}}/t$, (c) $x$-direction pairing, and (d) $y$-direction pairing of the Hamiltonian Eq.~(\ref{eq:HZM1}) (with bond operator parameters~Eq.~(\ref{eq:hijparams})) that commutes with the Majorana zero modes.}
    \label{fig:imagezms2d}
\end{figure}

To illustrate, we provide as input to our framework a pair of two spatially separated ``Majorana'' MZMs shaped according to a portrait of Ettore Majorana, shown in Fig.~\ref{fig:imagezms2d}(a). As we did above, we also lay down bond operators uniformly onto a 2D square lattice so that the nearest-neighbor hopping between sites is $-t$. The Hamiltonian that we find, whose parameters are shown in Fig.~\ref{fig:imagezms2d}(b)-(d), is both complicated and simple. It is complicated because of the non-trivial spatial distributions of the pairing and chemical potential terms, but it is also simple because of its locality and non-interacting nature.

\begin{figure}
    \begin{center}
    \includegraphics[width=0.48\textwidth]{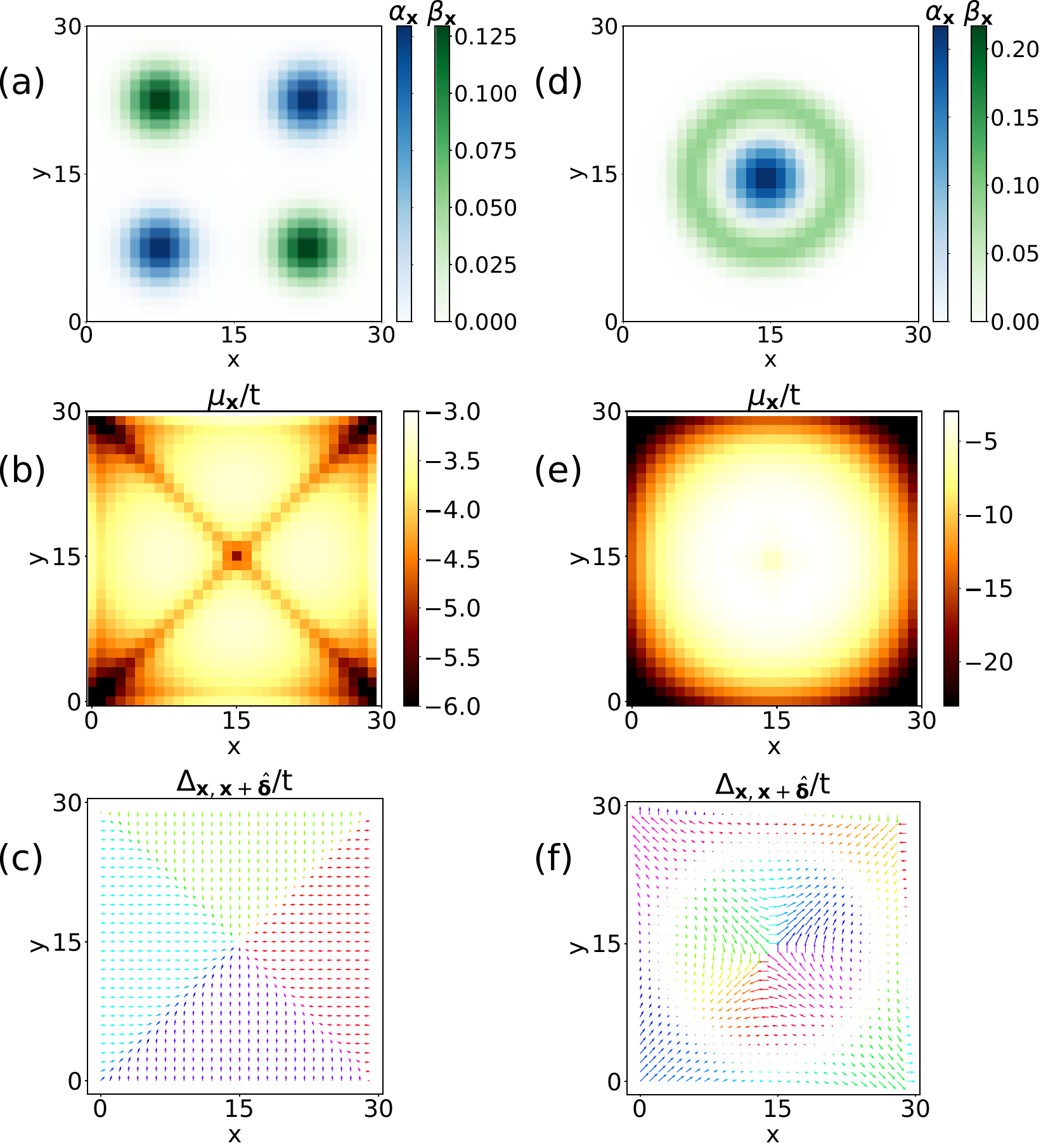}
    \end{center}
    \caption{Examples of Hamiltonians on a $30 \times 30$ square lattice that commute with exotic zero modes. (a) Two zero modes, each of which is split into two well-separated Gaussians. The (b) chemical potential $\mu_{\vec{x}}/t$ and (c) pairing, represented as a vector $(\Delta_{\vec{x},\vec{x}+\hat{\vec{x}}}/t, \Delta_{\vec{x},\vec{x}+\hat{\vec{y}}}/t)$, of the Hamiltonian Eq.~(\ref{eq:HZM1}) (with bond operator parameters~Eq.~(\ref{eq:hijparams})) that commutes with the two split zero modes. (d) A Gaussian-shaped zero mode surrounded by a ring-shaped zero mode. The (e) chemical potential and (f) pairing of the Hamiltonian that commutes with the Gaussian and ring zero modes. The colors of the vectors in (c) and (f) correspond to their angles.}
    \label{fig:nonmajoranazms2d}
\end{figure}

We also construct Hamiltonians that commute with exotic zero modes with non-trivial spatial distributions that possess some but not all of the properties of MZMs. Recall that a MZM is (1) spatially localized into a single location and (2) well separated from other MZMs. Here we consider two examples of zero modes that break the first of these two properties. As we did above, we consider a 2D square lattice geometry for our Hamiltonians and require constant hopping between sites. In our first example, shown in Fig.~\ref{fig:nonmajoranazms2d}(a), we provide as input a pair of zero modes that are spatially localized into \emph{two} locations but still well-separated from one another. The parameters of the resulting Hamiltonian are shown in Fig.~\ref{fig:nonmajoranazms2d}(b)-(c). In our second example, shown in Fig.~\ref{fig:nonmajoranazms2d}(d), we provide as input a pair of well-separated zero modes in which one of the modes is spatially delocalized into a ring surrounding the other zero mode. The parameters of the Hamiltonian that commutes with the pair of zero modes are shown in Fig.~\ref{fig:nonmajoranazms2d}(e)-(f). In both cases, the chemical potential distributions and pairing distributions required to produce these zero modes are quite non-trivial.

Note that for the Hamiltonians depicted in Figs.~\ref{fig:imagezms2d}~and~\ref{fig:nonmajoranazms2d}, we use the same notation (replacing the labels $i,j\rightarrow \vec{x},\vec{x}+\vec{\delta}$) as we did for the Gaussian zero modes, but the $\mu_{\vec{x}}, \Delta_{\vec{x},\vec{x}+\vec{\delta}}$ parameters are from the more general Eq.~(\ref{eq:hijparams}) instead of the special-case Eq.~(\ref{eq:gaussianparams}).

\noindent \textbf{Example: $s$-wave superconducting Hamiltonians with Majorana zero modes.} While we have restricted our attention to spinless fermions with $p$-wave superconductivity, it is also possible to use our framework to construct spinful Majorana zero mode Hamiltonians with different superconducting order parameters. However, building these models can come at the expense of breaking symmetries, such as spinful time-reversal symmetry and spin conservation in the $z$-direction; or the complication of employing spin-orbit coupling or applying a local magnetic field. To add spins to our models, we substitute our labels with $i,j\rightarrow i\sigma,j\sigma'$ where $i,j$ now correspond to sites and $\sigma,\sigma' \in \{\uparrow,\downarrow\}$ to spins.

Using our framework, we construct a 1D $s$-wave superconducting Hamiltonian with exponentially localized MZMs on the edges in two steps. (1) First, we construct two disconnected number-conserving Hamiltonians, one for the spin-up fermions and one for the spin-down fermions, that commute with four zero modes (two at each edge). (2) Second, we add a coupling between the up and down spins that does not commute with two of the four zero modes, leaving only one zero mode on each edge. Below we describe the two steps in detail.

(1) Consider the spin-up fermions. We first want to construct a Hamiltonian that commutes with a pair of spin-up exponentially decaying edges modes, both of which decay at rate $0<r<1$ from the left edge: $\hat{\gamma}_{\uparrow}^{(1)}\propto\sum_{j=1}^{L} r^j \hat{a}_{j\uparrow}, \hat{\gamma}_{\uparrow}^{(2)}\propto\sum_{j=1}^{L} r^j \hat{b}_{j\uparrow}$. For these two zero modes, the corresponding bond operator between neighboring sites in the chain is $\hat{h}_{j\uparrow,j+1\uparrow}=(\hat{c}^\dagger_{j\uparrow}\hat{c}_{j+1,\uparrow}+\textrm{H.c.})-r\hat{n}_{j\uparrow}-r^{-1}\hat{n}_{j+1\uparrow}$. Note that because the two zero modes are decaying in the same direction this operator has no superconducting pairing, only hopping and chemical potential. Now, consider the spin-down fermions. We would like to construct another Hamiltonian that commutes with two spin-down zero modes decaying exponentially at the same rate from the right edge: $\hat{\gamma}_{\downarrow}^{(1)}\propto\sum_{j=1}^{L} r^{-j} \hat{a}_{j\downarrow}, \hat{\gamma}_{\downarrow}^{(2)}\propto\sum_{j=1}^{L} r^{-j} \hat{b}_{j\downarrow}$. For these zero modes, the bond operator between neighboring sites is $\hat{h}_{j\downarrow,j+1\downarrow}=(\hat{c}^\dagger_{j\downarrow}\hat{c}_{j+1,\downarrow}+\textrm{H.c.})-r^{-1}\hat{n}_{j\downarrow}-r\hat{n}_{j+1\downarrow}$. We can add the spin-up and spin-down bond operators together to construct a 1D chain Hamiltonian $-t\sum_{j=1}^{L-1}\sum_{\sigma}\hat{h}_{j\sigma,j+1\sigma}$ that commutes with the four zero modes $\hat{\gamma}_{\uparrow}^{(1)},\hat{\gamma}_{\uparrow}^{(2)},\hat{\gamma}_{\downarrow}^{(1)},\hat{\gamma}_{\downarrow}^{(2)}$. 

(2) To ensure that only one zero mode persists at each edge, we add a perturbing bond operator $\hat{h}_{j\uparrow,j\downarrow}=i\hat{b}_{j\uparrow}\hat{a}_{j\downarrow} = -\hat{c}^\dagger_{j\uparrow}\hat{c}_{j\downarrow} - \hat{c}^\dagger_{j\uparrow}\hat{c}^\dagger_{j\downarrow}+\textrm{H.c.}$ that commutes with $\hat{\gamma}^{(1)}_{\uparrow}$ and $\hat{\gamma}^{(2)}_{\downarrow}$ but not with $\hat{\gamma}^{(1)}_{\downarrow}$ and $\hat{\gamma}^{(2)}_{\uparrow}$. 

Altogether, the Hamiltonian that we construct is
\begin{align}
\hat{H}_{s\textrm{-wave}} &= -t\sum_{j=1}^{L-1}\sum_{\sigma=\uparrow,\downarrow}\hat{h}_{j\sigma,j+1\sigma}+\Delta_s \sum_{j=1}^L \hat{h}_{j\uparrow,j\downarrow} \nonumber \\ 
&=-t\sum_{j=1}^{L-1}\sum_{\sigma=\uparrow,\downarrow}(\hat{c}^\dagger_{j\sigma}\hat{c}_{j+1\sigma}+\textrm{H.c.})-\mu\sum_{j=2}^{L-1} \hat{n}_{j} \nonumber \\
&\quad -\Delta_s \sum_{j=1}^{L}(\hat{c}^\dagger_{j\uparrow}\hat{c}_{j\downarrow}+\hat{c}^\dagger_{j\uparrow}\hat{c}^\dagger_{j\downarrow}+\textrm{H.c.}) + \hat{H}_{\textrm{edge}}, \label{eq:swavesc}
\end{align}
where $t$ and $\Delta_s$ are real free parameters, $\hat{n}_j \equiv \hat{n}_{j\uparrow}+\hat{n}_{j\downarrow}$, $\mu = -t(r+r^{-1})$, and the edge term is
\begin{align}
\hat{H}_{\textrm{edge}} &= t(r\hat{n}_{1\uparrow} + r^{-1}\hat{n}_{1\downarrow} + r^{-1}\hat{n}_{L\uparrow} + r\hat{n}_{L\downarrow}),
\end{align}
which involves a chemical potential and magnetic field. This Hamiltonian is an $s$-wave superconductor that commutes with only the two desired Majorana zero modes $\hat{\gamma}^{(1)}_{\uparrow}$ and $\hat{\gamma}^{(2)}_{\downarrow}$. In this case, this Hamiltonian breaks time-reversal symmetry, does not conserve $z$-magnetization, involves spin-orbit coupling of the same strength as the pairing, and requires a finely tuned magnetic field at the edge. This edge magnetic field, however, we do not expect to be essential and could possibly be removed by slightly modifying the spatial distributions of the input zero modes.

\noindent \textbf{Interacting Hamiltonians with zero modes.} Finally, we mention how to construct \emph{interacting} Hamiltonians that commute with particular zero modes. The main fact to note is that if Hermitian operators $\hat{A}$ and $\hat{B}$ commute with a zero mode $\hat{\gamma}^{(1)}$, then so do the Hermitian operators $i[\hat{A}, \hat{B}]$ and $\{\hat{A}, \hat{B}\}$ if they are non-zero. For example, for bond operators $\hat{h}_{ij}$ and $\hat{h}_{jk}$ that commute with $\hat{\gamma}^{(1)}$ and $\hat{\gamma}^{(2)}$, the operator $\{\hat{h}_{ij}, \hat{h}_{jk}\} \neq 0$ also commutes with $\hat{\gamma}^{(1)}$ and $\hat{\gamma}^{(2)}$. Using this observation, we can construct the class of fermion-parity-conserving interacting Hamiltonians
\begin{align*}
\hat{H}_{IZM} &= \sum_{ij} c_{ij} \hat{h}_{ij} + \sum_{ijkl} d_{ijkl} \{\hat{h}_{ij},\hat{h}_{kl}\} + \sum_{ijkl} e_{ijkl} i[\hat{h}_{ij},\hat{h}_{kl}] \\
&\quad\quad+ \sum_{ijklmn} f_{ijklmn} \{\hat{h}_{ij},\{\hat{h}_{kl},\hat{h}_{mn}\}\} + \cdots
\end{align*}
where the coefficients $c_{ij},d_{ijkl},\ldots$ form a connected graph. These Hamiltonians often contain complicated interacting terms such as $\hat{n}_i \hat{n}_j$, $\hat{n}_i \hat{c}^\dagger_j \hat{c}_k + \textrm{H.c.}$, $\hat{n}_i \hat{c}^\dagger_j \hat{c}^\dagger_k + \textrm{H.c.}$, etc.

As an example, consider the $s$-wave Hamiltonian Eq.~(\ref{eq:swavesc}) that we constructed above. To this Hamiltonian we can add an interacting term between neighboring sites of the form $\sum_j\{\hat{h}_{j\uparrow,j\downarrow}, \hat{h}_{j+1\uparrow,j+1\downarrow}\}/2 = -\sum_j\hat{b}_{j\uparrow}\hat{a}_{j\downarrow}\hat{b}_{j+1\uparrow}\hat{a}_{j+1\downarrow}$ and still have the resulting Hamiltonian commute with exactly the same two zero modes. When written in terms of complex fermions $\hat{c}_j$ and $\hat{c}_j^\dagger$, this term is a sum of eight quartic fermionic operators, many of which do not conserve particle number.

\section{$Z_2$ quantum spin liquid Hamiltonians} \label{sec:z2spinliquids}

\begin{figure}
    \begin{center}\includegraphics[width=0.44\textwidth]{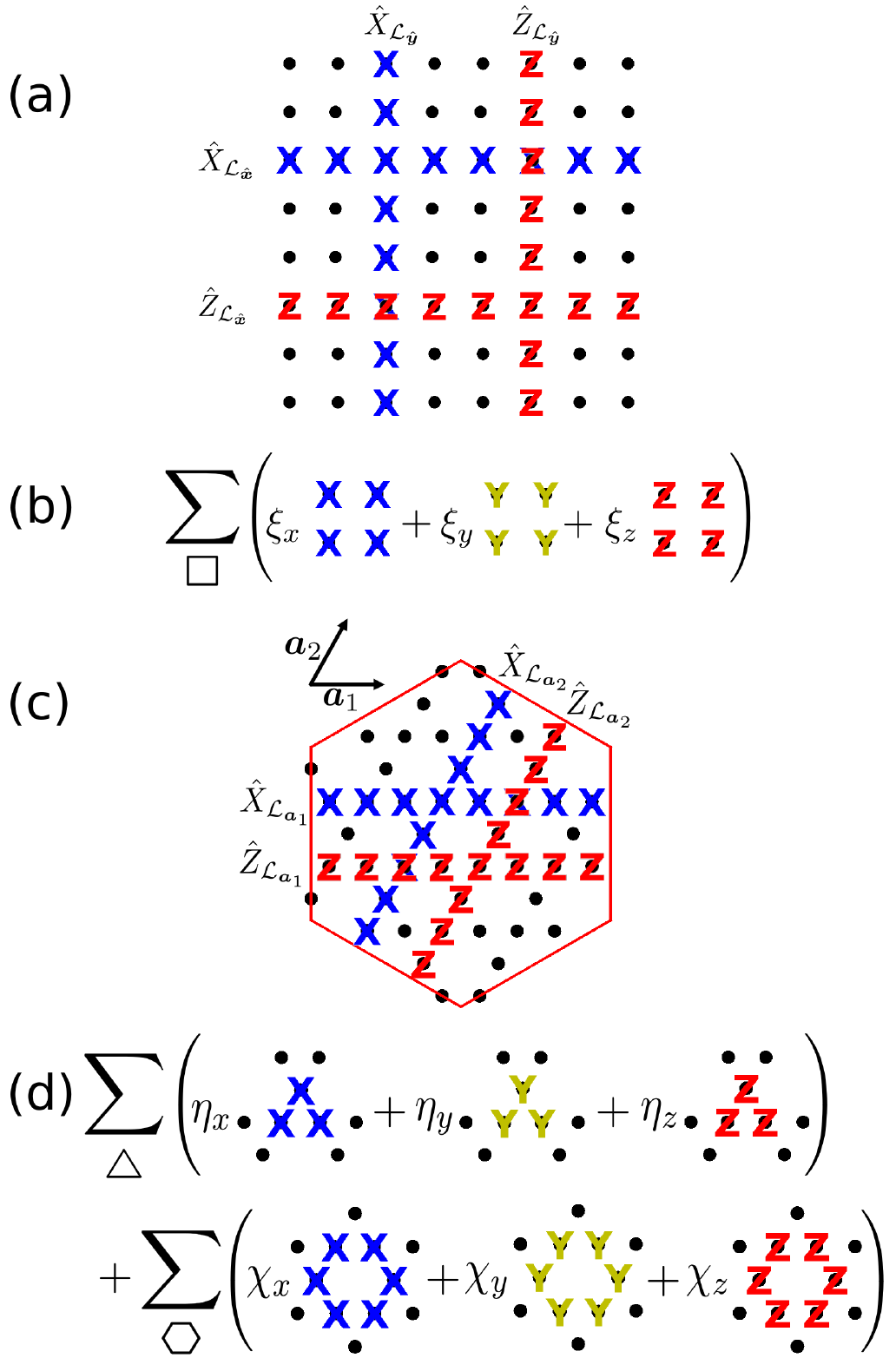} 
    \end{center}
    \caption{(a) The Wilson loops on the square lattice provided as input to the SHC method. (b) The Hamiltonians produced as output by SHC, which commute with the Wilson loops in (a) and obey the square lattice symmetries. (c) The Wilson loops on the kagome lattice provided as input to the SHC method. (d) The Hamiltonians produced as output by SHC, which commute with the Wilson loops in (c) and obey the kagome lattice symmetries. Note that the triangle summation includes all upward and downward facing triangles.}
    \label{fig:spin_liquid_summary}
\end{figure}

In this section, we use the SHC to numerically construct large classes of new $Z_2$ quantum spin liquid Hamiltonians on the square and kagome lattices. We discover many Hamiltonians that commute with the Wilson loops shown in Fig.~\ref{fig:spin_liquid_summary}. All of these Hamiltonians have at least four-fold degenerate ground states. We perform exact diagonalization on these Hamiltonians and determine that many have \emph{exactly} four-fold ground state degeneracy. For many of these Hamiltonians with four-fold ground states, we compute the modular $S$-matrix and find that it exactly matches the modular $S$-matrix of $Z_2$ spin liquids. Generically, the $Z_2$ spin liquid Hamiltonians that we find are not sums of commuting projectors, nor frustration-free. For some of these Hamiltonians, the Wilson loops are ``rigid'', i.e., they are only of a fixed length. In other Hamiltonians we find that some Wilson loops can be deformed, or extended to arbitrary length, like Wilson loops in the toric code. 

The Hamiltonians with deformable Wilson loops possess many local integrals of motion, while those with rigid loops do not. For none of these Hamiltonians do the integrals of motion that we identify form a complete mutually commuting set that fully specifies all of the eigenstates. While it is possible that there are integrals of motion we do not know about, we are able to explicitly rule out some types of integrals of motion.  An exhaustive numerical search rules out the existence of any additional truly local integrals of motion up to some range. A novel eigenstate clustering approach, discussed in Appendix~\ref{sec:levelspacing}, rules out the existence of a complete set of integrals of motion that commutes with a class of (Wilson-loop-preserving) perturbations of the Hamiltonian. Finally, we consider the level-spacing statistics of the $Z_2$ spin liquid Hamiltonians under these perturbations. For the energy levels of these perturbed Hamiltonians in individual quantum number sectors, we find GOE level-spacing statistics suggesting non-integrable behavior. 

\noindent \textbf{Background.} Wilson loops arise as integrals of motion in $Z_2$ quantum spin liquids. Consider a square lattice of spins wrapped into a torus so that there are periodic boundaries in both directions. On the torus, we can form two topologically inequivalent non-contractible loops $\mathcal{L}_{\hat{\vec{x}}}$ and $\mathcal{L}_{\hat{\vec{y}}}$ that span the entire system. Consider four non-local Wilson loop operators $\hat{X}_{\mathcal{L}_{\hat{\vec{x}}}}, \hat{X}_{\mathcal{L}_{\hat{\vec{y}}}}, \hat{Z}_{\mathcal{L}_{\hat{\vec{x}}}}$ and $\hat{Z}_{\mathcal{L}_{\hat{\vec{y}}}}$ with non-trivial support along these two loops. Suppose that these Wilson loop operators are integrals of motion of a Hamiltonian $\hat{H}$, that they square to identity (so that their eigenvalues are $\pm 1$), and that they obey the following set of commutation and anti-commutation relations
\begin{align}
[\hat{X}_{\mathcal{L}_{\hat{\vec{x}}}}, \hat{X}_{\mathcal{L}_{\hat{\vec{y}}}}] = [\hat{Z}_{\mathcal{L}_{\hat{\vec{x}}}}, \hat{Z}_{\mathcal{L}_{\hat{\vec{y}}}}] = 0 \nonumber \\
[\hat{X}_{\mathcal{L}_{\hat{\vec{x}}}},\hat{Z}_{\mathcal{L}_{\hat{\vec{x}}}}] = [\hat{X}_{\mathcal{L}_{\hat{\vec{y}}}},\hat{Z}_{\mathcal{L}_{\hat{\vec{y}}}}] = 0 \nonumber \\
\{\hat{X}_{\mathcal{L}_{\hat{\vec{x}}}},\hat{Z}_{\mathcal{L}_{\hat{\vec{y}}}}\} = \{\hat{Z}_{\mathcal{L}_{\hat{\vec{x}}}},\hat{X}_{\mathcal{L}_{\hat{\vec{y}}}}\} = 0. \label{eq:wilsonlooprelations}
\end{align}
The existence of Wilson loops satisfying these properties implies at least a four-fold degeneracy in each of the energy eigenstates of the Hamiltonian $\hat{H}$.

In this work, we will focus on a particularly simple set of Wilson loops of the form
\begin{align}
\hat{X}_{\mathcal{L}_{\hat{\vec{x}}}} = \prod_{j\in \mathcal{L}_{\hat{\vec{x}}}}\hat{\sigma}^x_j, \quad \hat{Z}_{\mathcal{L}_{\hat{\vec{x}}}} = \prod_{j\in \mathcal{L}_{\hat{\vec{x}}}}\hat{\sigma}^z_j, \nonumber \\
\hat{X}_{\mathcal{L}_{\hat{\vec{y}}}} = \prod_{j\in \mathcal{L}_{\hat{\vec{y}}}}\hat{\sigma}^x_j, \quad \hat{Z}_{\mathcal{L}_{\hat{\vec{y}}}} = \prod_{j\in \mathcal{L}_{\hat{\vec{y}}}}\hat{\sigma}^z_j, \label{eq:wilsonloops}
\end{align} 
where $\mathcal{L}_{\hat{\vec{x}}}$ and $\mathcal{L}_{\hat{\vec{y}}}$ are two \emph{straight-line} topologically distinct non-contractible loops across the torus that wind in the horizontal and vertical directions, respectively (see Fig.~\ref{fig:spin_liquid_summary}(a)). It can be verified that the Wilson loops of Eq.~(\ref{eq:wilsonloops}) square to identity and satisfy the properties of Eq.~(\ref{eq:wilsonlooprelations}). These particular Wilson loops are integrals of motion of the toric code, a well known solvable $Z_2$ spin liquid Hamiltonian that is a sum of commuting terms \cite{Kitaev1998}. Note that there are also $\hat{Y}_{\mathcal{L}} \propto \hat{X}_{\mathcal{L}} \hat{Z}_{\mathcal{L}}$ Wilson loops, though they are not independent from the ones in Eq.~(\ref{eq:wilsonloops}). Also, there are actually many straight-line Wilson loop operators whose loops are parallel to $\mathcal{L}_{\hat{\vec{x}}}$ and $\mathcal{L}_{\hat{\vec{y}}}$, though shifted by $j\hat{\vec{y}}$ or $k\hat{\vec{x}}$. We will refer to these shifted loops as $\mathcal{L}_{\hat{\vec{x}}}^{(j)}$ and $\mathcal{L}_{\hat{\vec{y}}}^{(k)}$ for $j,k=1,\ldots,L$. In addition to these Wilson loops on the square lattice, we also consider straight-line Wilson loops of the same form as Eq.~(\ref{eq:wilsonloops}) on the periodic kagome lattice. We refer to the kagome Wilson loops as $\hat{Z}_{\mathcal{L}_{\vec{a}_1}}, \hat{X}_{\mathcal{L}_{\vec{a}_1}}, \hat{Z}_{\mathcal{L}_{\vec{a}_2}}, \hat{X}_{\mathcal{L}_{\vec{a}_2}}$, where $\vec{a}_1=(1,0)$ and $\vec{a}_2=(1/2,\sqrt{3}/2)$ are the kagome lattice vectors (see Fig.~\ref{fig:spin_liquid_summary}(c)).

Finally, we would like to mention the form of the toric code on the kagome lattice, as it will be relevant to our later discussion. The toric code model can be defined on many lattices, such as the honeycomb lattice. While it is customary for the spins of the toric code to be on the links of the lattice, in this work we always consider spins to be on the sites. When the spins are defined on sites instead of links, the honeycomb toric code gets mapped to the kagome lattice. This model, which we will refer to as the kagome toric code model, is
\begin{align}
\hat{H}_{TC,\textrm{kagome}} = -\sum_{\triangle}\hat{X}_{\triangle} - \sum_{\hexagon} \hat{Z}_{\hexagon} \label{eq:toriccodekagome}
\end{align}
where the $\hat{X}_{\triangle}=\prod_{j \in \triangle} \hat{\sigma}^x_j$ and $\hat{Z}_{\hexagon}=\prod_{j \in \hexagon} \hat{\sigma}^z_j$ operators are three-spin and six-spin interactions defined on the triangles and hexagons, respectively, of the lattice. The model has the same essential features as the square lattice toric code: it is a sum of commuting terms, the $\hat{X}_{\triangle},\hat{Z}_{\hexagon}$ are local integrals of motion, and the model commutes with the straight-line Wilson loops $\hat{Z}_{\mathcal{L}_{\vec{a}_1}}, \hat{X}_{\mathcal{L}_{\vec{a}_1}}, \hat{Z}_{\mathcal{L}_{\vec{a}_2}}, \hat{X}_{\mathcal{L}_{\vec{a}_2}}$.

\begin{figure}
    \begin{center}
    \includegraphics[width=0.48\textwidth]{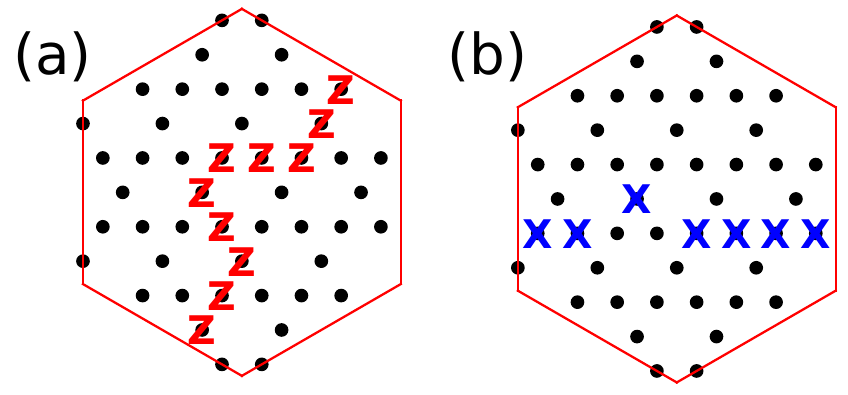} 
    \end{center}
    \caption{(a) A $Z$-Wilson loop deformed around a hexagon. (b) An $X$-Wilson loop deformed around a triangle.}
    \label{fig:deformed_wilson_loops}
\end{figure}

\subsection{$Z_2$ spin liquid Hamiltonians on the square lattice}

\noindent \textbf{SHC numerics.} To generate new $Z_2$ spin liquid Hamiltonians on the square lattice, we provided a list of desired symmetries as input to SHC: (1) the four straight-line Wilson loop operators $\hat{X}_{\mathcal{L}_{\hat{\vec{x}}}}, \hat{Z}_{\mathcal{L}_{\hat{\vec{x}}}}, \hat{X}_{\mathcal{L}_{\hat{\vec{y}}}}, \hat{Z}_{\mathcal{L}_{\hat{\vec{y}}}}$ (see Fig.~\ref{fig:spin_liquid_summary}(a)); and (2) the symmetry group of the square lattice, generated by translations of  lattice vectors $\hat{\vec{x}}, \hat{\vec{y}}$, $90^\circ$-rotation, and reflection about the side of a square. Note that using the $\hat{X},\hat{Z}$ pair of Wilson loops is an arbitrary choice and we could have instead used the $\hat{X},\hat{Y}$ or $\hat{Y},\hat{Z}$ pair. Given this input, one might expect that the SHC would produce the toric code as output, since it is an example of a square lattice Hamiltonian that commutes with Wilson loops of this form. However, it does \emph{not} because the toric code's translational symmetry is generated by translations of $2\hat{\vec{x}}$ and $2\hat{\vec{y}}$ instead of $\hat{\vec{x}}$ and $\hat{\vec{y}}$.

We performed the SHC calculations on a finite-size $N = 8\times 8 = 64$ site lattice. We used a basis of range-$R$ $k$-local Pauli strings, where $R=2$ and $k \in \{1,2,3,4,5\}$; this includes up to five-spin interactions on nearest, next-nearest, and next-next nearest neighbor sites on the square lattice. Before symmetrization by spatial symmetries, this basis was $67,584$-dimensional. After symmetrization, the basis was reduced to a $234$-dimensional basis of spatially-symmetric Hamiltonians. In this $234$-dimensional basis, we numerically constructed the commutant matrices $C_{\hat{X}_{\mathcal{L}_{\hat{\vec{x}}}}},C_{\hat{Z}_{\mathcal{L}_{\hat{\vec{x}}}}},\ldots$ for the four Wilson loops and found three vectors that were null vectors of all of these matrices. These three vectors correspond to the coupling constants of three Hamiltonians with all of the desired symmetries. This three-dimensional space of symmetric Hamiltonians takes the form
\begin{align}
\hat{H}_{\textrm{square}} = \sum_{\Square} \left( \xi_x\hat{X}_{\Square} + \xi_y\hat{Y}_{\Square} + \xi_z \hat{Z}_{\Square}\right), \label{eq:Hsquarexyz}
\end{align}
where $\hat{X}_{\Square}=\prod_{j \in \Square} \hat{\sigma}^x_j, \hat{Y}_{\Square}=\prod_{j \in \Square} \hat{\sigma}^y_j,$ and $\hat{Z}_{\Square}=\prod_{j \in \Square} \hat{\sigma}^z_j$ are four-spin interactions on the nearest neighbor squares of the lattice and $\xi_x,\xi_y,\xi_z$ are arbitrary real constants. These Hamiltonians are depicted in Fig.~\ref{fig:spin_liquid_summary}(b).

\noindent \textbf{Numerical checks of $Z_2$ order.} While the Hamiltonians of Eq.~(\ref{eq:Hsquarexyz}) commute with the Wilson loop operators, it is not guaranteed that they are $Z_2$ spin liquids. In this section, we numerically tested particular Hamiltonians in this space to check if they have $Z_2$ topological order. By construction, these Hamiltonians commute with Wilson loops satisfying Eq.~(\ref{eq:wilsonlooprelations}) and so are guaranteed to have eigenstates with degeneracies that are multiples of four. However, it is possible that they have \emph{greater} than four-fold degeneracy, either due to the existence of accidental degeneracy or additional symmetries that we did not require.

For each $\hat{H}_{\textrm{square}}$ that we tested, we used exact diagonalization (ED) to determine if the ground state was \emph{exactly} four-fold degenerate. If it was, we then calculated the modular $S$-matrix, a quantity that encodes the properties of anyons in a topologically ordered system \cite{Nayak2008,Zhang2012}. The modular $S$-matrix is an overlap matrix $S_{ij}\equiv \langle \Xi_i^{\hat{\vec{x}}}| \Xi_j^{\hat{\vec{y}}}\rangle$ between minimally-entangled-states (MES) $\ket{\Xi^{\hat{\vec{x}}}_i}$ and $\ket{\Xi^{\hat{\vec{y}}}_i}$ for $i=1,\ldots,4$ across loops $\mathcal{L}_{\hat{\vec{x}}}$ and $\mathcal{L}_{\hat{\vec{y}}}$ of the torus. The MES are particular linear combinations of the four degenerate ground states in the system that minimizes the Renyi entanglement entropy along the cuts defined by loops $\mathcal{L}_{\hat{\vec{x}}}$ and $\mathcal{L}_{\hat{\vec{y}}}$. For the toric code, the MES are $e$ and $m$ flux eigenstates along the two loops, i.e., they are eigenstates of the Wilson loop operators $\hat{Z}_{\mathcal{L}}, \hat{X}_{\mathcal{L}}$ \cite{Zhang2012}. Assuming this holds for the $\hat{H}_{\textrm{square}}$ Hamiltonians as well, we computed the MES for our Hamiltonians by finding the flux eigenstates of $\hat{Z}_{\mathcal{L}_{\vec{\delta}}}, \hat{X}_{\mathcal{L}_{\vec{\delta}}}$ for ${\vec{\delta}}=\hat{\vec{x}}, \hat{\vec{y}}$ from the four-fold degenerate ground states. In particular, for ${\vec{\delta}}=\hat{\vec{x}}$ and $\hat{\vec{y}}$, we built the set of four MES $\ket{\Xi_1^{\vec{\delta}}},\ldots,\ket{\Xi_4^{\vec{\delta}}}$ by computing the (unique) ground state of $\hat{H}_{\textrm{square}} - \kappa_e \hat{Z}_{\mathcal{L}_{\vec{\delta}}} - \kappa_m \hat{X}_{\mathcal{L}_{\vec{\delta}}}$ for $(\kappa_e,\kappa_m)=(+1,+1), (+1,-1), (-1,+1), (-1,-1)$, respectively. We numerically verified for the $\xi_x=\xi_y=\xi_z=-1$ Hamiltonian on the $4 \times 4$ lattice that the MES that we used did in fact minimize the Renyi entanglement entropy across the $\mathcal{L}_{\hat{\vec{x}}}$ and $\mathcal{L}_{\hat{\vec{y}}}$ cuts of the torus, as is expected \cite{Zhang2012}.

Using ED, we checked $27$ particular $\hat{H}_{\textrm{square}}$ Hamiltonians on a $4 \times 4$ square lattice to determine if they were $Z_2$ spin liquids. We considered the $3^3=27$ possible Hamiltonians with $\xi_\alpha$ set to $-1,0,+1$ for each $\alpha=x,y,z$. Six of these Hamiltonians (up to cyclic permutations of $\xi_x,\xi_y,\xi_z$), listed in Table~\ref{tab:z2squaremodel}, had exactly four degenerate ground states. For these six Hamiltonians, we computed the modular $S$-matrix and found it to be \footnote{Each MES is determined up to a phase. We chose those phases so that the first row and column of the modular $S$-matrix are real and positive.}
\begin{align}
S = \frac{1}{2}
\begin{pmatrix}
1 & 1 & 1 & 1 \\
1 & 1 & -1 & -1 \\
1 & -1 & 1 & -1 \\
1 & -1 & -1 & 1 
\end{pmatrix} \label{eq:smatrix}
\end{align}
up to numerical precision. This $S$-matrix corresponds to $Z_2$ topological order \cite{Zhang2012}.

\begin{table}

    \begin{center}
    \begin{tabular}{|c||c|c|c|c|c|c|}
    \hline
    $\xi_x$ & $0$ & $0$ & $0$ & $0$ & $+1$ & $-1$ \\ 
    %\hline
    $\xi_y$ & $+1$ & $+1$ & $-1$ & $-1$ & $+1$ & $-1$ \\ 
    %\hline
    $\xi_z$ & $+1$ & $-1$ & $+1$& $-1$ & $-1$ & $-1$ \\
    \hline
    \end{tabular}
    \end{center}
    
    \caption{Six square lattice Hamiltonians $\hat{H}_{\textrm{square}}$ of the form of Eq.~(\ref{eq:Hsquarexyz}) that have $Z_2$ topological order on the $4\times 4$ lattice. This table excludes cyclic permutations of $(\xi_x,\xi_y,\xi_z)$ that also have $Z_2$ topological order.}
    \label{tab:z2squaremodel}
\end{table}

\noindent \textbf{Symmetries of the discovered Hamiltonians.} Here we consider the symmetries of the Hamiltonians that we found. By construction, these models possess straight-line Wilson loops as integrals of motion and obey the spatial symmetries of the square lattice. However, it is possible that these Hamiltonians possess additional symmetries that we did not require.

We numerically determined that the $\hat{H}_{\textrm{square}}$ Hamiltonians listed in Table~\ref{tab:z2squaremodel} do not possess highly local integrals of motions. We verified this numerically by using the slow operator forward method, i.e., by computing the commutant matrices $C_{\hat{H}_{\textrm{square}}}$ for each of the Hamiltonians in Table~\ref{tab:z2squaremodel} with a local basis of operators. For the Hamiltonians tested, we found no integrals of motion with up to $9$-site terms on a local $3\times 3$ square cluster of sites in the lattice. The fact that these Hamiltonians do not possess such local integrals of motion suggests that the Wilson loops of these models are ``rigid,'' i.e., cannot be locally deformed. 

Even without any apparent local integrals of motion, we are able to identify a set of mutually commuting integrals of motion built from the straight-line Wilson loops. Consider an $L_x \times L_y$ square lattice. Without local integrals of motion, the $\hat{Z}_{\mathcal{L}_{\hat{\vec{x}}}^{(j)}}, \hat{Z}_{\mathcal{L}_{\hat{\vec{y}}}^{(k)}}$ Wilson loops for $j=1,\ldots,L_y; k=1,\ldots,L_x-1$ are all \emph{independent} conserved quantities \footnote{The $\hat{Z}_{\mathcal{L}_{\hat{\vec{x}}}^{(j)}}, \hat{Z}_{\mathcal{L}_{\hat{\vec{y}}}^{(k)}}$ operators satisfy $\prod_{j=1}^{L_y}\hat{Z}_{\mathcal{L}_{\hat{\vec{x}}}^{(j)}}=\prod_{k=1}^{L_x}\hat{Z}_{\mathcal{L}_{\hat{\vec{y}}}^{(k)}}$, so one of the operators is dependent on the rest.}. The same is true for the $X$ Wilson loops, but the $X$ Wilson loops do not commute with all of the $Z$ Wilson loops (see Eq.~(\ref{eq:wilsonlooprelations})). Nonetheless, products of \emph{two} $X$ loops do commute with all of the $Z$ loops. Therefore, the set of operators
\begin{align}
\{\hat{Z}_{\mathcal{L}_{\hat{\vec{x}}}^{(j)}}\}_{j=1}^{L_y}, \quad \{\hat{Z}_{\mathcal{L}_{\hat{\vec{y}}}^{(k)}}\}_{k=1}^{L_x-1}, \nonumber \\ \{\hat{X}_{\mathcal{L}_{\hat{\vec{x}}}^{(j)}}\hat{X}_{\mathcal{L}_{\hat{\vec{x}}}^{(j+1)}} \}_{j=1}^{L_y-1}, \quad \{ \hat{X}_{\mathcal{L}_{\hat{\vec{y}}}^{(k)}}\hat{X}_{\mathcal{L}_{\hat{\vec{y}}}^{(k+1)}}\}_{k=1}^{L_x-2} \label{eq:squareioms}
\end{align}
form a set of $2L_x+2L_y-4$ mutually commuting operators.

Generically, the square lattice $\hat{H}_{\textrm{square}}$ Hamiltonians of Eq.~(\ref{eq:Hsquarexyz}) also possess a global integral of motion that is a sum of Pauli strings. These Hamiltonians can be written as a sum of two commuting operators $\hat{H}_{\textrm{square}} = \hat{A} + \hat{B}$, where 
\begin{align}
\hat{A}&\equiv \sum_{\Square_A}(\xi_x \hat{X}_{\Square_A} + \xi_y \hat{Y}_{\Square_A} + \xi_z \hat{Z}_{\Square_A}) \\
\hat{B}&\equiv \sum_{\Square_B} (\xi_x \hat{X}_{\Square_B} + \xi_y \hat{Y}_{\Square_B} + \xi_z \hat{Z}_{\Square_B}), \label{eq:squareglobaliom}
\end{align}
and $\Square_A$ are the ``black squares'' and $\Square_B$ are the ``white squares'' of a black-white checkboard pattern laid down on the square lattice. Since $[\hat{A}, \hat{B}]=0$, we can consider one of these operators, say $\hat{B}$, as a global integral of motion, so that $[\hat{H}_{\textrm{square}}, \hat{B}]=0$. The $\hat{B}$ operator also commutes with the integrals of motion listed in Eq.~(\ref{eq:squareioms}).

Note that the $\hat{H}_{\textrm{square}}$ Hamiltonians of Eq.~(\ref{eq:Hsquarexyz}) are not sums of commuting terms nor frustration-free, making them difficult to solve analytically. The set of mutually commuting integrals of motion of Eqs.~(\ref{eq:squareioms})~and~(\ref{eq:squareglobaliom}) is not enough to fully diagonalize the $\hat{H}_{\textrm{square}}$ Hamiltonians, but can be used to block diagonalize them into quantum number sectors, allowing us to more effectively study these models numerically.

\noindent \textbf{Level-spacing statistics.} Typically, the level-spacing statistics of a quantum Hamiltonian are either distributed according to the Gaussian orthogonal ensemble (GOE) distribution or the Poisson distribution. If the level-spacing statistics are GOE distributed, then that is good evidence that the system is non-integrable. For Hamiltonians with many integrals of motion, such as the ones we found, the level-spacing statistics will generally appear Poisson when considering energies spread out over multiple quantum number sectors of the integrals of motion. However, it is possible for the statistics to be GOE in particular quantum number sectors.

\begin{figure}
    \begin{center}
    \includegraphics[width=0.48\textwidth]{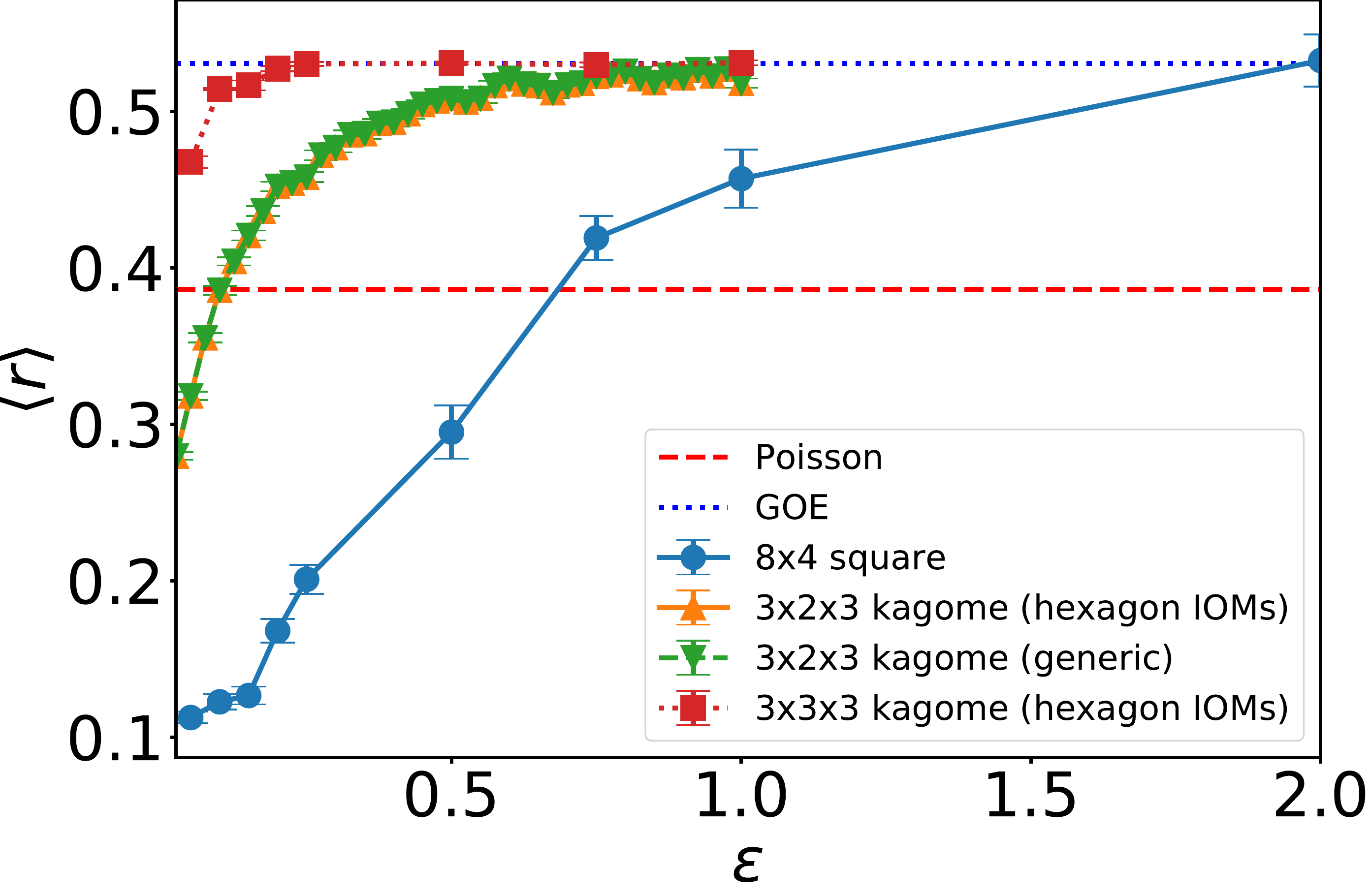}
    \end{center}
    \caption{The average level-spacing ratios versus disorder strength $\epsilon$ for the Hamiltonian Eq.~(\ref{eq:squareperturbed}) on an 32-site square lattice (8x4 square), the Hamiltonian Eq.~(\ref{eq:kagomeperturbed1}) on an 18-site kagome lattice (3x2x3 kagome (hexagon IOMs)), the Hamiltonian Eq.~(\ref{eq:kagomeperturbed2}) on an 18-site kagome lattice (3x2x3 kagome (generic)), and the Hamiltonian Eq.~(\ref{eq:kagomeperturbed1}) on a 27-site kagome lattice (3x3x3 kagome (hexagon IOMs)). The 18-site kagome lattice calculations were averaged over 100 random Hamiltonians, while the others were averaged over 10 random Hamiltonians. The energy levels considered were obtained in particular quantum number sectors, as described in Appendix~\ref{sec:levelspacing}.}
    \label{fig:levelspacingratios}
\end{figure}

Using ED, we numerically examined the level-spacing statistics of particular $\hat{H}_{\textrm{square}}$ Hamiltonians, i.e., the statistics of the level-spacing ratio $r_n=\min(\delta_n,\delta_{n-1})/\max(\delta_n, \delta_{n-1})$ where $\delta_n \equiv E_n - E_{n-1}$ and $E_n$ are the sorted energy eigenvalues of the Hamiltonian. For the Poisson distribution the average level-spacing ratio is expected to be $\langle r_{\textrm{Poisson}}\rangle =0.3863$, while for the GOE distribution it should be $\langle r_{\textrm{GOE}}\rangle =0.5307$. After accounting for the quantum number sectors we know about -- i.e., the ones listed in Eqs.~(\ref{eq:squareioms})~and~(\ref{eq:squareglobaliom}) -- we observed significant degeneracy in the spectrum, leaving only a small number of unique energy levels on the $8 \times 4$ square lattice that we considered. To break this degeneracy, we perturbed the sixth Hamiltonian of Table~\ref{tab:z2squaremodel} by a random perturbation, resulting in the Hamiltonian
\begin{align}
-\sum_{\Square}(\hat{X}_{\Square}+\hat{Y}_{\Square}+\hat{Z}_{\Square}) + \epsilon \delta \hat{H} \label{eq:squareperturbed}
\end{align}
where $\delta\hat{H}= \sum_{\Square}h_{\Square} \hat{Z}_{\Square}$ and $h_{\Square}$ are random numbers drawn from the uniform distribution between $-1$ and $1$ for each square and $\epsilon$ is the disorder strength. This particular perturbation breaks the spatial symmetries of the square lattice, but preserves the Wilson loop integrals of motion in Eq.~(\ref{eq:squareioms}) (and a modified global integral of motion Eq.~(\ref{eq:squareglobaliom})) and generically destroys the eigenstate degeneracy within the known quantum number sectors. 
After this perturbation, there are still eigenstates in a given quantum number sector that do not couple with each other.  We cluster these eigenstates by grouping together the connected set of states that couple through the perturbation to the Hamiltonian (see Appendix~\ref{sec:levelspacing}).  This coupling still leaves multiple eigenstates per cluster and suggests the existence of some ``hidden'' integrals of motion we have not explicitly identified. 
For 10 random realizations of Eq.~(\ref{eq:squareperturbed}) for $\epsilon$ from 0 to 6, we computed the average level-spacing ratio within these clusters. As shown in Fig.~\ref{fig:levelspacingratios}, as $\epsilon$ increases to 2 the average level-spacing ratio $\langle r \rangle$ approaches the GOE value. However, it does decrease slightly below that value for larger $\epsilon$, as shown in Fig.~\ref{fig:levelspacingratiossquare}. For the $8 \times 4$ square lattice, the eigenstate clusters we found were quite small, typically containing either 28 or 35 states. We also tested 10 random realizations of these perturbed Hamiltonians with disorder strengths $\epsilon=0.25,0.5,0.75,1$ on a $4 \times 4$ square lattice and observed that they always possessed exactly four-fold degenerate ground states, suggesting that the perturbed models are also $Z_2$ spin liquids.

\subsection{$Z_2$ spin liquid Hamiltonians on the kagome lattice}

\noindent \textbf{SHC numerics.} To construct new $Z_2$ quantum spin liquid Hamiltonians on the kagome lattice, we provided as input to SHC: (1) the four straight-line Wilson loop operators $\hat{X}_{\mathcal{L}_{\vec{a}_1}}, \hat{Z}_{\mathcal{L}_{\vec{a}_1}}, \hat{X}_{\mathcal{L}_{\vec{a}_2}}, \hat{Z}_{\mathcal{L}_{\vec{a}_2}}$ (see Fig.~\ref{fig:spin_liquid_summary}(c)); and (2) the symmetry group of the kagome lattice generated by translations of lattice vectors $\vec{a}_1$ and $\vec{a}_2$, $60^\circ$-rotation, and reflection.

We performed our SHC calculations on the finite-size $N = 48$ site symmetric cluster shown in Fig.~\ref{fig:spin_liquid_summary}(c). To construct Hamiltonians, we used a basis of range-$R$ $3$-local and $6$-local Pauli strings, where $R=2/\sqrt{3}$. Before symmetrization by spatial symmetries, this basis was $31,536$-dimensional. After symmetrization, the basis was reduced to a $220$-dimensional space. In this large space of local Hamiltonians, we found the following six-dimensional space of symmetric Hamiltonians that exactly obey all of the desired symmetries
\begin{align}
\hat{H}_{\textrm{kagome}} &= \sum_{\triangle}\left( \eta_x \hat{X}_{\triangle} + \eta_y \hat{Y}_{\triangle} + \eta_z \hat{Z}_{\triangle}\right) \nonumber \\
&\quad\quad+ \sum_{\hexagon} \left(\chi_x \hat{X}_{\hexagon} + \chi_y \hat{Y}_{\hexagon} + \chi_z \hat{Z}_{\hexagon}\right), \label{eq:Hkagomexyz}
\end{align}
where $\eta_\alpha,\chi_\alpha$ for $\alpha=x,y,z$ are arbitrary real constants, $\hat{X}_{\triangle}=\prod_{j \in \triangle} \hat{\sigma}^x_j, \hat{Y}_{\hexagon}=\prod_{j \in \hexagon} \hat{\sigma}^y_j,\ldots$ , and the summations are over all of the triangles (both upward and downward facing) and hexagons in the kagome lattice. These Hamiltonians are depicted in Fig.~\ref{fig:spin_liquid_summary}(d).

We also note that we performed the same SHC calculations with a basis of range-$R$ $k$-local Pauli strings with $R=2/\sqrt{3}$ and $k \in \{1,2,3,4,5,6\}$ and found six additional symmetric Hamiltonians, which involve four and five-site interactions. These Hamiltonians can be written in terms of products of triangle operators:
\begin{align*}
\sum_{\langle\triangle, \triangle'\rangle} \hat{X}_{\triangle}\hat{X}_{\triangle'}, \, \sum_{\langle\triangle, \triangle'\rangle} \hat{Y}_{\triangle}\hat{Y}_{\triangle'}, \, \sum_{\langle\triangle, \triangle'\rangle} \hat{Z}_{\triangle}\hat{Z}_{\triangle'}, \nonumber \\
\sum_{\langle\triangle, \triangle'\rangle} i(\hat{X}_{\triangle}\hat{Y}_{\triangle'} + \hat{X}_{\triangle'}\hat{Y}_{\triangle}), \, \sum_{\langle\triangle, \triangle'\rangle} i(\hat{X}_{\triangle}\hat{Z}_{\triangle'} + \hat{X}_{\triangle'}\hat{Z}_{\triangle}), \nonumber \\
\sum_{\langle\triangle, \triangle'\rangle} i(\hat{Y}_{\triangle}\hat{Z}_{\triangle'} + \hat{Y}_{\triangle'}\hat{Z}_{\triangle}),
\end{align*}
where the summations are over nearest-neighbor triangles $\triangle$ and $\triangle'$ that overlap at a single site. While we include these Hamiltonians for completeness, we will not examine their properties in the discussion below. We instead will focus on the Hamiltonians of Eq.~(\ref{eq:Hkagomexyz}).

\noindent \textbf{Numerical checks of $Z_2$ order.} Using ED on finite-size kagome lattices, we found Hamiltonians of the form of Eq.~(\ref{eq:Hkagomexyz}) that exhibit $Z_2$ topological order. On a $3\times 2\times 3=18$ site kagome lattice, we considered $\hat{H}_{\textrm{kagome}}$ Hamiltonians with all possible $3^3=27$ combinations of $\eta_\alpha=-1,0,+1$ for $\alpha=x,y,z$ with $(\chi_x,\chi_y,\chi_z)=(0,0,-1)$. We found that 24 of these 27 Hamiltonians had four-fold degenerate ground states and exactly the $Z_2$ modular $S$-matrix of Eq.~(\ref{eq:smatrix}), which we computed in the same way as described in the previous section. The three Hamiltonians that did not have these properties were the effectively classical $(\eta_x,\eta_y,\eta_z)=(0,0,-1),(0,0,0),(0,0,+1)$ Hamiltonians.

\noindent \textbf{Symmetries of discovered Hamiltonians.} By construction, all of the $\hat{H}_{\textrm{kagome}}$ Hamiltonians of Eq.~(\ref{eq:Hkagomexyz}) obey the symmetries of the kagome lattice and commute with the straight-line Wilson loops. Yet, generic $\hat{H}_{\textrm{kagome}}$ Hamiltonians do not possess any highly local integrals of motion. We checked this numerically using the slow operator forward method. In particular, we computed the commutant matrix $C_{\hat{H}_{\textrm{kagome}}}$ for $100$ random $\hat{H}_{\textrm{kagome}}$ Hamiltonians, with $\eta_\alpha,\chi_\alpha$ sampled uniformly from $[-1,1]$, and found that they had no null vectors for local bases of Pauli strings on the triangles, bowties, and hexagons of a $48$-site kagome lattice. For generic $\hat{H}_{\textrm{kagome}}$ Hamiltonians, these calculations suggest that there are no local integrals of motion and that the Wilson loops are rigid like they were for the square lattice Hamiltonians. For these generic Hamiltonians, we can identify the following set of mutually commuting integrals of motion
\begin{align}
\{\hat{Z}_{\mathcal{L}_{\vec{a}_1}^{(j)}}, \, \hat{Z}_{\mathcal{L}_{\vec{a}_2}^{(k)}}, \, \hat{X}_{\mathcal{L}_{\hat{\vec{a}_1}}^{(j)}}\hat{X}_{\mathcal{L}_{\hat{\vec{a}_1}}^{(j+1)}}, \, \hat{X}_{\mathcal{L}_{\hat{\vec{a}_2}}^{(k)}}\hat{X}_{\mathcal{L}_{\hat{\vec{a}_2}}^{(k+1)}}\}. \label{eq:kagome_genericioms}
\end{align}
However, as we discuss below, particular subspaces of these Hamiltonians possess different sets of mutually commuting integrals of motion, which do include local integrals of motion and deformable Wilson loops.

For example, the Hamiltonians in the four-dimensional subspace of Hamiltonians
\begin{align}
\sum_{\triangle}\left( \eta_x \hat{X}_{\triangle} + \eta_y \hat{Y}_{\triangle} + \eta_z \hat{Z}_{\triangle}\right) + \sum_{\hexagon} \chi_z \hat{Z}_{\hexagon}, \label{eq:hexagoniom_ham}
\end{align}
where $\chi_z\neq 0$, commute with $\hat{Z}_{\hexagon}$ for all hexagons. For these models, we can therefore build a large set of mutually commuting operators
\begin{align}
\{\hat{Z}_{\hexagon}, \, \hat{Z}_{\mathcal{L}_{\vec{a}_1}}, \, \hat{Z}_{\mathcal{L}_{\vec{a}_2}}, \, \hat{X}_{\mathcal{L}_{\hat{\vec{a}_1}}^{(j)}}\hat{X}_{\mathcal{L}_{\hat{\vec{a}_1}}^{(j+1)}}, \, \hat{X}_{\mathcal{L}_{\hat{\vec{a}_2}}^{(k)}}\hat{X}_{\mathcal{L}_{\hat{\vec{a}_2}}^{(k+1)}}\}. \label{eq:kagome_hexagonioms}
\end{align}
Moreover, for these models, the Wilson loops are partially deformable: no Wilson loops can be deformed around triangles, but the $Z$ Wilson loops can be deformed around hexagons. Similar subspaces with different hexagon operators, e.g., with $\chi_x\neq 0, \chi_y=\chi_z=0$, also exist and have the same properties.

Likewise, Hamiltonians of the form
\begin{align}
\sum_{\triangle} \eta_z \hat{Z}_{\triangle} + \sum_{\hexagon} \left(\chi_x \hat{X}_{\hexagon} + \chi_y \hat{Y}_{\hexagon} + \chi_z \hat{Z}_{\hexagon}\right), \label{eq:triangleiom_ham}
\end{align}
where $\eta_z\neq 0$, commute with $\hat{Z}_{\triangle}$ for all triangles. For these models, we can build an even larger set of mutually commuting operators
\begin{align}
\{\hat{Z}_{\triangle}, \, \hat{Z}_{\mathcal{L}_{\vec{a}_1}}, \, \hat{Z}_{\mathcal{L}_{\vec{a}_2}}, \, \hat{X}_{\mathcal{L}_{\hat{\vec{a}_1}}^{(j)}}\hat{X}_{\mathcal{L}_{\hat{\vec{a}_1}}^{(j+1)}}, \, \hat{X}_{\mathcal{L}_{\hat{\vec{a}_2}}^{(k)}}\hat{X}_{\mathcal{L}_{\hat{\vec{a}_2}}^{(k+1)}}\}, \label{eq:kagome_triangleioms}
\end{align}
since there are more triangles than hexagons in the kagome lattice. For these models, the $Z$ Wilson loops can be deformed around triangles, but no Wilson loops can be deformed around hexagons.

\begin{figure}
    \begin{center}
     \includegraphics[width=0.4\textwidth]{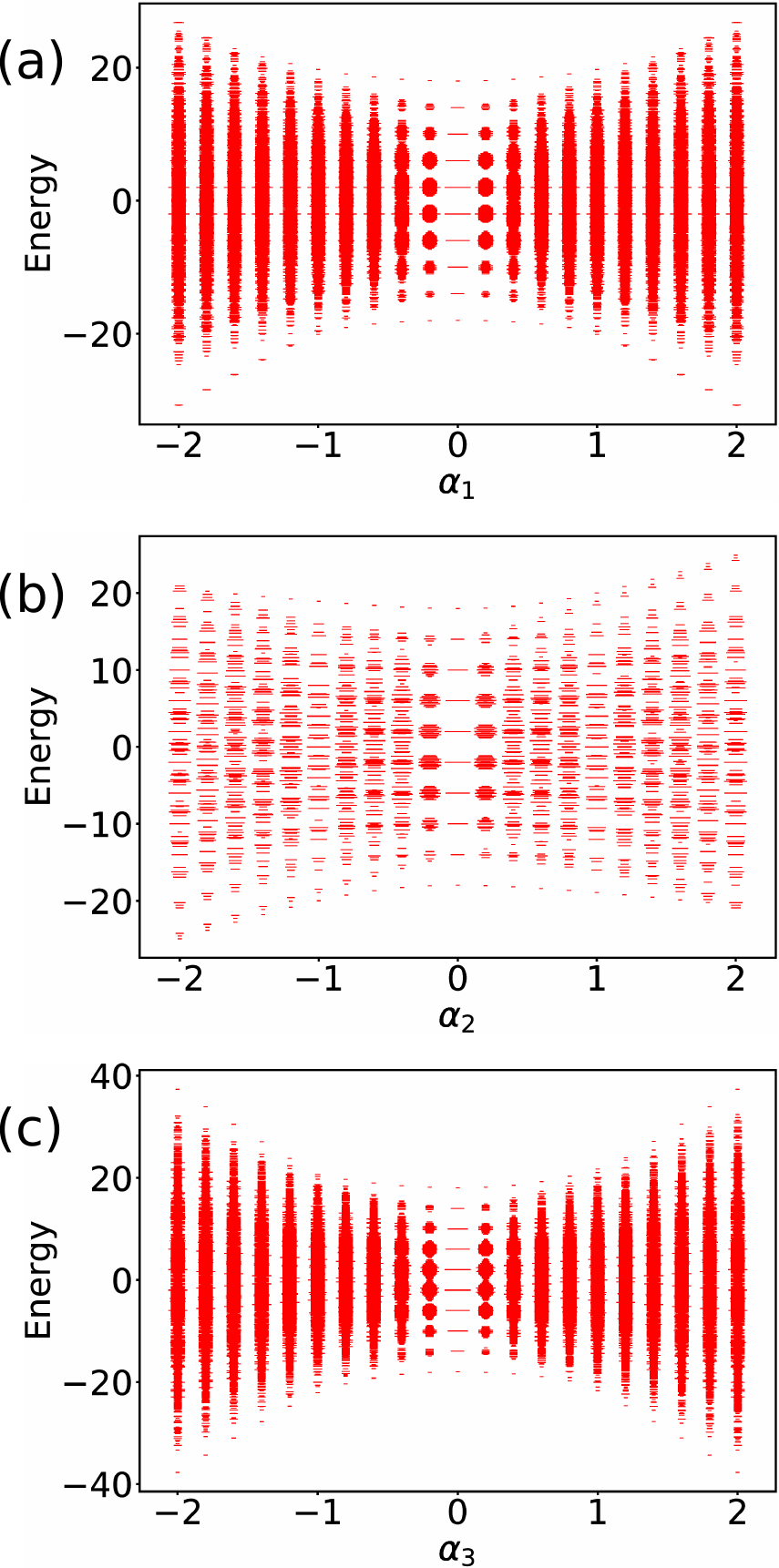} 
    \end{center}
    \caption{The full spectra of the Hamiltonians (a) $\hat{H}_{TC,\textrm{kagome}} + \alpha_1 \sum_{\triangle}\hat{Z}_{\triangle}$, (b) $\hat{H}_{TC,\textrm{kagome}} + \alpha_2 \sum_{\hexagon}\hat{X}_{\hexagon}$, and (c) $\hat{H}_{TC,\textrm{kagome}} + \alpha_3(\sum_{\triangle}\hat{Z}_{\triangle} + \sum_{\hexagon}\hat{X}_{\hexagon})$ on a $3 \times 2 \times 3$ kagome lattice. The width of each horizontal line corresponds to the degeneracy of the energy eigenstates, plotted on a log scale. The smallest degeneracy is four-fold, which occurs for all of the ground states of these models, and the largest degeneracy is about $80,000$-fold, which occurs for the eigenstates in the center of the toric code's spectrum ($\alpha_1=\alpha_2=\alpha_3=0$).}
    \label{fig:kagome_perturbed_toric_codes}
\end{figure}

Interestingly, the kagome lattice toric code of Eq.~(\ref{eq:toriccodekagome}) is a particularly special point, $(\eta_x,\eta_y,\eta_x,\chi_x,\chi_y,\chi_z)=(-1,0,0,0,0,-1)$, in the space of Hamiltonians. It is special in that \emph{both} $X$ triangles and $Z$ hexagons are local integrals of motion, making $X$ Wilson loops deformable about triangles and $Z$ Wilson loops deformable around hexagons (see Fig.~\ref{fig:deformed_wilson_loops}). Since the toric code is an exactly solvable model, it is interesting to consider perturbing it to better understand the other models in this space of Hamiltonians, which have different symmetry properties. Fig.~\ref{fig:kagome_perturbed_toric_codes} shows the full spectra of three families of Hamiltonians that all include the toric code as a special point. The Hamiltonians of Fig.~\ref{fig:kagome_perturbed_toric_codes}(a) are of the type described in Eq.~(\ref{eq:hexagoniom_ham}), the Hamiltonians of Fig.~\ref{fig:kagome_perturbed_toric_codes}(b) are of the type described in Eq.~(\ref{eq:triangleiom_ham}), and the Hamiltonians of Fig.~\ref{fig:kagome_perturbed_toric_codes}(c) are generic $\hat{H}_{\textrm{kagome}}$ Hamiltonians of the type described in Eq.~(\ref{eq:Hkagomexyz}). Notably, the ground state for each of the Hamiltonians shown stayed exactly four-fold degenerate, though the gap to the first-excited states did decrease with large enough perturbations. % "Since..." and figure move to appendix

Generically, the kagome lattice $\hat{H}_{\textrm{kagome}}$ Hamiltonians of Eq.~(\ref{eq:Hkagomexyz}), like the square lattice models, also possess a global integral of motion that is a sum of Pauli strings. A generic $\hat{H}_{\textrm{kagome}}$ Hamiltonian is a sum of two commuting terms, $\hat{H}_{\textrm{kagome}} = \hat{C} + \hat{D}$, where 
\begin{align}
\hat{C}&=\sum_{\triangle}(\eta_x \hat{X}_{\triangle} + \eta_y \hat{Y}_{\triangle} + \eta_z \hat{Z}_{\triangle}) \\
\hat{D}&=\sum_{\hexagon}(\chi_x \hat{X}_{\hexagon} + \chi_y \hat{Y}_{\hexagon} + \chi_z \hat{Z}_{\hexagon}) \label{eq:kagomeglobaliom}
\end{align}
are the terms only on the triangles or hexagons of the kagome lattice, respectively. We can take one of these operators, say $\hat{D}$, as a global integral of motion, since $[\hat{H}_{\textrm{kagome}},\hat{D}]=0$.

The $\hat{H}_{\textrm{kagome}}$ Hamiltonians of Eq.~(\ref{eq:Hkagomexyz}) are not sums of commuting terms nor frustration-free. Generically, the set of mutually commuting integrals of motion that we found do not fully specify the degrees of freedom of the model and cannot be used to fully diagonalize the $\hat{H}_{\textrm{kagome}}$ Hamiltonians, except at the special points corresponding to the toric code model. However, the set of mutually commuting integrals of motion Eq.~(\ref{eq:kagome_genericioms}) (or Eq.~(\ref{eq:kagome_hexagonioms})~or~(\ref{eq:kagome_triangleioms})) and Eq.~(\ref{eq:kagomeglobaliom}) can be used to block diagonalize the Hamiltonians into quantum number sectors, which allows us to more effectively study these models numerically.

\noindent \textbf{Level-spacing statistics.} We examined the level-spacing statistics of the Hamiltonians we discovered with the SHC to see if they obeyed GOE or Poisson statistics in their quantum number sectors. Initially, we computed the level-spacing statistics of the Hamiltonians in Fig.~\ref{fig:kagome_perturbed_toric_codes} for $3\times 2 \times 3=18$ and $3\times 3 \times 3=27$ site kagome lattices in the quantum number sectors of Eq.~(\ref{eq:kagome_genericioms}) (or Eq.~(\ref{eq:kagome_hexagonioms})~or~(\ref{eq:kagome_triangleioms})) and Eq.~(\ref{eq:kagomeglobaliom}). Like we observed for the square lattice Hamiltonians, the kagome lattice Hamiltonians possess large eigenstate degeneracy.  We again remove these degeneracies by considering perturbed Hamiltonians of the form
 \begin{align}
&-\sum_{\triangle} (\hat{X}_{\triangle}+\hat{Z}_{\triangle}) - \sum_{\hexagon} \hat{Z}_{\hexagon} + \epsilon\delta \hat{H}' \label{eq:kagomeperturbed1} \\
&-\sum_{\triangle} (\hat{X}_{\triangle}+\hat{Z}_{\triangle}) - \sum_{\hexagon} (\hat{X}_{\hexagon} + \hat{Z}_{\hexagon}) + \epsilon\delta \hat{H}' \label{eq:kagomeperturbed2}
 \end{align}
 where $\delta\hat{H}'=\sum_{\triangle} h_{\triangle} \hat{Z}_{\triangle}$ and $h_{\triangle}$ are random numbers drawn from the uniform distribution between $-1$ and $1$ for each triangle and $\epsilon$ is the disorder strength. These particular perturbations break the spatial symmetries of the kagome lattice, but still preserve the integrals of motion described in the previous section. 
 
 On 18-site lattices, for both Hamiltonians Eqs.~(\ref{eq:kagomeperturbed1})~and~(\ref{eq:kagomeperturbed2}), the $\delta\hat{H}'$ perturbation was sufficient to produce GOE statistics at intermediate $\epsilon$ for the 256 energy levels in the considered quantum number sectors, as shown in Fig.~\ref{fig:levelspacingratios}.  On the 27-site lattice (for which we only considered Hamiltonian Eq.~(\ref{eq:kagomeperturbed1})), this perturbation was not sufficient to produce GOE statistics; instead we found that there were additional unidentified ``hidden'' quantum number sectors that were affecting the level-spacing results. Using the eigenstate clustering approach described in Appendix~\ref{sec:levelspacing}, we identified (within one sector) the set of eigenstates that correspond to a hidden quantum number sector without identifying their corresponding integrals of motion. This procedure involved looking at the set of energy eigenstates coupled through the perturbation $\delta \hat{H}'$. After accounting for the hidden quantum numbers, our originally 8192-dimensional sector was reduced to four 2048-dimensional hidden sectors, suggesting that for the 27-site lattice there are likely two missing binary integrals of motion that we were unable to directly identify. Using the energy levels in the hidden sectors, we computed the average level-spacing ratio and observed that it indeed approached the GOE value as $\epsilon$ increased, in fact doing so much more rapidly than for the 18-site lattice (see Fig.~\ref{fig:levelspacingratios}). This seems to suggest that the non-GOE behavior for low $\epsilon$ could be a finite-size effect. We also numerically determined the ground state degeneracy of 10 random realizations of the Hamiltonians of Eqs.~(\ref{eq:kagomeperturbed1})~and~(\ref{eq:kagomeperturbed2}) with disorder strengths $\epsilon=0.25,0.5,0.75,1$ on an 18-site kagome lattice. For all random Hamiltonians considered, the ground states were exactly four-fold degenerate, suggesting that these perturbed models are still $Z_2$ spin liquids.

\section{Discussion and conclusions} \label{sec:conclusions}

We have introduced a new approach, the symmetric Hamiltonian construction (SHC), for constructing Hamiltonians with desired symmetries. 
This method is general and can be used to construct families of Hamiltonians that commute (or anti-commute) with desired integrals of motion, such as zero mode and Wilson loop operators, and are invariant (or anti-invariant) under discrete symmetry transformations, such as point-group symmetries.
In this work, we applied the SHC approach to design new topologically ordered Hamiltonians by providing as input to the method integrals of motion with topological properties and spatial symmetries.
We analytically determined large families of superconducting Hamiltonians with Majorana zero modes (MZMs) and numerically found new $Z_2$ spin liquid Hamiltonians on the square and kagome lattices.

Using the SHC, we developed a general framework for designing Hamiltonians that commute with a pair of zero mode operators. 
In this framework, we provide as input the spatial distribution of two zero modes and as output produce Hamiltonians that commute exactly with those two zero modes and no others. 
These Hamiltonians can be put onto arbitrary lattice geometries, e.g., square, kagome, tetrahedral, or quasicrystal lattices, or even arbitrary graphs.
Some examples of Hamiltonians that we designed with this framework are: a one-dimensional $s$-wave superconducting Hamiltonian that commutes with two exponentially-localized MZMs, a two-dimensional $p$-wave superconducting Hamiltonian that commutes with Gaussian-localized MZMs, and two-dimensional Hamiltonians that commute with exotic semi-localized zero modes.

Using the SHC numerically, we discovered new classes of $Z_2$ spin liquid Hamiltonians on the square and kagome lattices whose properties differ from known solvable models.
The Hamiltonians are not sums of commuting projectors nor are they frustration-free.
Generically, these Hamiltonians possess many integrals of motion, though not enough to fully diagonalize the Hamiltonian.
We observed that particular perturbations of these Hamiltonians, which break spatial symmetries but preserve the integrals of motion, exhibit GOE level-spacing statistics in their quantum number sectors.
Many of the Hamiltonians that we found, even with these perturbations, possess numerically exact four-fold ground state degeneracy in finite-size systems.

There are many future directions for using the SHC to study topological order.
Our general framework for designing zero mode Hamiltonians opens the door to the design of new experiments to search for MZMs.
Using our framework, it is now possible for experimentalists to design a Hamiltonian with MZMs that best fits their experimental constraints, rather than focusing on a particular idealized model such as the Kitaev chain. 
To realize these MZM Hamiltonians in the lab, these experiments need some degree of control over the spatial distribution of some combination of chemical potentials, magnetic fields, superconducting pairings, or hoppings, in fermionic superconducting systems. 
Such control can potentially be realized in existing experiments, such as in Josephson junction arrays \cite{Hassler2012}, superconductor-semiconductor heterostructures \cite{Lutchyn2018}, twisted bilayer graphene \cite{Yankowitz2019}, and strain-modulated superconductors \cite{Hicks2014,Bachmann2019}. 
Our framework also potentially allow theorists to design new model Hamiltonians with new exotic zero mode physics, such as zero modes that realize non-Ising non-Abelian anyons or zero modes displaying the physics of higher-order topological insulators and superconductors \cite{Benalcazar2017,Schindler2018}.
One can also use similar techniques as we used to discover $Z_2$ spin liquids in order to find other topologically ordered Hamiltonians. 
For example, by providing as input different topological symmetry operators, such as different types of Wilson loops, one can attempt to discover new model Hamiltonians with double semion or Fibonacci anyons.

Broadly speaking, the SHC is a tool that can be used to systematically study Hamiltonians with symmetries of interest.
For example, it can be used to find all local Hamiltonians consistent with particular crystallographic symmetries.
Once identified, these models can then be studied numerically or analytically to better understand their behavior.
The SHC method can also be used to generate (potentially interacting) realizations of Hamiltonians from well-known symmetry classifications, such as the ``ten-fold way'' classification of non-interacting topological insulators and superconductors \cite{Ryu2010}.
It could also be used to engineer Hamiltonians with properties that make them easier to study.
Certain classes of interacting fermionic Hamiltonians with specific symmetries are known to be sign-problem-free, a property that allows for their efficient numerical simulation using methods such as quantum Monte Carlo \cite{Wu2005,Zheng2011,Li2016}. 
The SHC method could be used to generate particular realizations of Hamiltonians with such symmetries, and thereby provide many new sign-problem-free Hamiltonians for numerical study.
The availability of new numerically solvable models could provide valuable insights into strongly correlated systems.

\section{Acknowledgements}

We acknowledge useful discussions with David Huse, Barry Bradlyn,  Roger Mong, Curt Von Keyserlingk, David Pekker, Dmitrii Kochkov, Ryan Levy, Jahan Claes, and Di Luo. We also acknowledge useful conversations with Shivaji Sondhi, including a suggestion to consider our spin-liquid's level statistics.  This work was supported by DOE DE-SC0020165.  BV acknowledges support from the Google AI Quantum team. 

\appendix

\newcommand{\beginsupplement}{%
        \setcounter{table}{0}
        \renewcommand{\thetable}{A\arabic{table}}%
        \setcounter{figure}{0}
        \renewcommand{\thefigure}{A\arabic{figure}}%
}
\beginsupplement
\section{Derivation of commutant matrix}

In this section, we derive Eq.~(\ref{eq:comNorm}), which expresses the commutator norm between Hamiltonian $\hat{H}$ and integral of motion $\hat{\mathcal{O}}$ in terms of the commutant matrix.

Consider a space of Hermitian operators spanned by a $d$-dimensional basis of Hermitian operators $\hat{\mathcal{S}}_a$ for $a=1,\ldots,d$. The commutator of two basis vectors in this space satisfies $[\hat{\mathcal{S}}_a,\hat{\mathcal{S}}_b]=-[\hat{\mathcal{S}}_b,\hat{\mathcal{S}}_a]=-[\hat{\mathcal{S}}_a,\hat{\mathcal{S}}_b]^\dagger$ and is therefore anti-Hermitian. An anti-Hermitian operator can be represented as an imaginary number times a Hermitian operator. So if this space of operators is large enough, then the operator $[\hat{\mathcal{S}}_a,\hat{\mathcal{S}}_b]$ can be represented as $[\hat{\mathcal{S}}_a,\hat{\mathcal{S}}_b]=\sum_{c=1}^d f_{ab}^c \hat{\mathcal{S}}_c$, where the expansion coefficients $f_{ab}^c$ are purely imaginary. Note that if this equation holds for all $a,b$ in the space of operators we are considering, then the space is a Lie algebra with the commutator as the Lie bracket and $f_{ab}^c$ as the (basis-dependent) structure constants of the Lie algebra.

Consider two operators $\hat{H}$ and $\hat{\mathcal{O}}$ expressed in our basis so that $\hat{H}=\sum_{a=1}^d J_a \hat{\mathcal{S}}_a$ and $\hat{\mathcal{O}}=\sum_{b=1}^d g_b \hat{\mathcal{S}}_b$ with real $J_a,g_b$. Their commutator is
\begin{align*}
[\hat{H},\hat{\mathcal{O}}] &= \sum_{a,b} J_a g_b [\hat{\mathcal{S}}_a,\hat{\mathcal{S}}_b] = \sum_{a,b,c} J_a g_b f_{ab}^c \hat{\mathcal{S}}_c \\
&= \sum_{a,c} J_a \bigg(\sum_b g_b f_{ab}^c \bigg)\hat{\mathcal{S}}_c = \sum_{a,c} {(L_{\hat{\mathcal{O}}})}_{ca}  J_a \hat{\mathcal{S}}_c
\end{align*}
where ${(L_{\hat{\mathcal{O}}})}_{ca}\equiv \sum_b g_b f_{ab}^c$. The norm of the commutator is then
\begin{align*}
&\norm[\big]{[\hat{H},\hat{\mathcal{O}}]}^2 = \norm[\big]{\sum_{a,c} J_a {(L_{\hat{\mathcal{O}}})}_{ca} \hat{\mathcal{S}}_c}^2 \\
&= \frac{1}{\tr{\hat{I}}}\textrm{tr}\bigg(\bigg( \sum_{a,c} J_a {(L_{\hat{\mathcal{O}}})}_{ca} \hat{\mathcal{S}}_c \bigg)^\dagger \bigg( \sum_{a',c'} J_{a'} {(L_{\hat{\mathcal{O}}})}_{c'a'} \hat{\mathcal{S}}_{c'} \bigg)\bigg) \\
&= \sum_{a,a'} J_a \bigg( \sum_{c,c'} {(L_{\hat{\mathcal{O}}})}_{ca}^* \frac{\tr{\hat{\mathcal{S}}_c^\dagger \hat{\mathcal{S}}_{c'}}}{\tr{\hat{I}}}  {(L_{\hat{\mathcal{O}}})}_{c'a'} \bigg) J_{a'} \\
&= \sum_{a,a'} J_a \bigg( \sum_{c,c'} {(L_{\hat{\mathcal{O}}})}_{ca}^* O_{cc'} {(L_{\hat{\mathcal{O}}})}_{c'a'} \bigg) J_{a'} \\
&= J^T {L_{\hat{\mathcal{O}}}}^\dagger O {L_{\hat{\mathcal{O}}}} J \\
&\equiv J^T C_{\hat{\mathcal{O}}} J
\end{align*}
where $C_{\hat{\mathcal{O}}} \equiv {L_{\hat{\mathcal{O}}}}^\dagger O L_{\hat{\mathcal{O}}}$ is the commutant matrix. The matrix $O_{cc'}=\tr{\hat{\mathcal{S}}_c^\dagger \hat{\mathcal{S}}_{c'}}/\tr{\hat{I}}=\langle \hat{\mathcal{S}}_c, \hat{\mathcal{S}}_{c'} \rangle$ is an overlap (or Gram) matrix, which is Hermitian and positive semi-definite, and is calculated by taking inner products between all operator strings. Since we require that the $\hat{\mathcal{S}}_a$ form a basis, they are linearly independent, making the overlap matrix $O$ non-singular and therefore positive definite. Since $O$ is Hermitian and positive definite, ${L_{\hat{\mathcal{O}}}}^\dagger O {L_{\hat{\mathcal{O}}}}$ is Hermitian and positive semi-definite. For orthonormal bases, $O_{cc'}=\delta_{cc'}$ and the commutant matrix simplifies to $C_{\hat{\mathcal{O}}} = {L_{\hat{\mathcal{O}}}}^\dagger L_{\hat{\mathcal{O}}}$.

\section{Properties of Pauli string basis} \label{sec:pauli_basis}

In this section, we describe some properties of the Pauli string basis, a complete basis for the space of Hermitian operators that can act on $n$ qubits, and discuss how to compute the structure constants $f_{ab}^c$ for this basis. 

The Pauli string operators, which are tensor products of the identity matrix and Pauli matrices, are defined in Eq.~(\ref{eq:pauliStringBasis}). These operators are Hermitian and unitary. These properties imply that the operators square to identity and have eigenvalues $\pm 1$. Moreover, except for the identity operator $\hat{I}$, the operators in this basis are traceless, implying that they have an equal number of $+1$ and $-1$ eigenvalues. Therefore, the Pauli strings $\hat{\mathcal{S}}_a$ are highly degenerate, with two $2^{n-1}$-dimensional degenerate subspaces.

The Pauli string basis is orthonormal with respect to the Frobenius inner product $\langle \hat{A} , \hat{B} \rangle \equiv \tr{\hat{A}^\dagger \hat{B}}/\tr{\hat{I}}$. To see this, consider two Pauli strings
\begin{align}
\hat{\mathcal{S}}_a &= \hat{\sigma}_1^{s_1} \otimes \cdots \otimes \hat{\sigma}^{s_n}_n \nonumber \\
\hat{\mathcal{S}}_b &= \hat{\sigma}_1^{t_1} \otimes \cdots \otimes \hat{\sigma}^{t_n}_n. \label{eq:twoPauliStrings}
\end{align}
The product of two $\hat{\sigma}_i^{s_i}$ operators on the same site $i$ obey the following relations 
\begin{align}
\hat{\sigma}_i^s \hat{\sigma}_i^t = 
\begin{cases}
\delta_{st}\hat{I} + i\epsilon_{stu}\hat{\sigma}_i^{u} & s,t \in \{1,2,3\} \\
\hat{\sigma}^{t}_i & s=0 \\
\hat{\sigma}^{s}_i & t=0
\end{cases}
\label{eq:pauliMatrixProduct}   
\end{align}
where $\epsilon_{stu}$ is the fully antisymmetric tensor. When multiplying two Pauli strings, we multiply operators site by site in the tensor product and apply Eq.~(\ref{eq:pauliMatrixProduct}):
\begin{align*}
\hat{\mathcal{S}}_a \hat{\mathcal{S}}_b &= (\hat{\sigma}_1^{s_1}\hat{\sigma}^{t_1}_1) \otimes \cdots \otimes (\hat{\sigma}^{s_n}_n \hat{\sigma}^{t_n}_n) \\
&= \delta_{s_1t_1}\cdots \delta_{s_n t_n}\hat{I} + \cdots.
\end{align*}
The term with the identity operator, which is non-zero only if $(s_1,\ldots,s_n)=(t_1,\ldots,t_n)$, i.e., only if $\hat{\mathcal{S}}_a=\hat{\mathcal{S}}_b$, is the only operator with non-zero trace in the expansion of the product. Therefore $\langle \hat{\mathcal{S}}_a, \hat{\mathcal{S}}_b \rangle = \tr{\hat{\mathcal{S}}_a \hat{\mathcal{S}}_b}/\tr{\hat{I}}=\delta_{ab}$ and the basis is orthonormal.

Next, we describe in more detail how to compute the product, and thereby the structure constants, of Pauli strings. The commutator of the two strings in Eq.~(\ref{eq:twoPauliStrings}) is
\begin{align*}
[\hat{\mathcal{S}}_a, \hat{\mathcal{S}}_b] &= \hat{\mathcal{S}}_a \hat{\mathcal{S}}_b - \hat{\mathcal{S}}_b \hat{\mathcal{S}}_a \\
&= (\hat{\sigma}_1^{s_1}\hat{\sigma}^{t_1}_1) \otimes \cdots \otimes (\hat{\sigma}^{s_n}_n \hat{\sigma}^{t_n}_n) \\
&\quad- (\hat{\sigma}^{t_1}_1 \hat{\sigma}_1^{s_1}) \otimes \cdots \otimes (\hat{\sigma}^{t_n}_n \hat{\sigma}^{s_n}_n) \\
&\equiv f_{ab}^c \hat{\mathcal{S}}_c.
\end{align*}
To compute the structure constants $f_{ab}^c$ in the final line for a particular $(a,b)$-pair in our basis, we need to apply Eq.~(\ref{eq:pauliMatrixProduct}). To precisely illustrate this calculation, we need to introduce some notation. We say that two Pauli strings ``agree'' on $p$ sites $i_1,\ldots,i_p$ when the operators on those sites match: $s_{i_1} = t_{i_1},\ldots,s_{i_p} = t_{i_p}$. They ``trivially disagree'' on $q$ sites $j_1,\ldots,j_q$ when the operators on those sites do not match, but one of the two operators on the site is identity: $s_{j_1} \neq t_{j_1},\ldots,s_{j_q} \neq t_{j_q}$ and $s_{j_m}=0$ or $t_{j_m}=0$ for all $m=1,\ldots,q$. Finally, the two strings ``non-trivially disagree'' on $r=n-p-q$ sites $k_1,\ldots,k_r$ when the operators do not match and are both non-identity operators: $s_{k_1} \neq t_{k_1},\ldots,s_{k_r} \neq t_{k_r}$ and $s_{k_1},t_{k_1},\ldots,s_{k_r},t_{k_r} \in \{1,2,3\}$. According to Eq.~(\ref{eq:pauliMatrixProduct}), the sites that agree become identity operators, while the sites that disagree, both trivially and non-trivially, become Pauli matrices. We then see that
\begin{align}
&[\hat{\mathcal{S}}_a, \hat{\mathcal{S}}_b] = \hat{\mathcal{S}}_a \hat{\mathcal{S}}_b - \textrm{H.c.} \nonumber \\
&= \prod_{m=1}^q (\delta_{s_{j_m},0}\hat{\sigma}^{t_{j_m}}_{j_m} + \delta_{t_{j_m},0}\hat{\sigma}^{s_{j_m}}_{j_m}) \prod_{l=1}^r \left(i\epsilon_{s_{k_l}t_{k_l}u_{k_l}} \hat{\sigma}^{u_{k_l}}_{k_l}\right) \nonumber\\
&\quad\quad- \textrm{H.c.} 
\end{align}
From this result, we see that if $r$ is even, then the two terms are purely real and cancel, leading to $[\hat{\mathcal{S}}_a,\hat{\mathcal{S}}_b] = 0$ and $f_{ab}^c = 0$ for all $c$. If $r$ is odd, then the two terms are purely imaginary and add, resulting in a single $c$ for which $f_{ab}^c \neq 0$. Therefore, the structure constant associated with $\hat{\mathcal{S}}_a$ and $\hat{\mathcal{S}}_b$ is
\begin{align}
f_{ab}^c = 
\begin{cases}
2i^r \prod_{l=1}^r \epsilon_{s_{k_l}t_{k_l}u_{k_l}} & r \textrm{ is odd} \\
0 & r \textrm{ is even}
\end{cases} \label{eq:pauliStringStructureConstant}
\end{align}
and $\hat{\mathcal{S}}_c = \prod_{m=1}^q (\delta_{s_{j_m},0}\hat{\sigma}^{t_{j_m}}_{j_m} + \delta_{t_{j_m},0}\hat{\sigma}^{s_{j_m}}_{j_m})\prod_{l=1}^r \hat{\sigma}^{u_{k_l}}_{k_l}$ when $r$ is odd.

The structure constants in Eq.~(\ref{eq:pauliStringStructureConstant}) can be computed efficiently, in time $O(q+r)$, for each pair of operator strings $\hat{\mathcal{S}}_a$ and $\hat{\mathcal{S}}_b$. Therefore, for a $d$-dimensional basis of $k$-local Pauli strings, all of the structure constants for the basis can be computed in time $O(kd^2)$. 

\section{Properties of fermion string basis} \label{sec:fermion_basis}

In this section, we describe some properties of the fermion string basis, a complete basis for the space of Hermitian operators that can act on $n$ fermions. A similarly defined basis was considered in Ref.~\onlinecite{Mierzejewski2018}.

Fermion strings, expressed in terms of the standard fermionic creation and anhillation operators $\hat{c}^\dagger_i$ and $\hat{c}_i$, are defined in Eq.~(\ref{eq:fermionStringBasis}). Unlike the Pauli strings, fermion strings are neither unitary, traceless, nor orthonormal according to the Frobenius inner product.

As shown in Eq.~(\ref{eq:fermionStringBasis}), we classify the fermion string operators into three types. 

The first type is of the form $\hat{\mathcal{S}}_a^{(1)}=\hat{c}^\dagger_{i_1} \cdots \hat{c}^\dagger_{i_m} \hat{c}_{i_m} \cdots \hat{c}_{i_1}=\hat{n}_{i_1}\cdots \hat{n}_{i_m}$ and can be interpreted as a product of number operators $\hat{n}_i\equiv \hat{c}_i^\dagger \hat{c}_i$. There are $\sum_{m=0}^n \binom{n}{m}=2^n$ ways to choose the $i_1,\ldots,i_m$ labels and so there are $2^n$ linearly independent $\hat{\mathcal{S}}_{a}^{(1)}$ operators including the identity operator $\hat{I}$. These operators are idempotent, so that $(\hat{\mathcal{S}}_{a}^{(1)})^2=\hat{\mathcal{S}}_{a}^{(1)}$, which means that they only have eigenvalues $0$ or $1$. Consider the Fock space (occupation number) basis states that span the fermionic Hilbert space: $\ket{\{n\}}\equiv \ket{n_1,\ldots,n_n}\equiv ((1-n_1)+n_1 \hat{c}^\dagger_1)\cdots((1-n_1)+n_n \hat{c}^\dagger_n)\ket{0}$ where $n_i \in \{0,1\}$ and $\hat{I}=\sum_{\{n\}}\ket{\{n\}}\bra{\{n\}}$. The $\hat{\mathcal{S}}_{a}^{(1)}$ operators are diagonal in this basis and possess non-zero trace.

The second type of operator we define is of the form $\hat{\mathcal{S}}_a^{(2)}=\hat{c}^\dagger_{i_1} \cdots \hat{c}^\dagger_{i_m} \hat{c}_{j_l} \cdots \hat{c}_{j_1} + \textrm{H.c.}$, where the indices are lexicographically ordered so that $(j_1,\ldots,j_l) < (i_1,\ldots,i_m)$. There are $\sum_{a=1}^{2^n}\sum_{b=1}^{a-1} 1 = 2^{n-1}(2^n - 1)$ ways to choose the labels in these operators and so there are this many linearly independent $\hat{\mathcal{S}}_a^{(2)}$ operators. By working in the Fock space basis, we can see that these operators are traceless. Note that the third type of operator, $\hat{\mathcal{S}}_a^{(3)}=i\hat{c}^\dagger_{i_1} \cdots \hat{c}^\dagger_{i_m} \hat{c}_{j_l} \cdots \hat{c}_{j_1} + \textrm{H.c.}$, has the same properties as the $\hat{\mathcal{S}}_a^{(2)}$ operators.

Altogether, from our counting, we see that the $\{\hat{\mathcal{S}}_a^{(1)},\hat{\mathcal{S}}_a^{(2)},\hat{\mathcal{S}}_a^{(3)}\}$ fermion string basis consists of $2^n + 2\times 2^{n-1}(2^n - 1) = 4^n$ linearly independent Hermitian operators and therefore spans the entire space of Hermitian operators.

Finally, we note that the product of two fermion strings, $\hat{\mathcal{S}}_a \hat{\mathcal{S}}_b$, and therefore the commutator, is non-trivial to calculate. Rather than working out the commutator and structure constants in the fermion string basis, we compute the structure constants in the Majorana string basis, which spans the same space of operators. We can then convert to and from the fermion string basis as needed by applying an invertible basis transformation, as discussed in the next section.

\section{Properties of Majorana string basis} \label{sec:majorana_basis}

In this section, we describe some properties of the Majorana string basis -- another complete basis for the space of Hermitian operators that can act on $n$ fermions. We also discuss how to compute the structure constants $f_{ab}^c$ for this basis and how to convert from the Majorana string basis to the fermion string basis. The Majorana string basis, while more difficult to interpret physically than the fermion string basis, is more amenable to the computation of structure constants, making it useful for our methods.

Majorana strings, which are products of the identity operator $\hat{I}$, Majorana fermion operators $\hat{a}_i,\hat{b}_i$, and the fermion parity operator $\hat{d}_i=-i\hat{a}_i \hat{b}_i$, are defined in Eq.~(\ref{eq:majoranaStringBasis}). Like Pauli strings, Majorana strings are Hermitian, unitary, and -- excluding the identity operator -- traceless. In fact, Majorana strings and Pauli strings can be directly related to one another via the Jordan-Wigner transformation
\begin{align*}
&\hat{a}_i = \left(\prod_{j=1}^{i-1} \hat{\sigma}_j^z\right) \hat{\sigma}^x_i, \quad \quad &\hat{b}_i = \left(\prod_{j=1}^{i-1} \hat{\sigma}_j^z\right) \hat{\sigma}^y_i, \quad \quad &\hat{d}_i = \hat{\sigma}_i^z, \\
&\hat{\sigma}_i^x = \left(\prod_{j=1}^{i-1} \hat{d}_j\right) \hat{a}_i, \quad \quad &\hat{\sigma}_i^y = \left(\prod_{j=1}^{i-1} \hat{d}_j\right) \hat{b}_i, \quad \quad &\hat{\sigma}^z_i = \hat{d}_i,
\end{align*}
and so the properties of Majorana strings can be thought of as being inherited from the Pauli strings.

To understand the properties of the Majorana string basis, it is important to understand the algebraic properties of the $\hat{a}_i,\hat{b}_i, \hat{d}_i$ operators. Operators with different labels $i \neq j$, satisfy the following commutation and anti-commutation relations:
\begin{align}
[\hat{a}_i,\hat{d}_j]=[\hat{b}_i,\hat{d}_j]=[\hat{d}_i,\hat{d}_j]=0 \nonumber \\
\{\hat{a}_i,\hat{a}_j\}=\{\hat{b}_i,\hat{b}_j\}=\{\hat{a}_i,\hat{b}_j\}=0,  \label{eq:majoranaOffsiteRelations}
\end{align}
which one can derive from the canonical fermionic anti-commutation relations $\{\hat{c}_i,\hat{c}_j^\dagger\}=\delta_{ij}$ and $\{\hat{c}_i, \hat{c}_j\}=0$. One can also show that operators with identical labels $i=j$ obey the relations:
\begin{align}
\hat{a}_i \hat{b}_i = i \hat{d}_i \quad \hat{a}_i \hat{d}_i = -i\hat{b}_i \quad \hat{b}_i \hat{d}_i = i\hat{a}_i. \label{eq:majoranaOnsiteRelations}
\end{align}
Using the $(\hat{\tau}_i^0,\hat{\tau}_i^1,\hat{\tau}_i^2,\hat{\tau}_i^3)=(\hat{I},\hat{a}_i,\hat{b}_i,\hat{d}_i)$ notation, the relations from Eq.~(\ref{eq:majoranaOnsiteRelations}) can be rewritten as 
\begin{align}
\hat{\tau}_i^s \hat{\tau}_i^t = 
\begin{cases}
\delta_{st}\hat{I} + i\epsilon_{stu}\hat{\tau}_i^{u} & s,t \in \{1,2,3\} \\
\hat{\tau}^{t}_i & s=0 \\
\hat{\tau}^{s}_i & t=0
\end{cases}
\label{eq:majoranaProduct}   
\end{align}
which are identical to the same-label algebraic relations, Eq.~(\ref{eq:pauliMatrixProduct}), of the Pauli string basis.

Next, we clarify the imaginary prefactor in Eq.~(\ref{eq:majoranaStringBasis}). Note that, unlike the Pauli string basis in Eq.~(\ref{eq:pauliStringBasis}), the Majorana string basis in Eq.~(\ref{eq:majoranaStringBasis}) involves products, not tensor products, of operators. This is because the ordering of the operators are important in our calculations due to the anti-commutation relations of the $\hat{a}_i,\hat{b}_i$ operators shown in Eq.~(\ref{eq:majoranaOffsiteRelations}). We now illustrate the effect of the operator ordering with an example. Suppose that, of the $n$ sites of a Majorana string $\hat{\mathcal{S}}_a$, $m_{AB}$ of the sites have $\hat{a}_i$ or $\hat{b}_i$ operators on them so that $m_{AB}=\sum_{i=1}^n (\delta_{t_i,1} + \delta_{t_i,2})$. For example, the Majorana string  $\hat{\mathcal{S}}_a=i^{\sigma_{a}}\hat{a}_2\hat{a}_3\hat{b}_4 \hat{a}_5$ has $m_{AB}=4$. For this $\hat{\mathcal{S}}_a$ to be Hermitian, we require that $\hat{\mathcal{S}}_a^\dagger = (-i)^{\sigma_{a}}\hat{a}_5\hat{b}_4\hat{a}_3\hat{a}_2=\hat{\mathcal{S}}_a$ upon the reordering of the anti-commuting $\hat{a}_i,\hat{b}_i$ operators. In general, reversing the order of these operators can be done with $\binom{m_{AB}}{2}=m_{AB}(m_{AB}-1)/2$ swaps, which multiplies the operator by a sign $(-1)^{m_{AB}(m_{AB}-1)/2}$. Therefore, we use the convention that $\sigma_a = m_{AB}(m_{AB}-1)/2 \textrm{ mod }2$ to make $\hat{\mathcal{S}}_a$ Hermitian. Note, from Eq.~(\ref{eq:majoranaOffsiteRelations}), that the identity operator and parity operators $\hat{d}_i$ commute with operators on different sites, so they do not contribute signs like the $\hat{a}_i$ and $\hat{b}_i$ operators do.

Now, we discuss how to compute a product of Majorana strings, which will demonstrate the orthonormality of the Majorana string basis. Consider a pair of length $n$ Majorana strings
\begin{align*}
\hat{\mathcal{S}}_a &= i^{\sigma_a} \hat{\tau}_1^{s_1} \cdots \hat{\tau}^{s_n}_n \nonumber \\
\hat{\mathcal{S}}_b &= i^{\sigma_b} \hat{\tau}_1^{t_1} \cdots \hat{\tau}^{t_n}_n
\end{align*}
where the $\sigma_a,\sigma_b \in \{0,1\}$ and $\hat{\tau}_i^{t_i}$ are as defined above and $s_i,t_i \in \{0,1,2,3\}$. The product of these two operators is
\begin{align}
\hat{\mathcal{S}}_a \hat{\mathcal{S}}_b &=  (i^{\sigma_a}\hat{\tau}_1^{s_1} \cdots \hat{\tau}^{s_n}_n)(i^{\sigma_b}\hat{\tau}_1^{t_1} \cdots \hat{\tau}^{t_n}_n) \nonumber \\
&= i^{\sigma_a + \sigma_b} s_{ab} (\hat{\tau}_1^{s_1} \hat{\tau}_1^{t_1}) \cdots (\hat{\tau}_n^{s_n} \hat{\tau}_n^{t_n}) \label{eq:majoranaStringProduct}
\end{align}
where $s_{ab} = \pm 1$ is a sign picked up from reordering the $\hat{a}_i$ and $\hat{b}_i$ operators. (In practice, we compute the sign $s_{ab}$ by sorting the $\hat{a}_i$ and $\hat{b}_i$ operators in the $\hat{\mathcal{S}}_a \hat{\mathcal{S}}_b$ string with a stable sorting algorithm and counting the number of swaps performed. An odd number of swaps leads to a minus sign.) Since $\hat{a}_i,\hat{b}_i,\hat{d}_i$, and strings of these operators are traceless, we see that the only possible term with non-zero trace in the final line is the identity operator, which occurs when $\hat{\mathcal{S}}_a=\hat{\mathcal{S}}_b$. Therefore, just like the Pauli string basis, the Majorana string basis is orthonormal, satisfying $\langle \hat{\mathcal{S}}_a, \hat{\mathcal{S}}_b \rangle = \tr{\hat{\mathcal{S}}_a \hat{\mathcal{S}}_b}/\tr{\hat{I}}=\delta_{ab}$.

Next, we discuss how to compute the structure constants. From Eq.~(\ref{eq:majoranaStringProduct}), we see that the commutator of two Majorana strings is
\begin{align}
[\hat{\mathcal{S}}_a,\hat{\mathcal{S}}_b] &= \hat{\mathcal{S}}_a \hat{\mathcal{S}}_b - \hat{\mathcal{S}}_b \hat{\mathcal{S}}_a \nonumber \\
&= i^{\sigma_a + \sigma_b} s_{ab} (\hat{\tau}_1^{s_1} \hat{\tau}_1^{t_1}) \cdots (\hat{\tau}_n^{s_n} \hat{\tau}_n^{t_n}) \nonumber \\
&\quad\quad -i^{\sigma_a + \sigma_b} s_{ba} (\hat{\tau}_1^{t_1} \hat{\tau}_1^{s_1}) \cdots (\hat{\tau}_n^{t_n} \hat{\tau}_n^{s_n}) \nonumber \\
&\equiv f_{ab}^c \hat{\mathcal{S}}_c. \label{eq:majoranaCommutatorNabc}
\end{align}

After the reordering of the Majorana operators, the calculation of $f_{ab}^c$ and $\hat{\mathcal{S}}_c$ parallels the one for the Pauli string basis. To compute the structure constants $f_{ab}^c$ in the final line of Eq.~(\ref{eq:majoranaCommutatorNabc}) for a particular $(a,b)$-pair in our basis, we need to apply Eq.~(\ref{eq:majoranaProduct}). To describe this calculation, we use the same notation as we used for the Pauli strings. We say that two Majoranas strings ``agree'' on $p$ sites $i_1,\ldots,i_p$ when the operators on those sites match: $s_{i_1} = t_{i_1},\ldots,s_{i_p} = t_{i_p}$. They ``trivially disagree'' on $q$ sites $j_1,\ldots,j_q$ when the operators on those sites do not match, but one of the two operators on the site is identity: $s_{j_1} \neq t_{j_1},\ldots,s_{j_q} \neq t_{j_q}$ and $s_{j_m}=0$ or $t_{j_m}=0$ for all $m=1,\ldots,q$. Finally, the two strings ``non-trivially disagree'' on $r=n-p-q$ sites $k_1,\ldots,k_r$ when the operators do not match and are both non-identity operators: $s_{k_1} \neq t_{k_1},\ldots,s_{k_r} \neq t_{k_r}$ and $s_{k_1},t_{k_1},\ldots,s_{k_r},t_{k_r} \in \{1,2,3\}$. According to Eq.~(\ref{eq:majoranaProduct}), the sites that agree become identity operators, while the sites that disagree, both trivially and non-trivially, become $\hat{a}_i,\hat{b}_i$ or $\hat{d}_i$ operators. We then see that
\begin{align}
&[\hat{\mathcal{S}}_a, \hat{\mathcal{S}}_b] = \hat{\mathcal{S}}_a \hat{\mathcal{S}}_b - \hat{\mathcal{S}}_b \hat{\mathcal{S}}_a  \nonumber \\
&= i^{\sigma_a + \sigma_b}s_{ab} \nonumber \\
&\quad\times \left\llbracket\prod_{m=1}^q (\delta_{s_{j_m},0}\hat{\tau}^{t_{j_m}}_{j_m} + \delta_{t_{j_m},0}\hat{\tau}^{s_{j_m}}_{j_m}) \prod_{l=1}^r \left(i\epsilon_{s_{k_l}t_{k_l}u_{k_l}} \hat{\tau}^{u_{k_l}}_{k_l}\right)\right\rrbracket \nonumber\\
&\quad- i^{\sigma_a + \sigma_b}s_{ba} \nonumber \\
&\quad\times \left\llbracket\prod_{m=1}^q (\delta_{s_{j_m},0}\hat{\tau}^{t_{j_m}}_{j_m} + \delta_{t_{j_m},0}\hat{\tau}^{s_{j_m}}_{j_m}) \prod_{l=1}^r \left(i\epsilon_{t_{k_l}s_{k_l}u_{k_l}} \hat{\tau}^{u_{k_l}}_{k_l}\right)\right\rrbracket \nonumber \\
&= i^{\sigma_a + \sigma_b + r}(s_{ab} - (-1)^{r} s_{ba}) \nonumber \\
&\quad\times \left\llbracket\prod_{m=1}^q (\delta_{s_{j_m},0}\hat{\tau}^{t_{j_m}}_{j_m} + \delta_{t_{j_m},0}\hat{\tau}^{s_{j_m}}_{j_m}) \prod_{l=1}^r \left(\epsilon_{s_{k_l}t_{k_l}u_{k_l}} \hat{\tau}^{u_{k_l}}_{k_l}\right)\right\rrbracket
\label{eq:majoranaCommutatorResult}
\end{align} 
where the notation $\llbracket \cdot\rrbracket$ indicates that the bracketed $\hat{\tau}_{i}^{t_i}$ operators are ordered by their labels $i$.

From Eq.~(\ref{eq:majoranaCommutatorResult}), we see that if $s_{ab} = (-1)^{r} s_{ba}$, then the two terms cancel, leading to $[\hat{\mathcal{S}}_a,\hat{\mathcal{S}}_b] = 0$ and $f_{ab}^c = 0$ for all $c$. If $s_{ab} = - (-1)^{r} s_{ba}$, then the two terms add, resulting in a single $\hat{\mathcal{S}}_c \neq 0$. Therefore, in this second case, the structure constant associated with $\hat{\mathcal{S}}_a$ and $\hat{\mathcal{S}}_b$ is
\begin{align*}
f_{ab}^c = 2i^{\sigma_a + \sigma_b - \sigma_c + r}s_{ab} \prod_{l=1}^r \epsilon_{s_{k_l}t_{k_l}u_{k_l}}
\end{align*}
which corresponds to
\begin{align*}
\hat{\mathcal{S}}_c = i^{\sigma_c}\left\llbracket\prod_{m=1}^q (\delta_{s_{j_m},0}\hat{\tau}^{t_{j_m}}_{j_m} + \delta_{t_{j_m},0}\hat{\tau}^{s_{j_m}}_{j_m})\prod_{l=1}^r \hat{\tau}^{u_{k_l}}_{k_l}\right\rrbracket.
\end{align*}

Finally, we discuss the conversion of Majorana strings to linear combinations of fermion strings. This conversion is done by applying the definitions $\hat{a}_i = \hat{c}_i + \hat{c}_i^\dagger$, $\hat{b}_i=-i\hat{c}_i + i\hat{c}^\dagger_i$, and $\hat{d}_i = \hat{I}-2\hat{c}^\dagger_i \hat{c}_i$. Inserting these relations into Eq.~(\ref{eq:majoranaStringBasis}) and expanding, we see that we obtain $2^k$ terms made of products of $\hat{c}_i$ and $\hat{c}_i^\dagger$, where $k$ is the number of non-identity terms in the Majorana string. These terms can cancel and can be combined to form non-diagonal fermion strings, which involve Hermitian conjugates. To correctly convert these terms to the fermion strings of Eq.~(\ref{eq:fermionStringBasis}), we need to normal order our expanded operators and reorder them so that they follow our label ordering convention. We implement this process algorithmically to build up a basis transformation matrix $B_{ab}$, which is invertible but not unitary, since the fermion string operators are not orthonormal. The construction of the $B$ matrix takes time $O(d \min(2^k,d))$.

\section{Representations in the operator string basis} \label{sec:representations}

Here we discuss how to compute the representations $D_g$ of a symmetry transformation $g \in G$ in the operator string basis for a few common discrete symmetries.

For fermionic systems, spatial unitary symmetry transformations can be represented by their action on the fermionic creation and anhillation operators $\hat{c}_i^\dagger$ and $\hat{c}_i$ \cite{Chiu2016}
\begin{align}
\hat{c}_i \rightarrow \hat{c}_i' \equiv g \cdot \hat{c}_i \equiv \hat{\mathcal{U}}_g \hat{c}_i \hat{\mathcal{U}}_g^{-1} = \sum_{j} U_{ji} \hat{c}_j \label{eq:Uij}
\end{align}
where $i,j$ label lattice site degrees of freedom and $U_{ji}$ is a unitary matrix. For example, for reflection and rotation symmetries, the matrix $U_{ji}$ is a permutation matrix that specifies how lattice site labels are permuted under the transformation. Eq.~(\ref{eq:Uij}) and its generalizations provide us with a rule for how to transform fermionic operator strings $\hat{\mathcal{S}}_a$. For Majorana fermions, Eq.~(\ref{eq:Uij}) can be re-expressed as
\begin{align}
\hat{\mathcal{U}}_g \hat{a}_i \hat{\mathcal{U}}_g^{-1} = \sum_{j} U_{ji} \hat{a}_j \quad\quad \hat{\mathcal{U}}_g \hat{b}_i \hat{\mathcal{U}}_g^{-1} = \sum_{j} U_{ji} \hat{b}_j \label{eq:Uijab}.
\end{align}
Now we can see that, for a Majorana string operator made of many Majorana fermion operators, these rules specify how the string transforms. For example, the string $i \hat{a}_i \hat{b}_j$ transforms as
\begin{align*}
\hat{\mathcal{U}}_g i\hat{a}_i \hat{b}_j \hat{\mathcal{U}}_g^{-1} &= i\hat{\mathcal{U}}_g \hat{a}_i \hat{\mathcal{U}}_g^{-1} \hat{\mathcal{U}}_g\hat{b}_j \hat{\mathcal{U}}_g^{-1} \\
&= \sum_{kl} U_{ki} U_{lj} i \hat{a}_k \hat{b}_l \\
&\equiv \sum_{(kl), (ij)} U_{(kl),(ij)}' i \hat{a}_k \hat{b}_l.
\end{align*}
For a space group symmetry, this is particularly simple and the $U_{ki}, U_{lj}, U_{(kl),(ij)}'$ matrices are all permutation matrices. For non-spatial symmetry transformations, such as charge-conjugation or time-reversal symmetry, the transformations also involve changes in sign in addition to permutations (see Tab.~\ref{tab:CT}). 

For Pauli strings, spatial symmetry transformations work the same way as for Majorana strings: they simply permute the labels of the Pauli matrices. The time-reversal symmetry transformation, on the other hand, involves an additional sign for every Pauli matrix, $\hat{\mathcal{T}} \hat{\sigma}^{\alpha}_i \hat{\mathcal{T}}^{-1}=-\hat{\sigma}^{\alpha}_i$.

\begin{table}[H]
\begin{center}
\begin{tabular}{c|c|c|c|c}
Symmetry $\hat{\mathcal{U}}$ & $\hat{\mathcal{U}} \hat{a}_j \hat{\mathcal{U}}^{-1}$ & $\hat{\mathcal{U}} \hat{b}_j \hat{\mathcal{U}}^{-1}$ & $\hat{\mathcal{U}} \hat{d}_j \hat{\mathcal{U}}^{-1}$ & $\hat{\mathcal{U}} i \hat{\mathcal{U}}^{-1}$ \\
\hline
$\hat{\mathcal{T}}$ & $\hat{a}_j$ & $-\hat{b}_j$ & $\hat{d}_j$ & $-i$ \\
$\hat{\mathcal{C}}$ & $\hat{a}_j$ & $-\hat{b}_j$ & $-\hat{d}_j$ & $i$ 
\end{tabular}
\end{center}
\caption{The effect of (spinless) time-reversal $\hat{\mathcal{T}}$ and charge-conjugation $\hat{\mathcal{C}}$ symmetries on Majorana fermion operators $\hat{a}_j,\hat{b}_j$, the fermion parity operator $\hat{d}_j=-i\hat{a}_j \hat{b}_j$, and the imaginary number $i$. }
\label{tab:CT}
\end{table}

\section{QOSY library}
\label{sec:qosy}

The Quantum Operators from Symmetry (QOSY) library is a \texttt{Python} package \cite{qosy} designed for finding operators that obey a desired list of symmetries. QOSY has convenient syntax for performing such calculations with operator strings. It supports the algebraic manipulation of operators that are sums of Pauli strings, Fermion strings, or Majorana strings. Using QOSY, one can, for example, take products, commutators, or anticommutators of such operators and convert operators into different operator string bases. The core functions of QOSY are numerical implementations of the methods discussed in Section~\ref{sec:methods}: methods for finding Hamiltonians that commute (or anti-commute) with desired operators and methods for finding Hamiltonians that are invariant (or anti-invariant) under desired discrete symmetry transformations. Altogether, QOSY provides a convenient, simple set of tools for designing operators with desired symmetry properties.

\section{Derivation of Hamiltonians that commute with zero modes}
\label{sec:derivation_zm}

Here we derive a large family of Hamiltonians that commute with desired zero modes. First, we look for two-site Hamiltonians that commute with a single zero mode. Then, we look for two-site Hamiltonians that commute with a pair of zero modes. Finally, we discuss how these two-site Hamiltonians can be used to construct many-body Hamiltonians with desired zero modes.

Suppose that we wish to find Hamiltonians that commute with a single zero mode of the form $\hat{\gamma}=\sum_j (\alpha_j \hat{a}_j + \beta_j \hat{b}_j)$. On the two sites $i$ and $j$, this zero mode has support $\hat{\gamma}_{ij} = \alpha_i \hat{a}_i + \beta_i \hat{b}_i + \alpha_j \hat{a}_j + \beta_j \hat{b}_j$. To find two-site fermion-parity-conserving Hamiltonians, we construct the commutant matrix $C_{\hat{\gamma}_{ij}}$ in the $7$-dimensional basis spanned by the Majorana strings
\begin{align}
\hat{\mathcal{S}}_1,\ldots,\hat{\mathcal{S}}_7=\hat{d}_i,\, \hat{d}_j,\, i\hat{a}_i \hat{a}_j,\, i\hat{a}_i \hat{b}_j,\, i\hat{b}_i \hat{a}_j,\, i\hat{b}_i \hat{b}_j,\, \hat{d}_i\hat{d}_j. \label{eq:twositeops}
\end{align}
In this basis, the $7\times 7$ commutant matrix is 
\begin{widetext}
\begin{align*}
C_{\hat{\gamma}_{ij}} = 4
\begin{pmatrix}
\alpha_i^2+\beta_i^2 & 0 & -\alpha_j\beta_i & -\beta_i\beta_j & \alpha_i\alpha_j & \alpha_i\beta_j & 0 \\
\cdot & \alpha_j^2+\beta_j^2 & \alpha_i\beta_j & -\alpha_i\alpha_j & \beta_i\beta_j & -\alpha_j\beta_i & 0 \\
\cdot & \cdot & \alpha_i^2+\alpha_j^2 & \alpha_j\beta_j & \alpha_i\beta_i & 0 & 0 \\
\cdot & \cdot & \cdot & \alpha_i^2+\beta_j^2 & 0 & \alpha_i\beta_i & 0 \\
\cdot & \cdot & \cdot & \cdot & \alpha_j^2+\beta_i^2 & \alpha_j\beta_j & 0 \\
\cdot & \cdot & \cdot & \cdot & \cdot & \beta_i^2+\beta_j^2 & 0 \\
\cdot & \cdot & \cdot & \cdot & \cdot & \cdot & \alpha_i^2+\beta_i^2 + \alpha_j^2+\beta_j^2
\end{pmatrix}
\end{align*}
\end{widetext}
where the lower triangle of this matrix is specified by the upper triangle since it is symmetric. We find that this matrix has three null vectors and four degenerate eigenvectors with positive eigenvalue $4(\alpha_i^2+\beta_i^2+\alpha_j^2+\beta_j^2) > 0$.

For $\alpha_i,\beta_i,\alpha_j,\beta_j \neq 0$, the three null vectors correspond to the following three operators that commute with $\hat{\gamma}_{ij}$
\begin{align*}
-\beta_i \beta_j i\hat{a}_i \hat{a}_j + \alpha_i \alpha_j i\hat{b}_i \hat{b}_j - \alpha_j \beta_j \hat{d}_i + \alpha_i \beta_i \hat{d}_j \nonumber \\
\beta_i i\hat{a}_i \hat{a}_j - \alpha_i i\hat{b}_i \hat{a}_j + \alpha_j \hat{d}_i \nonumber \\
\alpha_j i\hat{a}_i \hat{b}_j - \beta_j i\hat{a}_i \hat{a}_j + \alpha_i \hat{d}_j
\end{align*}
where $i\hat{a}_i \hat{a}_j = i\hat{c}_i^\dagger \hat{c}_j + i\hat{c}_i^\dagger \hat{c}^\dagger_j + \textrm{H.c.}, i\hat{a}_i \hat{b}_j = \hat{c}_i^\dagger \hat{c}_j - \hat{c}_i^\dagger \hat{c}^\dagger_j + \textrm{H.c.}, i\hat{b}_i \hat{a}_j = -\hat{c}_i^\dagger \hat{c}_j - \hat{c}_i^\dagger \hat{c}^\dagger_j + \textrm{H.c.}$, and $i\hat{b}_i\hat{b}_j = i\hat{c}_i^\dagger \hat{c}_j - i\hat{c}_i^\dagger \hat{c}^\dagger_j + \textrm{H.c.}$. In summary, when we looked for two-site Hamiltonians that commute with \emph{one} zero mode, we found three such Hamiltonians.

Now suppose that we wish to find Hamiltonians with \emph{two} particular zero modes, $\hat{\gamma}^{(1)}=\sum_j (\alpha_j^{(1)} \hat{a}_j + \beta_j^{(1)} \hat{b}_j)$ and $\hat{\gamma}^{(2)}=\sum_j (\alpha_j^{(2)} \hat{a}_j + \beta_j^{(2)} \hat{b}_j)$, with support $\hat{\gamma}^{(1)}_{ij}$ and $\hat{\gamma}^{(2)}_{ij}$ on sites $i,j$. To find the two-site Hamiltonians that commute with both of these zero modes, we examined the null space of the sum of their commutant matrices $C_{\hat{\gamma}^{(1)}_{ij}}+C_{\hat{\gamma}^{(2)}_{ij}}$ in the same $7$-dimensional basis as before. We found a one-dimensional null space of this matrix, which corresponds to a unique two-site Hamiltonian that commutes with \emph{both} of the desired zero modes. This unique Hamiltonian, converted into complex fermions, is
\begin{align}
\hat{h}_{ij}&=\left[(\tilde{t}^R_{ij} + i\tilde{t}^I_{ij})\hat{c}^\dagger_i \hat{c}_j + (\tilde{\Delta}^R_{ij} + i\tilde{\Delta}^I_{ij})\hat{c}^\dagger_i \hat{c}^\dagger_j + \textrm{H.c.}\right] \nonumber \\
&\quad\quad+ \tilde{\mu}_{i}^{(ij)} \hat{n}_i + \tilde{\mu}_{j}^{(ij)} \hat{n}_j \label{eq:generalhij}
\end{align}
where
\begin{align}
\tilde{t}^R_{ij} &\equiv  - \alpha_i^{(1)}  \beta_j^{(2)} + \beta_i^{(1)}\alpha_j^{(2)} - \alpha_j^{(1)}  \beta_i^{(2)}  + \beta_j^{(1)} \alpha_i^{(2)}  , \nonumber \\
\tilde{t}^I_{ij} &\equiv - \alpha_i^{(1)}  \alpha_j^{(2)} - \beta_i^{(1)}  \beta_j^{(2)} + \alpha_j^{(1)} \alpha_i^{(2)}  + \beta_j^{(1)} \beta_i^{(2)} , \nonumber \\
\tilde{\Delta}^R_{ij} &\equiv - \alpha_i^{(1)}  \beta_j^{(2)} -\beta_i^{(1)} \alpha_j^{(2)}  + \alpha_j^{(1)}  \beta_i^{(2)}  + \beta_j^{(1)} \alpha_i^{(2)} , \nonumber \\
\tilde{\Delta}^I_{ij} &\equiv + \alpha_i^{(1)}  \alpha_j^{(2)} - \beta_i^{(1)}  \beta_j^{(2)} - \alpha_j^{(1)} \alpha_i^{(2)}  + \beta_j^{(1)} \beta_i^{(2)} , \nonumber \\
\tilde{\mu}_{i}^{(ij)} &\equiv 2 (\alpha_j^{(1)}  \beta_j^{(2)} - \beta_j^{(1)} \alpha_j^{(2)}), \nonumber \\
\tilde{\mu}_{j}^{(ij)} &\equiv 2 (\alpha_i^{(1)}  \beta_i^{(2)} - \beta_i^{(1)} \alpha_i^{(2)}). \label{eq:generalhijparams}
\end{align}
Note that if we consider the zero mode coefficients as vectors $\vec{\gamma}^{(1)} \equiv (\alpha_i^{(1)}, \beta_i^{(1)}, \alpha_j^{(1)}, \beta_j^{(1)})^T$ and $\vec{\gamma}^{(2)} \equiv(\alpha_i^{(2)}, \beta_i^{(2)}, \alpha_j^{(2)}, \beta_j^{(2)})^T$, then each of the parameters of the Hamiltonian are indefinite quadratic forms, e.g., $\tilde{t}_{ij}^R = {\vec{\gamma}^{(1)}}^T A \vec{\gamma}^{(2)}$ for a particular antisymmetric matrix $A$.

Next, we checked if the operator $\hat{h}_{ij}$ commuted with zero modes other than $\hat{\gamma}^{(1)}$ and $\hat{\gamma}^{(2)}$. We performed this check by computing the commutant matrix $C_{\hat{h}_{ij}}$ in the four-dimensional basis spanned by the Majorana fermions $\hat{a}_i,\hat{b}_i,\hat{a}_j,\hat{b}_j$. For the bond operator specified by Eqs.~(\ref{eq:generalhij})~and~(\ref{eq:generalhijparams}), the commutant matrix has a null space that is exactly two-dimensional and spanned precisely by the $\hat{\gamma}^{(1)}_{ij}$ and $\hat{\gamma}^{(2)}_{ij}$ zero modes. The two remaining eigenstates of the $4 \times 4$ $C_{\hat{h}_{ij}}$ matrix are degenerate with eigenvalue $4\Delta\varepsilon_{ij}$, where
\begin{align}
\Delta\varepsilon_{ij} &= ({\alpha_i^{(1)}}^2+{\beta_i^{(1)}}^2+{\alpha_j^{(1)}}^2+{\beta_j^{(1)}}^2) \nonumber \\
&\quad\quad\times({\alpha_i^{(2)}}^2+{\beta_i^{(2)}}^2+{\alpha_j^{(2)}}^2+{\beta_j^{(2)}}^2) \nonumber\\
&\quad -({\alpha_i^{(1)}}{\alpha_i^{(2)}}+{\beta_i^{(1)}}{\beta_i^{(2)}}+{\alpha_j^{(1)}}{\alpha_j^{(2)}}+{\beta_j^{(1)}}{\beta_j^{(2)}})^2 \nonumber \\
&= ||\hat{\gamma}^{(1)}_{ij}||^2||\hat{\gamma}^{(2)}_{ij}||^2 - |\langle \hat{\gamma}^{(1)}_{ij} , \hat{\gamma}^{(2)}_{ij} \rangle|^2 \label{eq:gap}
\end{align}
This eigenvalue gap is non-negative ($\Delta\varepsilon_{ij} \geq 0$) by the Cauchy-Schwarz inequality and is positive as long as $\hat{\gamma}^{(1)}_{ij}\not \propto \hat{\gamma}^{(2)}_{ij}$. It is largest when the two zero modes are orthogonal on sites $i$ and $j$.

Finally, we discuss how the $\hat{h}_{ij}$ bond operator can be used as a building block to construct Hamiltonians that commute with desired zero modes. Since $\hat{h}_{ij}$ is even in fermionic operators, $[\hat{h}_{ij}, \hat{a}_k]=[\hat{h}_{ij}, \hat{b}_k]=0$ for $k \neq i,j$. The above derivation showed that $[\hat{h}_{ij},\hat{\gamma}^{(1)}_{ij}]=[\hat{h}_{ij},\hat{\gamma}^{(2)}_{ij}]=0$. Together, these two facts imply that $[\hat{h}_{ij},\hat{\gamma}^{(1)}]=[\hat{h}_{ij},\hat{\gamma}^{(2)}]=0$ for the entire zero modes $\hat{\gamma}^{(1)}$ and $\hat{\gamma}^{(2)}$.

Let us examine what happens when we attempt to construct an $N$-site Hamiltonian made of only the bond operators
\begin{align}
\hat{H}_{ZM} = \sum_{ij} J_{ij} \hat{h}_{ij}.
\end{align}
In this procedure, imagine that we have beforehand decided on a desired pair of zero modes, $\hat{\gamma}^{(1)}, \hat{\gamma}^{(2)}$, so that we have specified the $\alpha_k^{(1)}, \beta_k^{(1)}, \alpha_k^{(2)}, \beta_k^{(2)}$ parameters for all $k=1,\ldots,N$. This in turn specifies all of the $\hat{h}_{ij}$ operators. To avoid the case where $\Delta\varepsilon_{ij}=0$, we also require that $\hat{\gamma}^{(1)}_{ij}\not \propto \hat{\gamma}^{(2)}_{ij}$ for the $(i,j)$ pairs that we consider below.

Suppose that we start building our Hamiltonian from the zero operator, so that $J_{ij}=0$ for all $i,j$. In this case, there are $2N$ zero modes that (trivially) commute with $\hat{H}_{ZM}=0$: $\hat{a}_k,\hat{b}_k$ for $k=1,\ldots,N$. Note that each site has two zero modes and each bond has four. Next, suppose that we turn on a bond $J_{lm}\neq 0$ for a particular pair of sites $l,m$. On sites $l,m$, the bond operator gaps out two of the four zero modes, so that only \emph{two} zero modes $\hat{\gamma}^{(1)}_{lm}, \hat{\gamma}^{(2)}_{lm}$ on sites $l,m$ commute with the Hamiltonian. After laying down the first bond, there are $2N-2$ zero modes. Moreover, the zero modes on sites $l,m$ are now constrained to locally match $\hat{\gamma}^{(1)}$ and $\hat{\gamma}^{(2)}$. Now, consider iterating the procedure and laying down one bond at a time. If we connect the bonds together, e.g., so that $J_{lm},J_{mn},J_{np}\neq0$, then each bond we lay down eliminates two zero modes from the system and acts as a constraint that forces the zero modes on those sites to match our desired zero modes. If we think of the non-zero $J_{ij}$ as being the edges of a graph, then we can see that we are building a connected component into the graph and that the only zero modes that commute with the Hamiltonian on that connected component are constrained to match $\hat{\gamma}^{(1)}$ and $\hat{\gamma}^{(2)}$. In general, if we build the $\hat{H}_{ZM}$ Hamiltonian to have $N_C$ connected components, then there will be $2 N_C$ zero modes, each of which are ``pieces'' of the $\hat{\gamma}^{(1)}$ and $\hat{\gamma}^{(2)}$ zero modes. If $N_C=1$, i.e., the graph is connected, then the only two zero modes are exactly the entire $\hat{\gamma}^{(1)}$ and $\hat{\gamma}^{(2)}$ operators.

\section{Level-spacing statistics of perturbed spin Hamiltonians}
\label{sec:levelspacing}

The Hamiltonians discussed in Sec.~\ref{sec:z2spinliquids} have many integrals of motion. Using the eigenstates of these integrals of motion, we block diagonalized the Hamiltonians according to their quantum number sectors and analyzed the level-spacing statistics in particular sectors. Many of these Hamiltonians have significant eigenstate degeneracy in these sectors, which makes analysis of the level-spacing statistics unreliable or inconclusive. We observed that particular perturbations of these Hamiltonians possess the same integrals of motions as the unperturbed Hamiltonians, though at the cost of breaking spatial symmetries. These perturbed Hamiltonians, however, have almost no degenerate eigenstates, which allows us to gather more reliable level-spacing statistics. Below, we describe how we computed the level-spacing statistics of these perturbed Hamiltonians while accounting for as many integrals of motion as possible.

On the square lattice, we analyzed the perturbed Hamiltonian of Eq.~(\ref{eq:squareperturbed}). Like the unperturbed model, the perturbed model commutes with straight-line $Z$ Wilson loops and products of two straight-line $X$ loops, which are listed in Eq.~(\ref{eq:squareioms}). 

We diagonalized these Hamiltonians in a basis of states that are eigenstates of the known integrals of motion. Here we describe how we found these eigenstates. The $+1$ eigenstates of the $\hat{Z}_{\mathcal{L}}$ integrals of motion, for example, are simply product state spin configurations $\ket{S}$ that satisfy $\hat{Z}_{\mathcal{L}}\ket{S}=+\ket{S}$, i.e., spin configurations that have an even number of down spins on the sites of loop $\mathcal{L}$. The $+1$ eigenstates $\ket{S'}$ of the $\hat{X}_{\mathcal{L}}\hat{X}_{\mathcal{L}'}$ operators can be constructed from spin configurations $\ket{S}$ in the following way: $\ket{S'}=\frac{1}{\sqrt{2}}(\hat{I} + \hat{X}_{\mathcal{L}}\hat{X}_{\mathcal{L}'})\ket{S}$. Using these two facts, we are able to construct all $+1$ eigenstates of the $\hat{Z}_{\mathcal{L}}$ and $\hat{X}_{\mathcal{L}}\hat{X}_{\mathcal{L}'}$ integrals of motion (with slight modification, we can build $-1$ eigenstates as well). In addition to the $Z$ and $X$ loop integrals of motion, there is also a global integral of motion $\hat{B}$, slightly modified from the operator in Eq.~(\ref{eq:squareglobaliom}), which is a sum of the terms of Hamiltonian Eq.~(\ref{eq:squareperturbed}) that lie on $B$-sublattice squares. We determine the eigenstates of $\hat{B}$ by perturbing the Hamiltonian by $\hat{B}$. In particular, instead of diagonalizing $\hat{H}_{\textrm{square}} + \epsilon \delta \hat{H}$, we diagonalize $\hat{H}_{\textrm{square}} + \epsilon \delta \hat{H} + \lambda \hat{B}$ for a large constant $\lambda$. The $\hat{B}$-perturbation separates out the eigenstates of $\hat{B}$ so that they are far away from one another in energy. This allows us to numerically identify eigenstates with the same eigenvalue of $\hat{B}$.

We also attempted to account for integrals of motion that we were not able to identify directly. Such unknown integrals of motion, if left unaccounted for, would give rise to Poisson level-spacing statistics. This would occur when neighboring energy levels $E_n, E_{n+1}$ are in different quantum number sectors of these integrals of motion. Note that here we are considering a family of perturbed Hamiltonians, $\hat{H}_0 + \epsilon \delta\hat{H}$, where $[\hat{H}_0,\delta\hat{H}]\neq 0$. If there is a hidden integral of motion $\hat{\mathcal{O}}$ that commutes with $\hat{H}_0 + \epsilon \delta\hat{H}$ for all $\epsilon$, then $[\hat{H}_0,\hat{\mathcal{O}}]=[\delta\hat{H},\hat{\mathcal{O}}]=0$. This means that one can block diagonalize $\delta \hat{H}$ according to the eigenstates of $\hat{\mathcal{O}}$ and that eigenstates $\ket{\psi_1},\ket{\psi_2}$ of $\hat{\mathcal{O}}$ with different eigenvalues satisfy $\langle \psi_2 |\delta\hat{H}|\psi_1\rangle = 0$. Moreover, energy eigenstates $\ket{\psi_j}$ of $\hat{H}_0 + \epsilon \delta\hat{H}$ that are non-degenerate are also eigenstates of $\hat{\mathcal{O}}$. Using these observations, we computed the perturbation overlaps $\delta H_{ij}=\langle \psi_i |\delta\hat{H}|\psi_j\rangle$ in the basis of $\{\ket{\psi_j}\}$ obtained with ED. We then reordered the states so as to block diagonalize the $\delta H_{ij}$ matrix. When block diagonalizing $\delta H_{ij}$, we considered entries of the matrix smaller than $10^{-6}$ to be zero. The eigenstates within the same block were considered as a  ``sector'' of the hidden integrals of motion. We performed our level-spacing ratio calculations using the energies in these sectors, which are also contained within the known integral of motion quantum number sectors mentioned above.

\begin{figure}
    \begin{center}
    \includegraphics[width=0.48\textwidth]{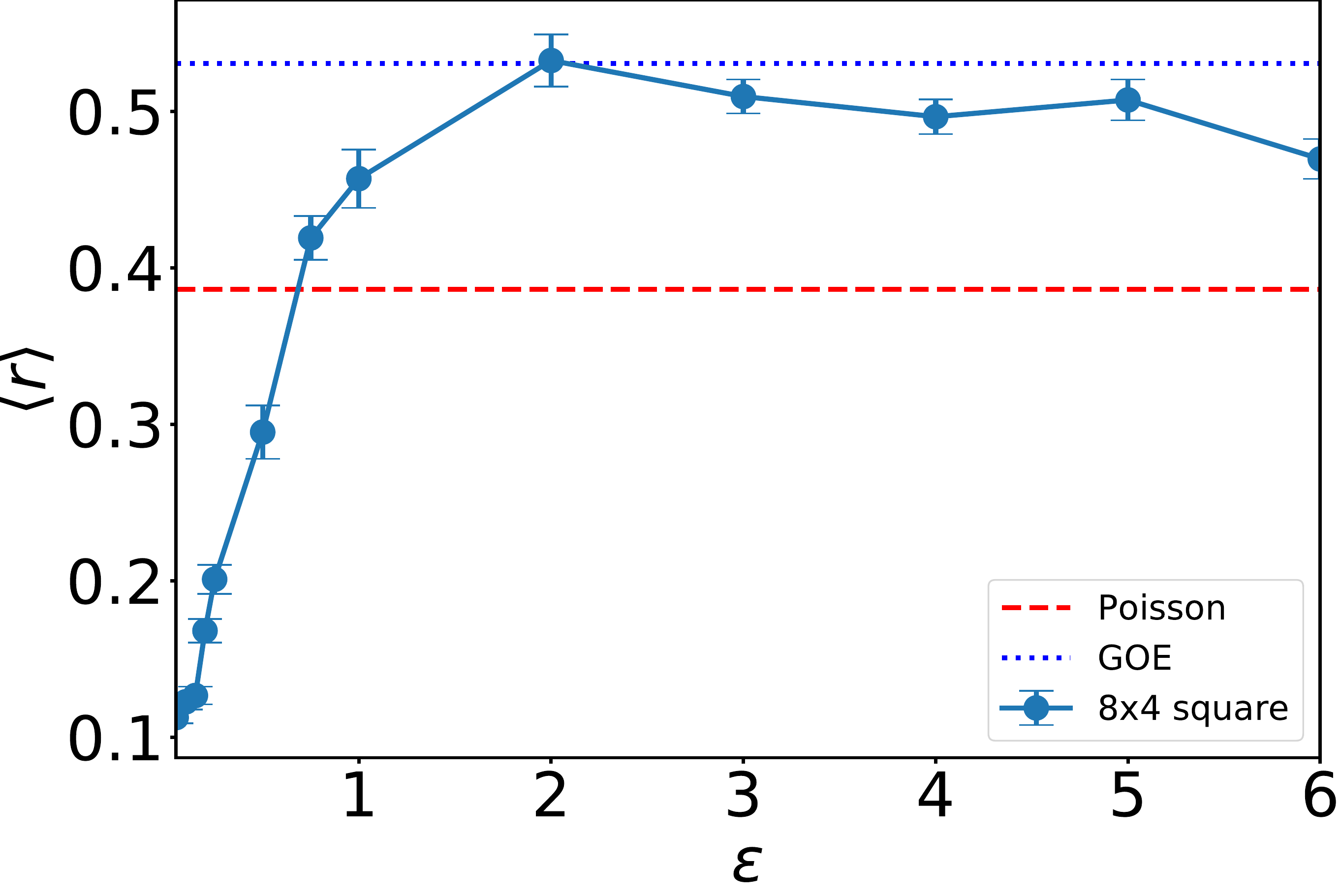}
    \end{center}
    \caption{The average level-spacing ratios versus disorder strength $\epsilon$ for the Hamiltonian Eq.~(\ref{eq:squareperturbed}) on an 32-site square lattice (8x4 square) averaged over 10 random Hamiltonians. The energy levels considered were obtained in particular quantum number sectors, as described in the main text.}
    \label{fig:levelspacingratiossquare}
\end{figure}

Accounting for both known and unknown integrals of motion as described above, we computed the level-spacing ratios of $4096$ eigenvalues in the $+1$ sectors of the $Z$ loops and $X$ product loops for the Hamiltonians of Eq.~(\ref{eq:squareperturbed}) on an $8 \times 4$ square lattice. We did this for 10 random realizations of the perturbed Hamiltonians for $\epsilon$ from $0.05$ to $2$. After accounting for the global integral of motion, the energy eigenstates typically separated into sectors of 64 states. After accounting for the hidden integrals of motion, these 64 states separated further into three sectors containing 28 states, 35 states, and 1 state. We computed the level-spacing ratios between states of neighboring energies in the 28 and 35-dimensional sectors and averaged the results over all such sectors and over random realizations of the Hamiltonians. The resulting average level-spacing ratios for $\epsilon$ from 0 to 6 are depicted in Fig.~\ref{fig:levelspacingratiossquare}.

On the kagome lattice, we analyzed the perturbed Hamiltonians of Eqs.~(\ref{eq:kagomeperturbed1})~and~(\ref{eq:kagomeperturbed2}). Both of these Hamiltonians commute with $Z$ Wilson loops and products of two $X$ loops. However, the Hamiltonian of Eq.~(\ref{eq:kagomeperturbed1}) also commutes with the local hexagon integrals of motion $\hat{Z}_{\hexagon}$, while Eq.~(\ref{eq:kagomeperturbed2}) is a generic Hamiltonian that does not. 

Using the same techniques described above to account for known and unknown integrals of motion, we performed ED on the Hamiltonians of Eq.~(\ref{eq:kagomeperturbed1})-(\ref{eq:kagomeperturbed2}) for $3\times 2 \times 3=18$ site and $3 \times 3 \times 3 = 27$ site kagome lattices. For the Eq.~(\ref{eq:kagomeperturbed1}) Hamiltonians on the 18-site lattice, we performed ED in a 256-dimensional basis corresponding to the $+1$ quantum number sector of each of the Wilson loops and hexagon local integrals of motion. We were not able to identify any ``hidden'' quantum number sectors in this 256-dimensional space and found that all energies in this sector were unique. For the Eq.~(\ref{eq:kagomeperturbed1}) Hamiltonians on the 27-site lattice, we performed ED in a 8192-dimensional basis. In this case, we did find ``hidden'' integral of motion sectors that were 2048-dimensional and contained no degeneracies. For the Eq.~(\ref{eq:kagomeperturbed2}) Hamiltonians on the 18-site lattice, we performed ED in a 1024-dimensional basis. In this case, the global integral of motion (a modified version of operator $\hat{D}$ from Eq.~(\ref{eq:kagomeglobaliom}) that only contains the hexagon interactions of the Hamiltonian) split the space into three sectors, which are 256, 256, and 512-dimensional. Interestingly, while the first two sectors have no degeneracies, the third sector is doubly-degenerate in each of its energy eigenstates. In this case, we ignored the degeneracy when computing the level-spacing statistics. We did not find any hidden integral of motion sectors for these Hamiltonians. For the Hamiltonians of Eq.~(\ref{eq:kagomeperturbed1})-(\ref{eq:kagomeperturbed2}) at a particular disorder strength $\epsilon$, we averaged the level-spacing ratios $\langle r \rangle$ of the sectors over many random realizations of the random variables $h_{\triangle}$ (100 realizations for the 18-site kagome lattice and 10 realizations for the 27-site lattice). The average level-spacing ratios as a function of $\epsilon$ are shown in Fig.~\ref{fig:levelspacingratios}.

\bibliography{refs}

\end{document}